\documentclass[12pt]{article}
\pdfoutput=1
\usepackage{jheppub}
\usepackage{graphics}
\usepackage{graphicx}
\usepackage{pstricks}
\usepackage{multirow}

\newcommand{\nutau}{{\nu_{\tau}}}

\newcommand{\gev}{{\rm\ GeV}}
\newcommand{\be}{\begin{eqnarray}}
\newcommand{\beq}{\begin{equation}}
\newcommand{\eeq}{\end{equation}}
\newcommand{\ee}{\end{eqnarray}}

\newcommand{\bmp}{\noindent\begin{minipage}{16cm}}
\newcommand{\emp}{\end{minipage}\vskip 7mm} 

\newcommand{\gsim} {\buildrel > \over {_\sim}}
\usepackage{bm}
\usepackage{amsmath}
\usepackage{amsfonts}
\usepackage{bbm}

\title{Prompt atmospheric neutrino fluxes: perturbative QCD models
and nuclear effects}

\author[a,b]{Atri Bhattacharya,}
\author[c]{Rikard Enberg,}
\author[d,e]{Yu Seon Jeong,}
\author[d]{C. S. Kim,}
\author[f]{\\ Mary Hall Reno,}
\author[a,g]{Ina Sarcevic,}
\author[h]{Anna Stasto}

\affiliation[a]{Department of Physics, University of Arizona, 1118 E.\ 4th St.\ Tucson, AZ 85704}
\affiliation[b]{Space sciences, Technologies and Astrophysics Research (STAR) Institute,
                Universit\'{e} de Li\`{e}ge, B\^{a}t.~B5a, 4000 Li\`{e}ge,
                Belgium}
\affiliation[c]{Department of Physics and Astronomy, Uppsala University,
Uppsala, Sweden}
\affiliation[d]{Department of Physics and IPAP, Yonsei University, Seoul 03722, Korea}
\affiliation[e]{National Institute of Supercomputing and Networking, KISTI, Daejeon 34141, Korea}
\affiliation[f]{Department of Physics and Astronomy, University of Iowa, Iowa City, Iowa 52242}
\affiliation[g]{Department of Astronomy, University  of Arizona, 933 N.\ Cherry Ave.,
  Tucson, AZ 85721}
\affiliation[h]{Department of Physics, The Pennsylvania State University, University Park, PA 16802}

\emailAdd{a.bhattacharya@ulg.ac.be}
\emailAdd{rikard.enberg@physics.uu.se}
\emailAdd{ysjeong@kisti.re.kr}
\emailAdd{cskim@yonsei.ac.kr}
\emailAdd{mary-hall-reno@uiowa.edu}
\emailAdd{ina@physics.arizona.edu}
\emailAdd{ams52@psu.edu}

\date{\today}
\abstract{
We evaluate the prompt atmospheric neutrino flux at high energies using three different frameworks for
calculating the heavy quark production cross section in QCD: NLO perturbative QCD,
$k_T$ factorization including low-$x$ resummation, and the dipole model including
parton saturation. We use QCD parameters, the value for the
charm quark mass and the range for the
factorization and renormalization scales that provide the
best description of the
total charm cross section measured at fixed target experiments,
 at RHIC and at LHC.  Using these parameters we calculate
 differential cross sections for charm and bottom production and compare
with the latest data on forward charm meson production from
LHCb at $7$ TeV and at $13$ TeV, finding good agreement with the data.
In addition, we investigate the role of nuclear shadowing by
including nuclear parton distribution functions (PDF) for the
target air nucleus using two different nuclear PDF schemes.
Depending on the scheme used, we find the reduction of the flux
due to nuclear effects varies from $10\%$ to $50 \%$ at
the highest energies.
Finally, we compare our results with the IceCube
limit on the prompt neutrino flux, which is already providing
 valuable information about some of the QCD models.}

\begin{document}
\maketitle
\flushbottom

\section{Introduction}

Measurements of high-energy extraterrestrial neutrinos by the IceCube Collaboration \cite{Aartsen:2013bka,IceCube} have heightened
interest in other sources of high-energy neutrinos. A background to neutrinos from astrophysical sources are neutrinos produced
in high energy cosmic ray interactions with  nuclei in the Earth's atmosphere. While pion and kaon production and decay dominate the low energy ``conventional'' neutrino flux
\cite{Honda:2006qj,Barr:2004br,Gaisser:2014pda}, short-lived charmed hadron decays to neutrinos dominate the ``prompt'' neutrino flux
\cite{Gondolo:1995fq, Pasquali:1998ji,Pasquali:1998xf,Martin:2003us,ERS,Bhattacharya:2015jpa,Garzelli:2015psa,Gauld:2015yia,Gauld:2015kvh} at high energies.
The precise cross-over energy where the prompt flux dominates the conventional flux depends on the zenith angle and  is somewhat obscured by
the large uncertainties in the prompt flux. The astrophysical flux
appears to dominate the atmospheric flux at an energy of $E_\nu\sim 1$
PeV.
Atmospheric neutrinos come from  hadronic
interactions which occur at much higher energy.  With the prompt neutrino carrying about a third of the
parent charm energy $E_c$, which in turn carries about 10\%  of the
incident cosmic ray nucleon energy $E_{CR}$, the relevant center of
mass energy
for the $pN$ collision that produces $E_\nu\sim 1$ PeV is $\sqrt{s}\sim 7.5$ TeV, making a connection to LHC experiments, e.g., \cite{Aaij:2013mga,Aaij:2015bpa}.

There are multiple approaches to evaluating the prompt neutrino flux. The standard approach is to  use NLO perturbative QCD (pQCD) in the collinear approximation with the integrated  parton distribution functions (PDFs) and to evaluate the  heavy quark pair production which is dominated by the elementary gluon fusion process
\cite{Nason:1987xz,Nason:1989zy,Mangano:1991jk}.
Such calculations were performed in  \cite{Pasquali:1998ji,Pasquali:1998xf,Martin:2003us} (see also \cite{Gondolo:1995fq}).
Recent work to update these predictions using modern PDFs and
models of the incident cosmic ray energy spectrum and composition
appears in \cite{Bhattacharya:2015jpa},
and including accelerator physics Monte Carlo interfaces, in
\cite{Garzelli:2015psa,Gauld:2015yia,Gauld:2015kvh}.
Using $x_c=E_c/E_{CR}\sim 0.1$ for charm production,
one can show that high energies require gluon PDF with longitudinal momentum fractions $x_1\sim x_c$ and $x_2\sim 4m_c^2/(x_c s ) \ll x_1$. For a factorization scale $M_F\sim
0.5-4 m_c$, this leads to large uncertainties. In addition, due to the small $x$ of the gluon PDFs in the target  one may need to address the  resummation of large logarithms at low $x$.

In particular, comparisons with LHCb data at $7$ TeV \cite{Aaij:2013mga} were used in
ref.~\cite{Gauld:2015kvh} to reduce uncertainties in pQCD
calculation  (see also ref.~\cite{Cacciari:2015fta}).
Using FONLL \cite{Cacciari:1993mq,Cacciari:1998it,Cacciari:2001td,Nason:2004rx}
predictions for
the $p_T$ distribution of charm mesons obtained with
different PDFs, they have shown that LHCb data for
$D$ mesons and $B$ mesons can reduce the theoretical uncertainty
due to the choice of scales to $10\%$ and
the uncertainty due to the PDF by as much as a factor of
2 at high energies in the region of large rapidity and small
$p_T$. Still, the uncertainty due to the low $x$ gluon PDF remains relatively large.

Given the fact that the gluon PDF is probed at very small values of $x$, it is important to investigate approaches that resum large logarithms $\ln (1/x)$ and that can incorporate other novel effects in this regime, such as parton saturation.  Such effects are naturally incorporated in the so-called dipole model approaches \cite{Nikolaev:1990ja,
mueller, Nikolaev:1995ty, gbw, forshaw, sgbk, bgbk, Kopeliovich:2002yv, Raufeisen:2002ka, kowalski,
iim, Goncalves:2006ch,soyez,albacete,aamqs,gkmn, Albacete:2009fh, Albacete:2010sy, Ewerz:2011ph, Jeong:2014mla, Block:2014kza}
and within the $k_T$ (or high energy) factorization framework \cite{Catani:1990eg,Collins:1991ty,Levin:1991ya,Ryskin:1995sj}.

There is another major source of uncertainty in the low $x$ region.
The target air nuclei have an average nucleon number of $\langle A\rangle=14.5$.
Traditionally in the perturbative approach, the nuclear effects are entirely neglected
and a linear scaling with $A$ is used  for the cross section.
Nuclear shadowing effects, however, may be not negligible at very low $x$ and low factorization scale.

In the present  paper, we expand our previous work (BERSS)
\cite{Bhattacharya:2015jpa} to include nuclear effects in the target and analyze the impact of the low $x$ resummation and saturation effects on the prompt neutrino flux.

We incorporate nuclear effects in the target PDFs
by using in our perturbative calculation two different sets of nuclear parton distribution
functions: nCTEQ15 \cite{Kovarik:2015cma} and EPS09 \cite{Eskola:2009uj}.
As there is no nuclear data in the relevant energy regime, these nuclear PDFs are largely unconstrained in the low $x$ region ($x<0.01$) and there is a substantial uncertainty associated with nuclear effects.
Nevertheless, for charm
production, the net effect is a suppression of the  cross section and the corresponding neutrino flux. At $E_\nu=10^6$ GeV, the central values of the nCTEQ PDF yields a flux as low as
$\sim 73\%$ of the flux evaluated with free nucleons in the target,
while the corresponding reduction from the EPS09 PDF is at the level of $ 10\% $.

We also show our results using the dipole approach, with
significant theoretical  improvements with respect to our previous
work (ERS) \cite{ERS}.  These include models of the dipole cross sections that are updated to include more precise experimental data. Furthermore, we calculate the prompt
neutrino flux in the $k_T$ factorization approach, using unintegrated gluon distribution functions with low $x$ resummation and also with saturation effects. We compare these calculations to the dipole and NLO pQCD results.

 Overall we find that for all calculations, there is a consistent
description of the total charm cross section at high energies,
for $pp$ and $pN$ production of $c\bar{c}$.  We also evaluate
the  $b\bar{b}$ cross section and the contribution of
beauty hadrons to the atmospheric lepton flux.
For each approach we find that our choice for theoretical parameters is 
in agreement with the
 latest  LHCb data \cite{Aaij:2013mga,Aaij:2015bpa} on charm transverse
momentum and
rapidity distributions in the  forward region,
and the total cross sections at $7$ TeV and at $13$ TeV.

In addition to including nuclear and low $x$ effects, we also
consider four different cosmic ray fluxes \cite{Gaisser:2012zz,Gaisser:2013bla,Stanev:2014mla}  and show
how the prompt neutrino flux strongly depends on the choice of
the primary cosmic ray flux.

The present paper is organized as follows.
In the next section we present calculations of the
total and differential charm cross section.
We present comparisons of all three approaches, pQCD,
dipole model and $k_T$ factorization, and we show the impact of
nuclear effects on the total charm cross sections.
We show comparisons of our theoretical results
with the rapidity distributions measured at
LHCb energies. In sec.~3 we compute neutrino fluxes
for muon and tau neutrinos and compare them with
the IceCube limit. Finally, in sec.~4 we state our conclusions.
Detailed formulas concerning the fragmentation
functions and meson decays are collected in the Appendix.

\section{Heavy quark cross sections}
%
%
\subsection{NLO perturbation theory}

We start by expanding our recent work on
heavy quark cross section with NLO
perturbation theory \cite{Bhattacharya:2015jpa} to constrain QCD
parameters by comparison with RHIC and LHC data,  and by
including nuclear effects.  In particular, we shall compare the
results of the calculation with the latest LHCb data on forward
production of charm mesons \cite{Aaij:2013mga,Aaij:2015bpa}. Gauld
et~al.~\cite{Gauld:2015yia,Gauld:2015kvh}
have evaluated charm forward production to constrain gluon PDFs and to
compute the prompt atmospheric lepton flux.
 Garzelli et~al.~\cite{Garzelli:2015psa} have  recently
evaluated the total charm
production cross section at NNLO and
used the NLO differential charm cross section
to evaluate the prompt
atmospheric lepton flux. Below, we discuss differences between our
approaches to evaluating first charm production, then the prompt fluxes.

We use the HVQ computer program to evaluate the energy distribution of the
charm quark at NLO in pQCD \cite{Nason:1987xz,Nason:1989zy,Mangano:1991jk}.
The resummation of  logarithms associated with large transverse
momentum $p_T$ as incorporated by the FONLL calculation
\cite{Cacciari:1993mq,Cacciari:1998it,Cacciari:2001td}  is not
necessary
for this application since the low $p_T$ kinematic region dominates
the cross section.

For heavy quark production, one important parameter is the charm quark mass.
In ref.~\cite{Garzelli:2015psa},
neutrino fluxes were evaluated using NLO
QCD on free nucleon targets with $m_c=1.40$ GeV taken as the
 central choice of charm quark mass, based on the
pole mass value of $m_c=1.40\pm 0.15$ GeV.  Values of
$m_c=1.5\pm 0.2$ GeV are used in
refs.\ \cite{Gauld:2015yia,Gauld:2015kvh}.
In our work, we use
the running charm quark mass of $m_c=1.27$ GeV, which is consistent with the average value quoted in  \cite{Agashe:2014kda}, $m_c(m_c)=1.275
\pm 0.025$ GeV. A direct translation between the pole mass and running
mass is
not possible because of poor convergence of the perturbative series,
as discussed in, e.g., ref.~\cite{Marquard:2015qpa}.
By using $m_c=1.27$ GeV,
we can  make use of the data-constrained analysis of the factorization and
renormalization scale dependence discussed in
ref.~\cite{Nelson:2012bc}.

The mass dependence enters through the renormalization and
factorization scale dependence as well as through the kinematic
threshold. By
keeping the values of the
factorization and renormalization scales
fixed and only varying the charm mass dependence in the matrix
element and phase space integration,
one can show that there is a strong dependence on mass at
low incident beam energies,
but at higher energies, the mass dependence is much weaker. For
example,
keeping the renormalization and factorization scales fixed at
$M_R=M_F=2.8$ GeV,
the cross section $\sigma(pp\to c\bar{c}X)$  with $m_c=1.27$ GeV  is a
factor of only 1.26--1.16 larger than the cross section with $m_c=1.4$
GeV for incident proton beam energies of
$10^6-10^{10}$ GeV. The uncertainties due to the choice of
scales are
larger than those due to the mass variation. We discuss below the impact of the
scale variations on both the cross section and prompt fluxes.

For the NLO pQCD $b\bar{b}$ contribution to the prompt flux, we use a fixed
value of $m_b=4.5$ GeV and consider the same range of scale factors as
for
$c\bar{c}$ production.

In the perturbative calculation of the heavy quark pair
production cross section in cosmic ray interactions with air nuclei
with $\langle A\rangle = 14.5$,
one has to take into account  the fact that the
nucleons are bound in nuclei, as opposed to free nucleons.
Nuclear effects can result in both suppression and
enhancement of the nuclear parton distribution functions (nPDF) relative to the free nucleon PDF, depending on the kinematic variables $(x,Q)$. The extraction of nPDF at NLO has been done by several groups,  among them
Eskola, Paukkunen and Salgado (EPS09) \cite{Eskola:2009uj} and Kovarik et al.\ (nCTEQ15)
\cite{Kovarik:2015cma}. The nuclear PDFs in the
EPS framework \cite{Eskola:2009uj}
 are defined by a nuclear modification factor multiplying the free
 proton PDFs. For example, the up quark PDF for the quark in a proton
 bound
in nucleus $A$ is
\begin{equation}
u_A(x,Q) = R_u^A(x,Q)\, u_p(x,Q) \, ,
\end{equation}
where $R_u^A(x,Q)$ is the nuclear modification factor to the free proton PDF.
For our calculations, we use the central CT14 NLO \cite{Dulat:2015mca}
PDF for free protons and to approximate nitrogen targets, the EPS09
NLO results for oxygen.
The recent nCTEQ15 PDF sets \cite{Kovarik:2015cma} instead provide
directly the parton distribution functions for partons in protons
bound
in the nucleus, e.g.,
$u_A(x,Q)$. As usual, one uses isospin symmetry to account for
neutrons bound in nuclei.
For the calculation in this work we take as our standard free proton PDFs those
of
\cite{Kovarik:2015cma}, labeled here as nCTEQ15-01, and
PDFs for nucleons in nitrogen, labeled here as nCTEQ15-14.

\begin{figure}
\centering
\includegraphics[width=0.7\textwidth]{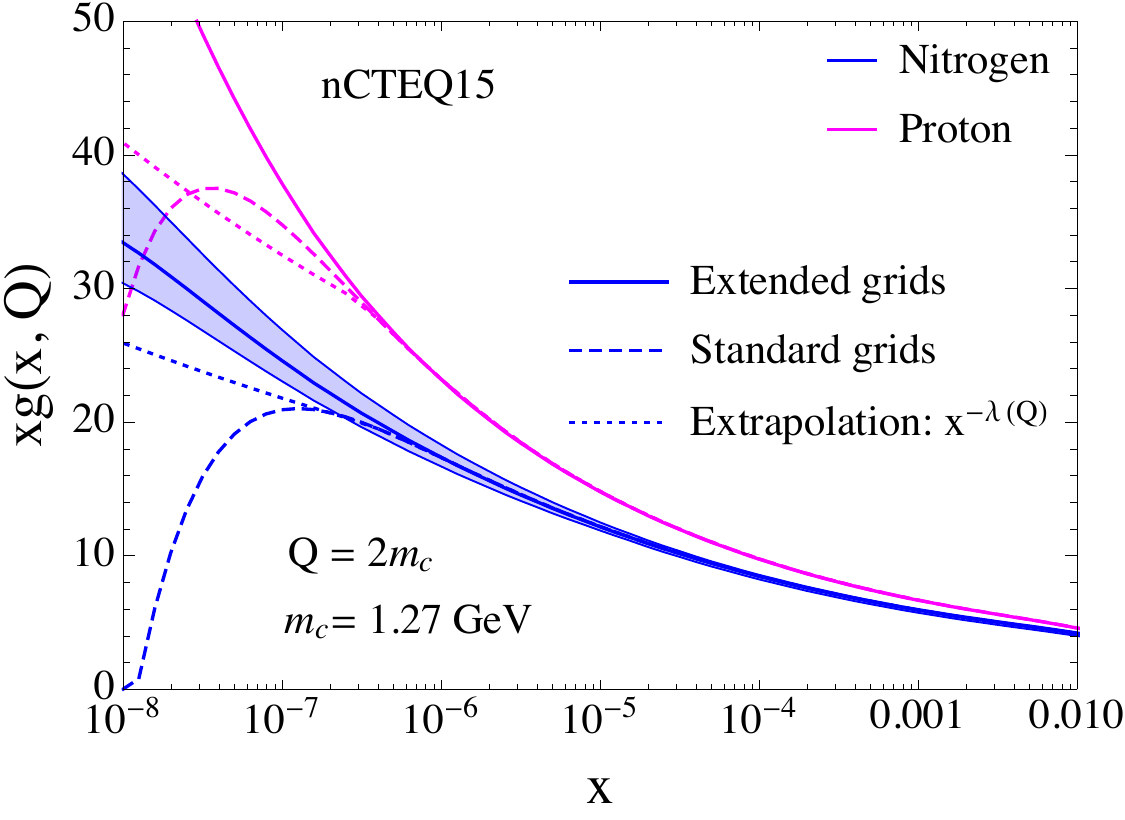}
\caption{The gluon distribution functions for free protons (upper, magenta) and
  isoscalar nucleons bound in nitrogen (lower, blue) in the
nCTEQ15 PDF sets \cite{Kovarik:2015cma} with $Q=2m_c$. The standard
distribution of the PDF sets are shown with dashed lines. Small-$x$
extrapolations with $xg(x,Q)\sim x^{-\lambda(Q)}$ for $x<10^{-6.5}$
are shown with dotted lines. The solid lines show PDFs with grids
extended to treat the small-$x$ regime \cite{Olness-email}, with a
shaded band to show the range of predictions for the 32 sets for nitrogen, likely an underestimate of the uncertainty
since the fits were made for $x>0.01$. }
\label{fig:gluonpdfncteq}
\end{figure}

\begin{figure}
\centering
\includegraphics[width=0.78\textwidth]{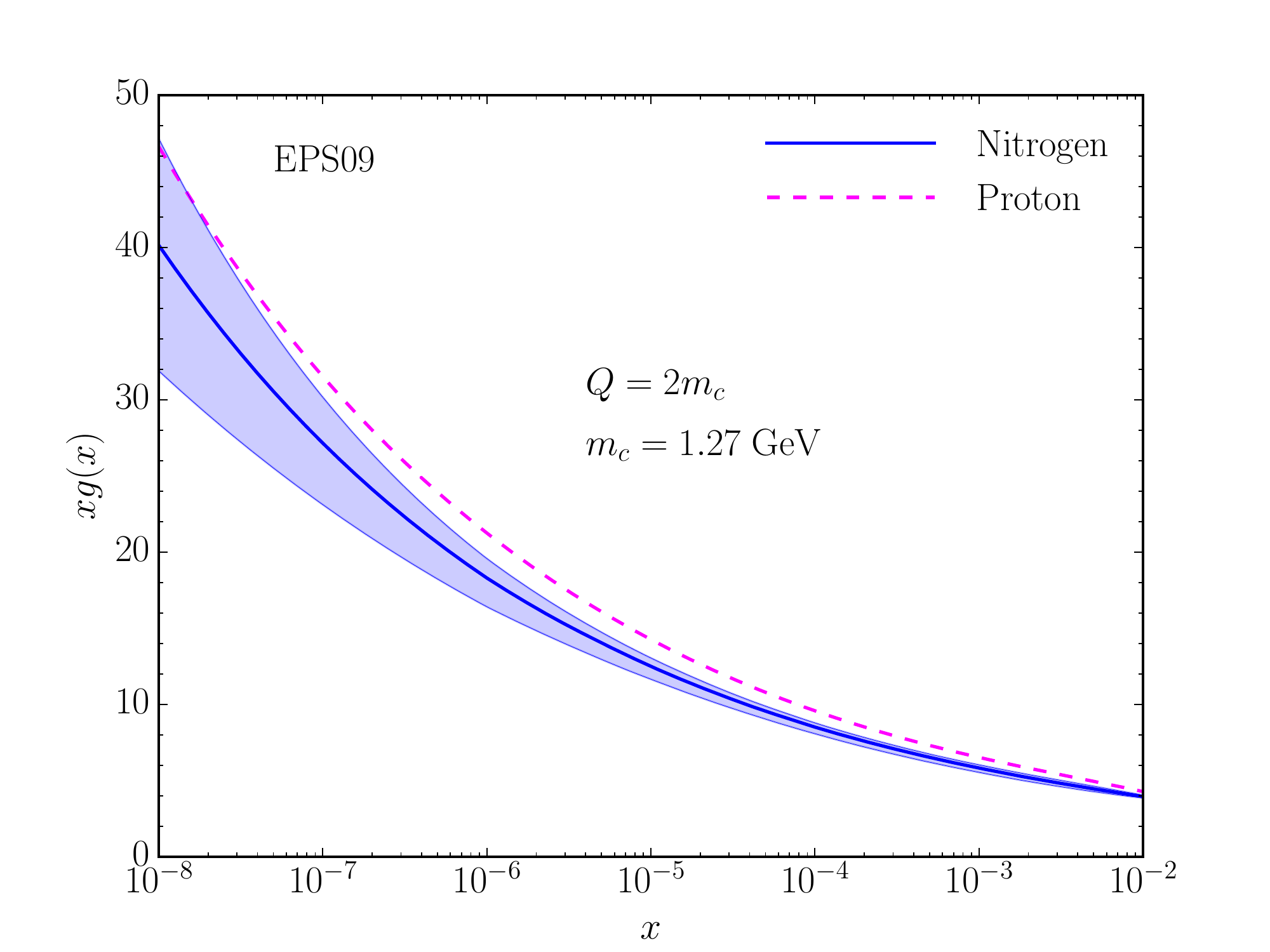}
\caption{The gluon distribution functions for free protons and
  isoscalar nucleons bound in nitrogen in the
  EPS09 sets \cite{Eskola:2009uj} with $Q=2m_c$ with CT14 PDFs \cite{Dulat:2015mca}.
  The uncertainty band (blue shaded) around the central nuclear gluon distribution is
  obtained by combining the maximal uncertainties from the proton CT14NLO 
PDFs sets and
  those from the different EPS09 nuclear modification factors. Set 55
  of CT14NLO PDFs is not included here.}
\label{fig:gluonpdfeps}
\end{figure}

In figs.~\ref{fig:gluonpdfncteq} and \ref{fig:gluonpdfeps} we show
the impact of nuclear modification on the gluon  distribution
in the small $x$ region using nCTEQ15 and EPS PDFs respectively.  In the standard distribution of the nCTEQ15
grids, low $x$ extrapolations must be used to avoid the unphysical
behavior
shown by the dashed lines in fig.\ \ref{fig:gluonpdfncteq}. The dotted
lines show a power law extrapolation $xg(x,Q)\sim x^{-\lambda(Q)}$
below $x_{\rm min}=10^{-6.5}$. The solid lines in
fig.\ \ref{fig:gluonpdfncteq}
show the nCTEQ15 results
with grids extended to low $x$ \cite{Olness-email}. The shaded band
shows the range of nuclear PDF uncertainties in the 32 sets
provided. We use
the corresponding lower and upper curves (sets 27 and 28) to quantify
the nuclear PDF uncertainty, which is likely underestimating the uncertainty
given the lack of data in this kinematic regime.
Similarly, for the EPS09 gluon distribution in fig.~\ref{fig:gluonpdfeps},
we also show the uncertainty band, which is now computed as the maximal
deviation from the central band due to a combination of uncertainties
from the 57 different members of the base proton CT14NLO PDF set and those from the
different members of EPS09 modification factors themselves.
As a result of incorporating PDF uncertainties from both the proton PDF and
nuclear modification factors,
the net uncertainty bands at low $ x $ in results obtained using the
EPS09 scheme are generally larger than those from the nCTEQ15 PDF's. Overall, we find that within the nuclear PDF sets used here, the uncertainty is rather modest, which is due to the constraints stemming from the parametrization. We note that the real uncertainty for the nuclear PDFs can be much larger in the low $x$ region. \footnote{Set number 55
from the CT14NLO PDF leads to total cross-sections that significantly exceed experimental upper limits from
ALICE and LHCb results at $ \sqrt{s} = 7 $ TeV, even when using our central values of the
factorization and renormalization scales. Consequently, it has been excluded when
computing the uncertainty bands throughout this work.}

Depending on the observable, the nuclear effects 
 in the nCTEQ15 and EPS09 frameworks
can be sizeable.
For the total
cross section, the dominant contribution comes from the symmetric configuration of partons' longitudinal momenta, i.e., $x_{1,2}\sim 2 m_c/\sqrt{s}$.
On the other hand,  the
differential distribution in outgoing charm energy fraction
$x_c=E_c/E_{CR}$, for the forward production is dominated by asymmetric configurations $x_c\sim x_1 \gg x_2$, and thus probes deeper into the shadowing region of the target nucleus. We will show
below that the impact of shadowing on the total charm cross section is less
significant than it is on the neutrino flux, which is dominated by the forward charm production.

In table~\ref{table:sigccb2}, we show the cross sections $\sigma(pp\to c\bar{c}X)$
and $\sigma(pA\to c\bar{c}X)/A$ using the nCTEQ15 PDFs
for our central set of factorization
and
renormalization scale factors
$(N_F,N_R)=(2.1,1.6)$ such that $(M_F,M_R)=(N_F,N_R)m_T$ and
$(M_F,M_R)=(N_F,N_R)m_c$. Here, $m_T$ is the
 transverse mass, $m_T^2=p_T^2 + m_c^2$. For an incident beam energy
$E_p=10^6$ GeV,  $\sigma(M_{F,R}\propto m_c)$ is larger than
$\sigma(M_{F,R}\propto m_T)$ by a factor of $ 1.16-1.17$, while at
$10^8$ GeV, the cross sections are nearly equal.
These choices of factorization and renormalization scales proportional
to $m_c$ are the
central values constrained by the data in an analysis using NLO pQCD charm cross section 
calculation in ref.~\cite{Nelson:2012bc} and used in ref.~\cite{Bhattacharya:2015jpa}. As noted, we find similar results for the
cross sections for scales
proportional to $m_T$.
Scale variations of $(M_F,M_R)=(1.25,1.48)m_T$ and
$(M_F,M_R)=(4.65,1.71)m_T$ bracket the results of
ref.~\cite{Nelson:2012bc}, and we use this range here as well. While the total cross section requires extrapolations of the fiducial
to inclusive phase space for data comparisons with theory, we show
below that our choices of scales  are consistent with forward
charm measurements at LHCb.

\begin{table}[tb]
\begin{center}
\vskip 0.5in
\begin{tabular}{|c|c|c|c|c|c|c|}
 \hline
 \multirow{2}{*}{$ E_p $} &
\multicolumn{2}{c|} {$\sigma (pp\rightarrow c \bar{c} X)$ [$\mu$b]} 
& \multicolumn{2}{c|} {$\sigma (pA\rightarrow c \bar{c} X) /A$ [$\mu$b]} 
& \multicolumn{2}{c|} {$[\sigma_{pA}/A] / [\sigma_{pp}]$} \\
 \cline{2-7}
 & $M_{F,R}\propto m_T$ & $M_{F,R}\propto m_c$  & $M_{F,R}\propto m_T$ &  $M_{F,R}\propto m_c$ 
 & $M_{F,R}\propto m_T$ &  $M_{F,R}\propto m_c$  \\
  \hline
 $10^2$ & 1.51 & $1.87 $ &1.64& $1.99$ &1.09 & 1.06\\
 \hline
 $10^3$ & $3.84\times 10^1$  & $4.72\times 10^1$ & $4.03\times 10^1$ & $4.92\times 10^1$ & 1.05 & 1.04\\
 \hline
 $10^4$ & $2.52\times 10^2$ & $3.06\times 10^2$ & $2.52\times 10^2$ & $3.03\times 10^2$  & 1.00 & 0.99\\
 \hline
  $10^5$ & $8.58\times 10^2$ & $1.03\times 10^3$ & $8.22\times 10^2$ & $9.77\times 10^2$ & 0.96 & 0.95\\
 \hline
 $10^6$ &  $2.25\times 10^3$ & $2.63\times 10^3$ & $2.10\times 10^3$ & $2.43\times 10^3$  & 0.93 & 0.92\\
 \hline
 $10^7$ & $5.36\times 10^3$  & $5.92\times 10^3$ & $4.90\times 10^3$ & $5.35\times 10^3$ & 0.91 & 0.90\\
 \hline
 $10^8$ &  $1.21\times 10^4$ & $1.23\times 10^4$ & $1.08\times 10^4$ & $1.09\times 10^4$  & 0.89 & 0.89\\
 \hline
 $10^9$ & $2.67\times 10^4$  & $2.44\times 10^4$ & $2.35\times 10^4$ & $2.11\times 10^4$  &0.88 & 0.86\\
 \hline
 $10^{10}$ & $5.66\times 10^4$  & $4.67\times 10^4$ & $4.94\times 10^4$ & $3.91\times 10^4$ & 0.87 & 0.84\\
 \hline
\end{tabular}
\end{center}
\caption{
The NLO pQCD total cross section per nucleon [$\mu$b] for charm pair production as a function of incident energy
[GeV] for scale factors $(N_F,N_R)=(2.1,1.6)$ (the central values for
charm production) for protons incident on isoscalar nucleons. The PDFs are for free nucleons (nCTEQ15-01)
and the target nucleons bound in nitrogen (nCTEQ15-14) using the low-$x$ grids. For these calculation, we use
$\Lambda_{\rm QCD} = 226$ MeV, $N_F=3$ and $m_c=1.27$ GeV. }
\label{table:sigccb2}
\end{table}%

\begin{figure}
\centering
\includegraphics[width=0.49\textwidth]{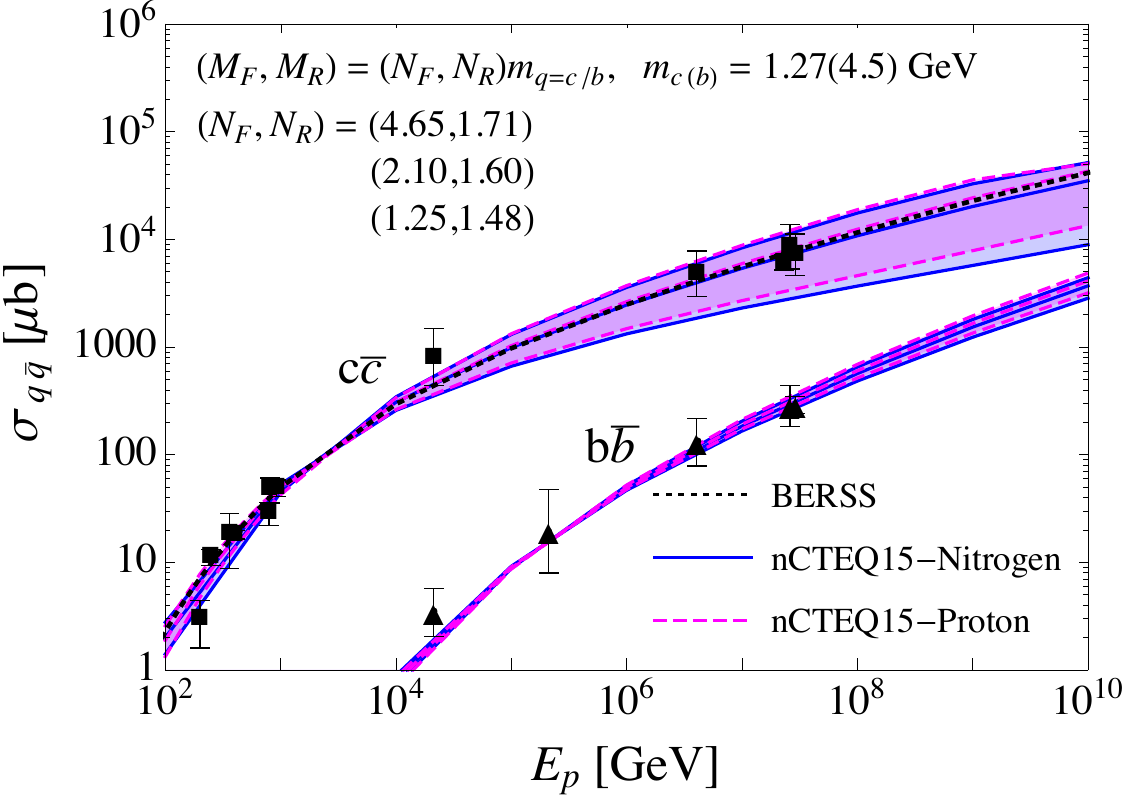}
\includegraphics[width=0.49\textwidth]{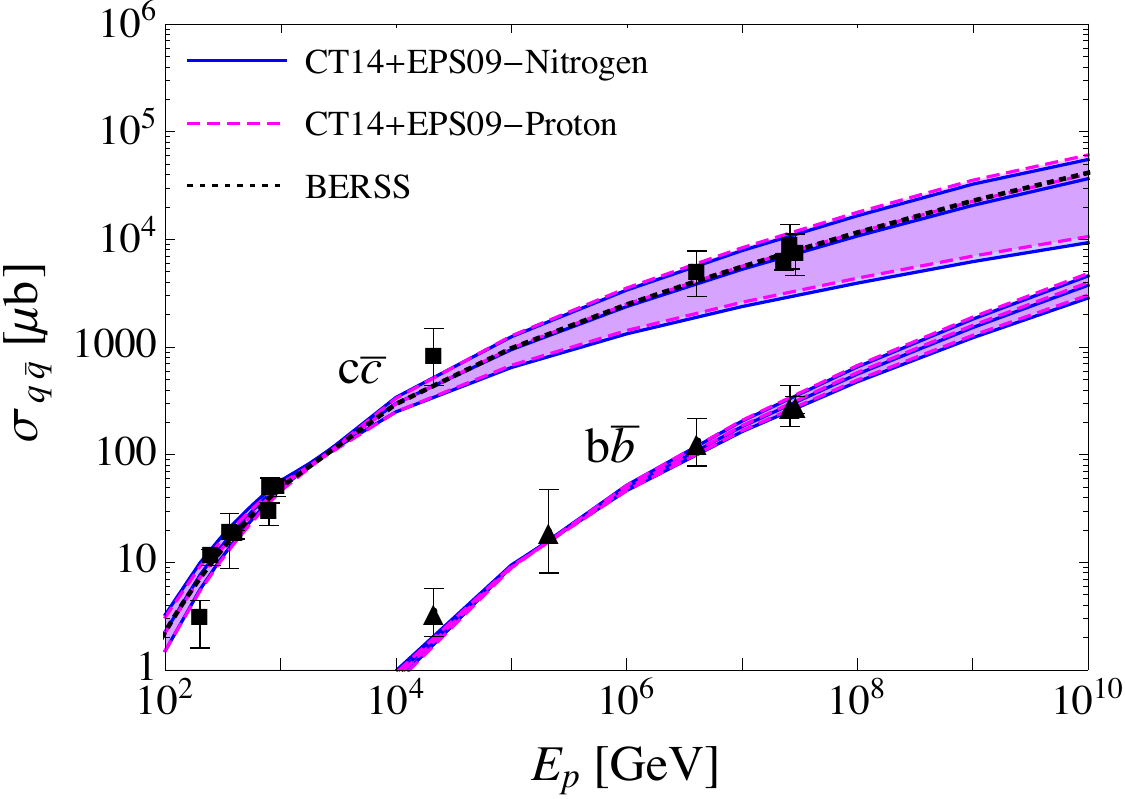}
   \caption{
   Left: Energy dependence of the total nucleon-nucleon charm and
bottom cross section obtained in NLO pQCD approach
using the nCTEQ15-01
   PDFs for protons incident on a free proton target (dashed red
   curves) and
nCTEQ15-14 for an isoscalar nucleon target bound in nitrogen (solid blue curves). The central curves are for $(M_F,M_R)=(2.1,1.6)m_Q$, while the upper and lower curves
   are for scaling with factors of (1.25,1.48) and (4.65,1.71)
   correspondingly. The dashed black curve is the BERSS result
   \cite{Bhattacharya:2015jpa}. The data points for the total charm
   cross section from $pp$ collisions at RHIC and LHC energies are
   from refs.
   \cite{Abelev:2012vra,Aaij:2013mga,Aad:2015zix,Adare:2006hc,Adamczyk:2012af,Abt:2007zg,Adare:2014iwg,Abelev:2014hla,Abelev:2012gx,Albajar:1990zu,Aaij:2011jh},
   while the lower energy data are from
 a compilation of fixed target data in ref.~\cite{Lourenco:2006vw}.
  Right: Energy dependence of the charm and bottom total cross section in
 nucleon-nucleon collision obtained in NLO pQCD approach using
 NLO CT14 PDFs and the
 EPS09 NLO
   nuclear modification factor $R_i^A$  (solid blue curve)
   \cite{Eskola:2009uj} and $(M_F,M_R)=(2.1,1.6)m_Q$.  The upper
   and lower curves correspond to the same variation of the
factorization and
renormalization scales as in the
   left panel. }
\label{fig:sigpert}
\end{figure}

The total charm and bottom cross sections per nucleon in $pp$
and $pA$ collisions as
functions
of incident proton energy are shown in the left panel of fig.~\ref{fig:sigpert} for nCTEQ15
PDFs for free nucleons (dashed-magenta curves)  and for the case when
 nucleons are bound in nitrogen (solid blue curves). The range of
curves reflects the uncertainty in the cross section due to the
scale dependence.
The dependence of the cross section on the nuclear PDFs is on the order of a few percent at the highest energies when one uses the
32 sets of nCTEQ15-14 PDFs.
The right panel of fig.~\ref{fig:sigpert} shows with the solid blue curve the total charm cross section per nucleon,
$\sigma(pA\to c\bar{c} X)/A$,  for nitrogen  with the EPS09 nuclear
modification factor.  
For each fixed set of scales, the maximal deviation from the central cross-section
due to uncertainties from the different members of EPS09 and CT14NLO PDFSets is at the level of 30\%
at energies of $ 10^{10} $ GeV.
The cross section with nitrogen (per nucleon) falls within the data constrained QCD scale uncertainties (shaded blue area) evaluated for the isoscalar nucleon
cross sections in ref.~\cite{Bhattacharya:2015jpa}.
In fig.~\ref{fig:sigpert}, we vary the factorization scale from
 $M_F= 1.25 m_c$ to $4.65 m_c$ and the renormalization scale from
$M_R=1.48 m_c$ to $1.71 m_c$.
The data points for the total charm cross section in proton-proton
collisions at RHIC and LHC energies in the figures are from
 \cite{Abelev:2012vra,Aaij:2013mga,Aad:2015zix,Adare:2006hc,Adamczyk:2012af,Abt:2007zg,Adare:2014iwg,Abelev:2014hla,Abelev:2012gx,Albajar:1990zu,Aaij:2011jh},
 while the lower energy data are
from a compilation of fixed target data in \cite{Lourenco:2006vw}.

The nCTEQ15-01 free nucleon sets yield slightly larger
isoscalar nucleon cross section for charm production than the
CT10 evaluation of BERSS \cite{Bhattacharya:2015jpa} which are shown by the
black dotted lines in fig.\ \ref{fig:sigpert}.
The nuclear
corrections to the CTEQ15-01 set decrease the cross section relative to the
BERSS  evaluation using
CT10, with a net decrease relative to CT10 of
10\% at the highest energies, where the differences in the small $x$ distribution of the PDFs
are most important.  The EPS09 parametrizations incorporate less
nuclear shadowing at small $x$ than the nCTEQ15 nuclear corrected
PDFs.

\begin{figure}
\centering
 \includegraphics[width=0.7\textwidth]{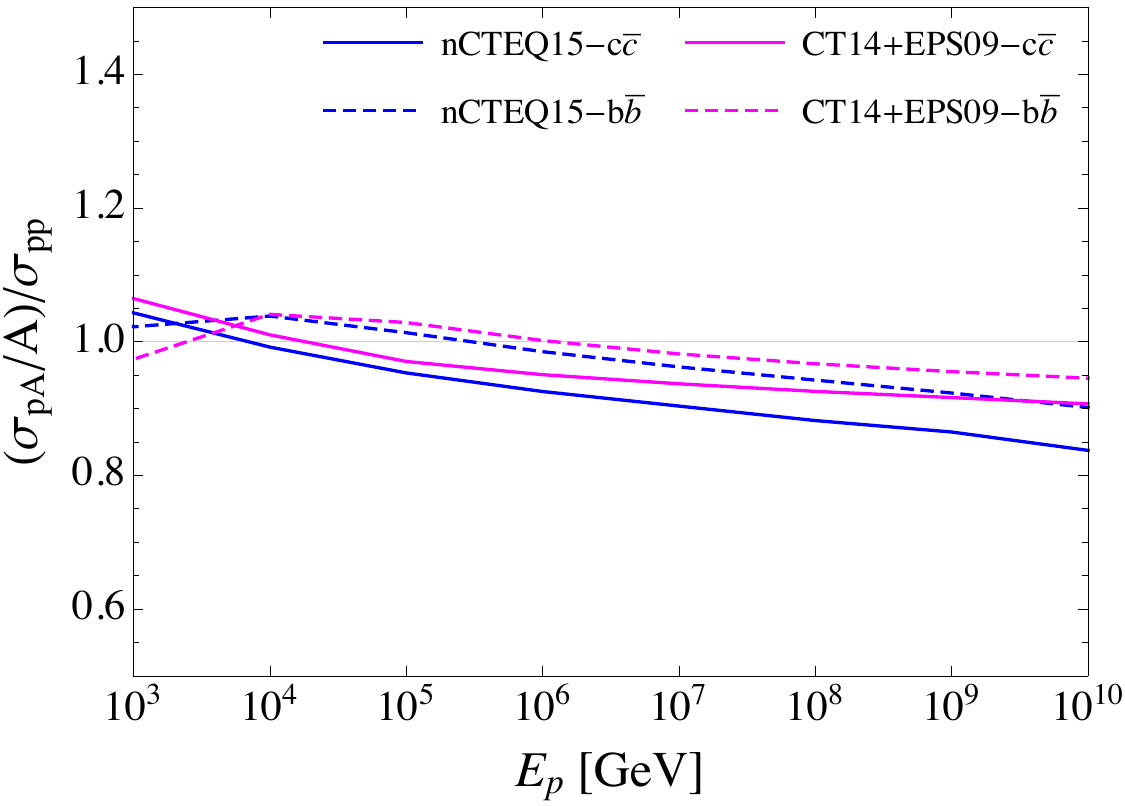}
\caption{The ratio of the NLO pQCD charm (solid curves) and bottom
  (dashed curves)
total cross sections per nucleon
  with partons in nitrogen and partons in free nucleons for
nCTEQ15 (red curves) and for the EPS09 (blue curves) nuclear
modifications to the {CT14}
PDFs. Here, the factorization and renormalization scales are set to be
 $(M_F,M_R)=(2.1,1.6)m_Q$ for $m_c=1.27$ GeV and $m_b=4.5$ GeV.}
\label{fig:nucrat}
\end{figure}

Fig.~\ref{fig:nucrat} shows the cross section ratio for $(\sigma(pA\to
Q\bar{Q} X)/A)
/\sigma(pN\to Q\bar{Q} X))$ for $Q=c$ (solid lines) and $Q=b$ (dashed
lines)
for isoscalar target $N$ and $A=14$.
The ratio of the cross section per nucleon for partons in nitrogen and
free
nucleons for $(M_F,M_R)=(2.1,1.6)m_c$ using nCTEQ15 PDFs are shown in
blue curves in fig.
\ref{fig:nucrat}, and for EPS09 with CT14 free proton PDFs using the
magenta curves.
 At low energies, where the cross section is quite small due to threshold effects, the anti-shadowing dominates, however for the energy range of interest, shadowing is more important, resulting in a 20\% (10\%) decrease in the cross section at high energies for $c\bar{c}$ production with the nCTEQ15 (EPS09) PDF. For
$b\bar{b}$ production, the cross section is decreased by $\sim 6\%$--$10\%$ at $E=10^{10}$ GeV
depending on the choice of nuclear PDF.

So far, we have considered calculations based on the standard integrated parton distribution functions and the collinear framework.
However as discussed above, neutrino production at high energy probes the region of very small values of $x$ of the gluon distribution,
which is not very well constrained at present. The standard DGLAP evolution, which is based on the resummation of large logarithms of scale, does not provide constraints on the small $x$ region.   Therefore, it is worthwhile to explore other approaches which  resum the potentially large
logarithms $\alpha_s \ln 1/x$.  There are two approaches at present, the dipole model and the $k_T$ factorization. The dipole model \cite{Nikolaev:1990ja,mueller,Nikolaev:1995ty,gbw,sgbk,bgbk,iim,soyez,albacete,aamqs,forshaw,kowalski,gkmn,Jeong:2014mla,Block:2014kza,Ewerz:2011ph,Albacete:2010sy,Albacete:2009fh} is particularly convenient
for including corrections due to parton saturation. Parton saturation
in this approach is taken into account as multiple rescatterings
of the dipole as it passes through the nucleus. The dynamics is encoded in the dipole cross section, which can be either parametrized or obtained from the nonlinear
evolution equation.  Below we shall explore improvements to the previous calculation based on the dipole model \cite{ERS}, which include using more
modern parametrizations for the dipole scattering cross section.
Another approach to evaluating the prompt neutrino flux is
based on $k_T$ factorization \cite{Catani:1990eg,Collins:1991ty,Levin:1991ya,Ryskin:1995sj}.
In this approach the dynamics of the gluon evolution is encoded in the unintegrated parton densities, which include information about the transverse momentum dependence
of the gluons in addition to the longitudinal components. We shall be using the unified BFKL-DGLAP evolution approach, with nonlinear effects, to compute the evolution of the unintegrated PDFs, which
should provide for a reliable dynamical extrapolation of the gluon density towards the small $x$ regime.

\subsection{Dipole model}

The color dipole model \cite{Nikolaev:1990ja,mueller,Nikolaev:1995ty,Raufeisen:2002ka,Kopeliovich:2002yv,Goncalves:2006ch}
is an alternative approach to evaluating the
heavy quark pair production cross section. The advantage of this framework is that gluon saturation at small $x$ can be included in a relatively straightforward way, as a unitarization of the dipole-proton scattering amplitude.
The partonic interaction cross section of the gluon with the target can be described in the regime of high energy by a two-step process.
First, a gluon fluctuation into a $q\bar{q}$ pair is accounted by a wave function squared, then
this dipole interacts with the target with a dipole cross section. In this framework, the partonic cross
section for $q\bar{q}$ production can be written as \cite{Nikolaev:1990ja}
\begin{align}
\sigma^{g p \to q\bar q X}(x,M_R,Q^2) = \int dz \, d^2\vec r \,
|\Psi^q_g(z,\vec r,M_R,Q^2)|^2 \sigma_{d}(x,\vec r)\ ,
\label{eq:dipolexsec}
\end{align}
for gluon momentum squared $Q^2$ and renormalization scale $M_R$.
The wave function squared, for pair separation $\vec{r}$ and
fractional momentum $z$ for $q=c$ and $q=b$, is
\begin{align}
|\Psi^q_g(z,\vec r,M_R,Q^2=0)|^2 =
\frac{\alpha_s(M_R)}{(2\pi)^2}
\left[ \left( z^2 + (1-z)^2 \right)
 m_q^2 K_1^2(m_q r) + m_q^2 K_0^2(m_q r)\right],
\label{eq:ps2}
\end{align}
in terms of the modified Bessel functions $K_0$ and $K_1$. The dipole cross section $\sigma_{d}$ can
be written in terms of the color singlet dipole $\sigma_{d, em}$ applicable to electromagnetic
scattering \cite{Nikolaev:1995ty,Kopeliovich:2002yv}
\begin{align}
\sigma_{d} (x,\vec r) &= \frac{9}{8} \left[\sigma_{d,em}(x,z\vec r)
+\sigma_{d,em}(x,(1-z)\vec r)\right]
-\frac{1}{8}\sigma_{d,em}(x,\vec r)\ .
\label{eq:sigdg}
\end{align}
Using
eqs.\ (\ref{eq:ps2},\ref{eq:sigdg}) in the expression
given by eq. (2.2), the
 heavy quark rapidity distribution in proton-proton
scattering is given by \cite{ERS}
\begin{equation}
\frac{d\sigma (pp\to q\bar{q}X)}{dy}\simeq x_1 g(x_1,M_F)\sigma^{gp\to q\bar{q}X}
(x_2,M_R,Q^2=0)\ ,
\end{equation}
where we use
\begin{equation}
x_{1,2}=\frac{2m_q}{\sqrt{s}}e^{\pm y}\ .
\end{equation}
Similarly, the Feynman $x_F$ distribution in the dipole model is given
by \cite{ERS},
\begin{equation}
\frac{d\sigma (pp\to q\bar{q}X)}{dx_F}\simeq \frac{x_1}{\sqrt{x_F^2 + \frac{4M_{q\bar q}^2}{s}}}
 g(x_1,M_F)\sigma^{gp\to q\bar{q}X}
(x_2,M_R,Q^2=0)\ ,
\end{equation}
in terms of the $q\bar{q}$ invariant mass squared $M_{q\bar{q}}^2$ and
center of mass energy squared $s$.
A LO gluon PDF is used for the value of large $x_1$, while the dipole
cross section encodes the information about  the small $x$ dynamics of the target, including the saturation
effects.

In ref.\ \cite{ERS}, the dipole cross section parametrized by Soyez \cite{soyez} was used
to evaluate the prompt atmospheric lepton flux. This parametrization was  based
on the form discussed by Iancu, Itakura and Munier \cite{iim} which
approximated
the solution to the nonlinear Balitsky-Kovchegov (BK) \cite{Balitsky:1995ub,Kovchegov:1999yj}  evolution equation.
In the present calculation we use updated PDFs and dipole model
parametrizations. In the flux evaluations and comparisons with LHCb data, we have updated the
fragmentation fractions (see Appendix \ref{sec:appendix}) compared to earlier work \cite{ERS}.
For $g(x_1,M_F)$, we use
the CT14 LO PDF \cite{Dulat:2015mca}.
There has been significant progress in the extractions of  the dipole scattering amplitudes  by including the
running coupling constant (rcBK), and now, more recently, full NLO
corrections to the BK equation. Here, we use Albacete et al.'s AAMQS
dipole cross section result that includes heavy quarks in the rcBK formalism
\cite{aamqs} which has been fitted to the inclusive HERA data.
We   compare the cross section and flux calculations using this parametrization   with the  calculations based on the dipole cross section by Soyez.

Finally, we use a third dipole model that is phenomenologically based. Starting from a parametrization of the electromagnetic
structure function $F_2(x,Q^2)$ guided by unitarity considerations by
Block et al.~\cite{Block:2014kza}, one can show that the dipole cross
section for electromagnetic scattering
is approximately
\begin{equation}
\label{eq:blockdip}
\sigma_{d,em}(x,r)\simeq \pi^3 r^2 Q^2\frac{\partial F_2}{\partial
Q^2}\Big\vert_{Q^2=(z_0/r)^2} \, ,
\end{equation}
for $z_0\simeq 2.4$
\cite{Ewerz:2011ph,Jeong:2014mla}.
We refer to this approximate form with the parametrization of $F_2$ from ref.\ \cite{Block:2014kza} as the
``Block dipole.'' This  dipole  cross section does well in describing electromagnetic, weak interaction and hadronic
cross sections \cite{Arguelles:2015wba}, and
 yields a flux similar to the AAMQS and Soyez calculations.

\begin{figure} [t]
\centering
\includegraphics[width=0.7\textwidth]{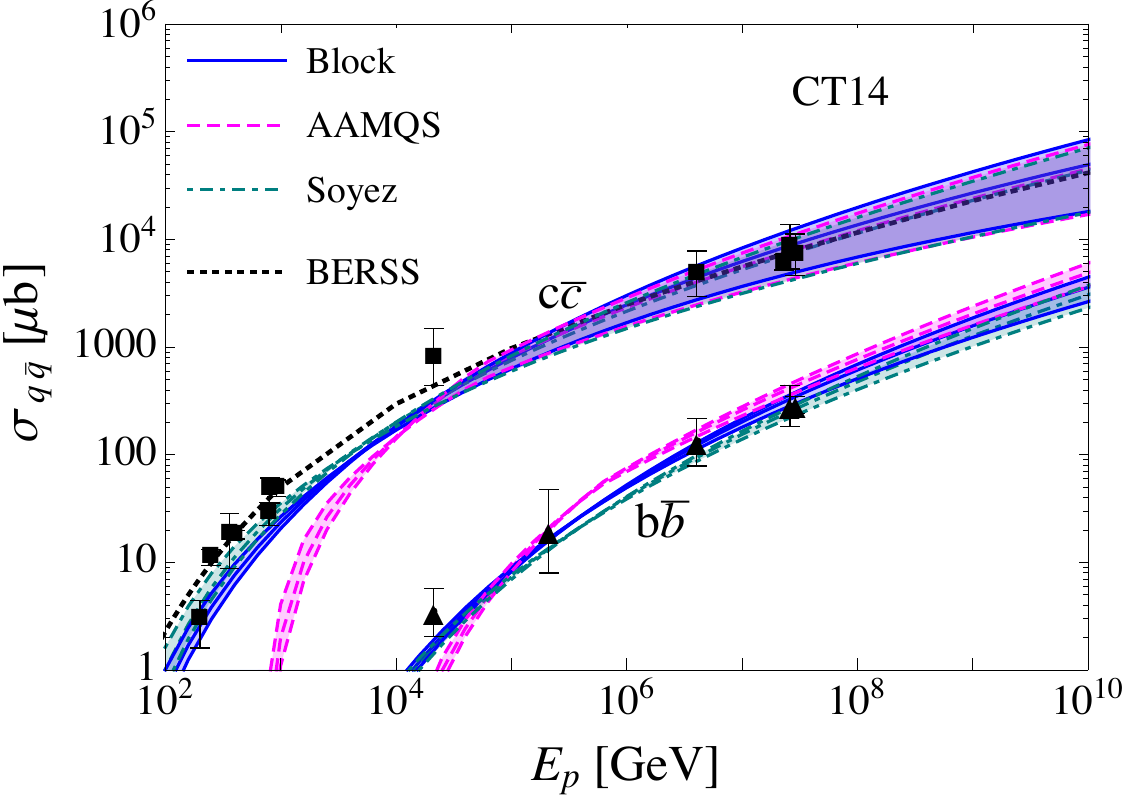}
  \caption{The $q\bar{q}$ production cross section in $pp$ collisions
  from the dipole model for $q=c$ and $q=b$.
   The cross sections use the following charm quark and
bottom quark mass, Soyez model: $m_c=1.4$ GeV, $m_b=4.5$ GeV;
 Block and AAMQS models:
   $m_c=1.27$ GeV, $m_b=4.2$ GeV 
   and fixed values of $\alpha_s$: $\alpha_s=0.373$ for charm,
   $\alpha_s=0.215$ for bottom quark production. The cross section is evaluated using the CT14 LO PDFs with a range of factorization scales
   $M_F=m_c$ to $4m_c$.
    We also show the experimental data and BERSS results, as in
   fig.~\ref{fig:sigpert}.}
\label{fig:aamqs-xmax}
\end{figure}

Fig.~\ref{fig:aamqs-xmax} shows the cross sections for charm and bottom
pair production from $pp$ interactions  calculated from the various
dipole models introduced above with the gluon factorization scale
varied between $m_c$ and $4 m_c$.
While all the $c \bar{c}$ cross sections are comparable at $E \gsim 10^6 \gev$,
for the $b \bar{b}$ cross sections, there is a difference by a factor of 1.8 (1.6) at $E=10^6 (10^8) \gev$
between the Soyez (lowest) and the AAMQS (highest) results.

The dipole cross section is applicable for $x_2\ll 1$. The AAMQS
dipole cross section is provided for
$x<0.008$. The unphysical sharp increase as a function of energy for the AAMQS result near $E=10^3$ GeV is an artifact of this
cutoff in $x$.
We checked that this default $x_{max}$ of AAMQS has no important
effect for $E > 10^6 \gev$, relevant energies where the prompt fluxes are dominant.
As in previous work \cite{ERS}, we fix $\alpha_s$ at
$\alpha_s=0.373$ for charm and
   $\alpha_s=0.215$ for bottom quark production. These values of $\alpha_s$ come from taking $M_R\sim m_q$. We take a central factorization scale equal to
   $M_F=2 m_q$. These choices give reasonable cross
   sections as
   fig.~\ref{fig:aamqs-xmax} shows. In this approach, the cross section scales linearly
   with $\alpha_s$.  Within the constraints of the cross section
   measurements and other experimental results, e.g., LHCb, $\alpha_s$
   can be varied with different renormalization scales. Rather than
   make this scale variation, we keep $\alpha_s$ fixed for all the
   dipole model calculations presented here.

The comparison shown in fig.~\ref{fig:aamqs-xmax} of the
total charm and total bottom cross sections in $pp$ collisions
 shows good agreement with the data at high energies.
However, at low energies dipole models underestimate the
cross section
because of the aforementioned
limitation of $x_{max}$, and the fact that
quark and anti-quark contributions to the
cross section are not included in this model.
At high energies, initial state gluons dominate,
while at low energies, this is not the case.

Nuclear effects can be incorporated in the dipole model by modifying the saturation scale. This approach, as discussed
by Armesto, Salgado and Wiedemann (ASW) \cite{Armesto:2004ud},
involves a relative scaling of the free proton saturation scale by an $A$ dependent ratio $(AR_p^2/R_A^2)^{1/\delta}$ where
the power $\delta=0.79$ is a phenomenological fit to $\gamma^*A$ data \cite{Armesto:2004ud}
and $R_A = 1.12\, A^{1/3}-0.86/A^{1/3}$ fm is the nuclear radius and
$R_p$ is the proton radius.
This method is used in ref.\ \cite{ERS}, where the Soyez dipole cross section is described in terms of the
saturation scale
which depends on $r$ and $x$, however, the ASW approach cannot be used
if the dipole is not parametrized in terms of a saturation scale.
The method used here
is the Glauber-Gribov formalism, where
\begin{eqnarray}
\sigma_d^A(x,r)&=&\int d^2\vec{b}\, \sigma_d^A(x,r,b)\, ,\\
\sigma_d^A(x,r,b) &=& 2\Biggl[ 1-\exp\Biggl( -\frac{1}{2} AT_A(b)\sigma_d^p(x,r)\Biggr)\Biggr]\ .
\end{eqnarray}
The nuclear profile function $T_A(b)$ depends on the nuclear density $\rho_A$ and is normalized to unity:
\begin{eqnarray}
T_A(b) &=& \int dz\rho_A(z,\vec{b})\, ,\\
\int d^2 \vec{b}\, T_A(b)&=& 1\ .
\end{eqnarray}
We use a Gaussian distribution for nuclear density,
\begin{equation}
\rho_A (z,b) = \frac{1}{\pi^{3/2} a^3} e^{-r^2/a^2} \quad {\rm for}\;  r^2=z^2+\vec{b}^{\, 2}\, , \
\end{equation}
with $a^2=2 R_A^2/3$.
This agrees well with a three parameter Fermi fit \cite{DeJager:1987qc}, used in other studies \cite{Armesto:2002ny,Goncalves:2003kp}.

The nuclear corrections in the dipole model are smaller than in the NLO
pQCD approach with nCTEQ15-14 PDFs, however, they are similar to the
EPS09
nuclear corrections.
For the Block dipole model nuclear effects range from
 1\% at $10^3$ GeV to about 88\% at $10^{10}$ GeV, while for
the AAMQS
at $10^{10}$ GeV it is $93\%$, and for the Soyez dipole model
it is approximately
90\%.
The nuclear corrected cross sections for charm and bottom pair production
are presented in fig.~\ref{fig:sigcompare} with the cross sections
from other approaches.

\subsection{$k_T$ factorization}

\begin{figure}
\centering
\includegraphics[width=0.7\textwidth]{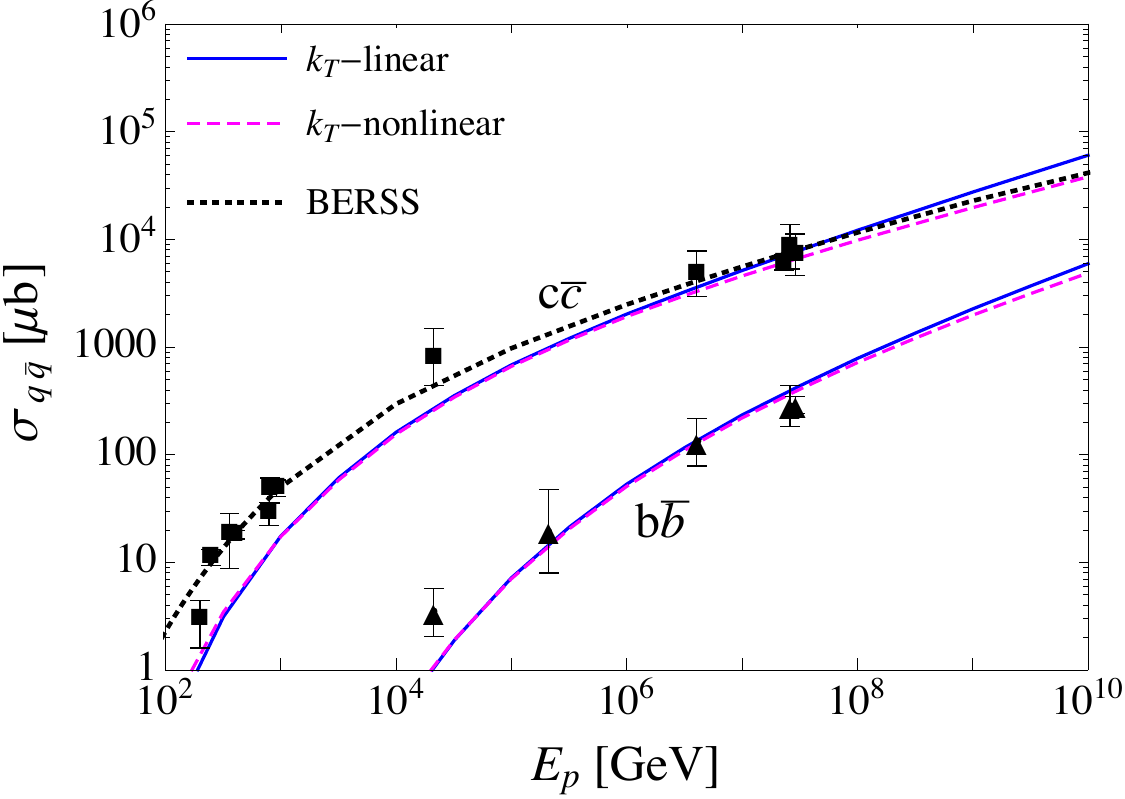}
\caption{The integrated charm cross section in $pp$ collisions from $k_T$ factorization, using the unintegrated gluon from  linear evolution  from resummed BFKL (solid  blue, upper curve),
and non-linear evolution (dashed magenta, lower curve). Both calculations were based on the unintegrated gluon PDFs taken from \cite{Kutak:2012rf}.  Shown for comparison is the perturbative
cross section from ref.~\cite{Bhattacharya:2015jpa} (black short-dashed curve) and data points as in fig.~\ref{fig:sigpert}.
}
\label{fig:ktcrosssection}
\end{figure}

In this subsection we  discuss the calculation of the heavy-quark production cross section using the approach of $k_T$ factorization. As mentioned previously, since the kinematics of the process is such that the dominant contribution to the neutrino flux comes from forward production of the heavy quark, the values of the longitudinal momenta of partons in this process are highly asymmetric. The longitudinal momentum fraction $x$ of the parton participating in this process from the target side (the air nucleus) is very small, and hence one needs to extrapolate the parton densities beyond the region in which they are currently constrained by experimental data. On the other hand, we know that in the regime of small $x$ and relatively low scales, one should take into account potentially large logarithms $\alpha_s \, \ln 1/x$. Such contributions   are resummed in the framework of $k_T$ factorization and BFKL evolution~\cite{Lipatov:1976zz,Kuraev:1976ge,Kuraev:1977fs,Balitsky:1978ic}. The
$k_T$ factorization approach to heavy quark production in hadron-hadron collisions  has been formulated in \cite{Catani:1990eg,Collins:1991ty}. The framework involves  matrix elements for the $gg\rightarrow Q\bar{Q}$ process with off-shell incoming gluons. For the forward kinematics relevant here, we shall be using an approximation in which the large $x$ parton from the incoming cosmic ray particle is on-shell and the low $x$ parton from the target is off-shell. This is   referred to as the hybrid formalism, in which on one side the integrated collinear parton density is used, and on the other side the unintegrated gluon density with explicit $k_T$ dependence is used (for a recent calculation in the color glass condensate framework of the hybrid factorization see \cite{Altinoluk:2015vax}). The $k_T$ factorization formula for the single inclusive heavy quark production with one off-shell gluon reads
\begin{equation}
\frac{d\sigma}{d x_{F}}(s,m_Q^2) = \int \frac{dx_1}{x_1} \frac{dx_2}{x_2} \,  dz \, \delta(z x_1-x_F) \, x_1 g(x_1,M_F)\int \frac{dk_T^2}{k_T^2} \hat{\sigma}^{\rm off} (z,\hat{s},k_T) \, f(x_2,k_T^2) \; .
\end{equation}
In the above  formula, $x_F$ is the Feynman variable for the produced heavy quark, $x_1g(x_1,M_F)$ is the integrated gluon density on the projectile side, $\hat{\sigma}^{\rm off}(z,\hat{s},k_T)$ is the partonic cross section for the process $gg^*\rightarrow Q\bar{Q}$, where $g^*$ is the off-shell gluon on the target side, and   $f(x_2,k_T^2)$ is the unintegrated gluon density.
For the unintegrated gluon density, we have used the resummed version of the BFKL evolution which includes important subleading effects due to DGLAP evolution and the kinematical constraint \cite{Kwiecinski:1997ee,Kutak:2003bd,Kutak:2004ym}.  These terms are relevant since they correspond to the resummation of subleading terms in the small $x$ expansion. As a result, the calculation with resummation  should be more reliable than the calculation based on purely LL or NLL small $x$ terms. We have used the latest fits, where the unintegrated parton density has been fitted to high precision experimental data on deep inelastic scattering from HERA \cite{Kutak:2012rf}.  In addition, we have considered two cases, with or without parton saturation effects included for $\sigma_{c\bar{c}}$ and $\sigma_{b\bar{b}}$, shown in fig.\ \ref{fig:ktcrosssection}. Parton saturation was included through a nonlinear term in the parton density in the evolution \cite{Kutak:2003bd,Kutak:2004ym,Kutak:2012rf}.
Both calculations of the total integrated charm cross section, as compared with the BERSS calculation,  are consistent with the perturbative calculation for high energies $ \ge 10^4 \; {\rm GeV}$.  The calculation without parton saturation effects is higher than with saturation. At low energies, the calculation within $k_T$ factorization tends to be below the NLO perturbative calculation within the collinear framework. This is understandable as the $k_T$ factorization can be thought of as a higher order computation with respect to the LO collinear framework, but only in the region of high energies. At low energies the $\ln 1/x$  resummation is not effective anymore, and  $k_T$ factorization becomes closer to the LO collinear calculation. In order to match to NLO collinear in this region one would need to include other NLO  effects in the calculation or supplement the $k_T$ factorization calculation with the energy dependent $K$ factor.

We also analyzed the impact of  nuclear corrections in the $k_T$ factorization approach. The nuclear effects in this approach are encoded in the unintegrated gluon parton density through the nonlinear term in the evolution equation as described in \cite{Kutak:2012rf}. The strength of the nonlinear term in the nuclear case is enhanced by the factor $A^{1/3}$ with respect to the proton case. 

\subsection{QCD predictions for charm and bottom quark total and differential cross section}

\begin{figure}
\centering
\includegraphics[width=0.7\textwidth]{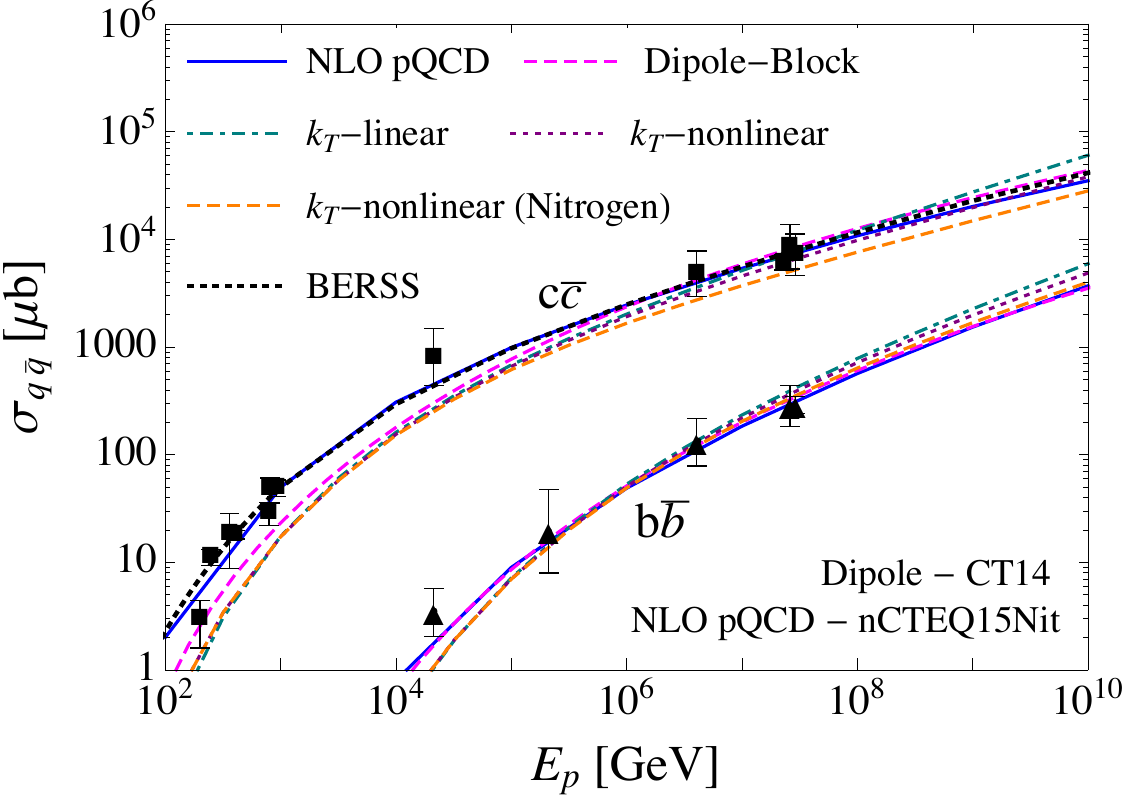}
\caption{ Total $c\bar{c}$ and
  $b\bar{b}$ cross sections as a function of the incident proton energy. The different curves correspond to:
 NLO perturbative (solid blue) obtained with nCTEQ15 parton distributions,
 dipole model
 calculation based on the Block parametrization (dashed-magenta),
 $k_T$ factorization with unintegrated PDF from linear evolution
 (dashed-dotted green), $k_T$ factorization with unintegrated PDF from
 non-linear evolution for nucleon (short-dashed violet) and $k_T$ factorization with unintegrated PDF from
 non-linear evolution for nitrogen (dashed orange). Comparison is made with
 the results from previous NLO calculation, denoted by BERSS
 (short-dashed black curve), ref.~\cite{Bhattacharya:2015jpa} and data
 points as in fig.~\ref{fig:sigpert}.}
\label{fig:sigcompare}
\end{figure}

In fig.~\ref{fig:sigcompare}, we show results for the energy
dependence of the total charm
and bottom cross sections obtained using the three different
QCD models:\ perturbative, dipole and $k_T$ factorization (linear
evolution, nonlinear evolution). For comparison we also show our calculation based on $k_T$ factorization with nuclear effects included.  We find
good agreement with the
experimental data with all models for LHC energies. However, at
lower energies only the  perturbative NLO approach gives a good
agreement with the data.
In the calculation of the prompt neutrino flux, the higher energy
cross sections are relevant. In order to evaluate the neutrino flux, one needs to
 convolute the differential cross section for charm
production with the steeply falling (with energy) cosmic ray flux.
This evaluation is sensitive to charm production in the
forward region. The
 differential charm quark distributions are different for different QCD models,
 some being in better agreement with the data than others,
 as we discuss below.

Similar to the procedure in
refs.~\cite{Garzelli:2015psa}
and \cite{Gauld:2015kvh}, we constrain our QCD parameters by
comparing our results for charm production in $pp$ collisions with
LHCb measurements in the forward region at $\sqrt{s}=7$ TeV and $\sqrt{s}=13$ TeV.
In refs.~\cite{Aaij:2015bpa,Aaij:2013mga}, the  charm cross
section has been measured by the LHCb experiment in
 the rapidity  range  $2.0\leq y\leq 4.5$ and for the charmed hadron transverse momentum  $p_T\leq 8$ GeV.
 In table \ref{table:lhcb}, we show the experimental values for the total
charm cross section measured by the LHCb collaboration
compared with our theoretical calculations.
 The experimental results from
LHCb
\cite{Aaij:2015bpa,Aaij:2013mga} for the charm cross section are obtained by taking their measured values of, for example, $D^0$ and
its charge conjugate, and dividing by two times the fragmentation
fraction for $c\to D^0$ to account for the inclusion of the charge
conjugate state in the measurement.
Fragmentation functions do not appreciably change the
rapidity
distributions at fixed $\sqrt{s}$.
The theoretical calculations using NLO perturbative QCD and  nCTEQ15-01 PDFs were performed with
$(N_F,N_R)=(2.1,1.6)$ scaling $m_T$ and $m_c$ for the central values,
with constraining the charm rapidity and transverse momentum to
correspond to the experimental
 kinematic
restrictions, scaled by
fragmentation fractions.
 The theoretical uncertainty band corresponds to
the scale variation in the range of
$(N_F,N_R)=(4.65,1.71)$ (upper limit)  and $(N_F,N_R)=(1.25,1.48)$ (lower
limit).  The NLO pQCD results using nCTEQ15-01 are
listed in the first two columns in table \ref{table:lhcb}.

We also show the calculation using the dipole model (DM)
and $k_T$ factorization.
The dipole model uncertainty band comes from the three
different dipole models and the scale variation in the gluon PDF
from $M_F=m_c$ to $M_F=4m_c$.
For the central values, we take the average of the cross sections
for all the models considered, with $M_F=2m_c$.
The upper limit of the uncertainty band corresponds to
 the Block dipole with $M_F=4m_c$ while the lower one is
 the Soyez dipole with $M_F=1m_c$.
Our results which include theoretical uncertainties are in
agreement with the LHCb rapidity distributions at $7$ TeV and
at $13$ TeV.

\begin{table}[tb]
\vskip 0.5in
\begin{center}
\begin{tabular}{|c|c|c|c|c|c|}
\hline
\multirow{2}{*}{$ \sqrt{s} $} &
\multicolumn{5}{c|} {$\sigma (pp\rightarrow c \bar{c} X)$ [$\mu$b]}\\
 \cline{2-6}
 & NLO ($\mu\propto m_T$) & NLO ($\mu\propto m_c$)  & DM & $k_T$ &Experiment  \\
\hline
7 TeV & $1610^{+480}_{-620}$ &  $1730^{+900}_{-1020}$ & $1619^{+726}_{-705}$ & $1347\div1961$ &
$1419\pm 134$\\
\hline
13 TeV & $ 2410^{+700}_{-960}$ &  $2460^{+1440}_{-1560}$ &$ 2395^{+1276}_{-1176}$& $ 2191\div3722$&
$2940\pm 241$ \\
\hline
\end{tabular}
\end{center}
\caption{The total cross section for $pp\to c\bar{c}X$ for the rapidity
  range limited to  $2\leq y\leq
  4.5$. In the NLO pQCD evaluation, we take
$p_T\leq 8$ GeV and we use scales, $(M_F,M_R)=(2.1,1.6)m_T$
and  $(M_F,M_R)=(2.1,1.6)m_c$, with error bars according to upper and
lower scales. The dipole model result shows the central value with
the uncertainty band obtained by varying the factorization scale
between $M_F=m_c$ and $M_F=4m_c$. The values of
$\alpha_s$ in dipole models are held fixed. Also shown are the
ranges for cross sections in the $k_T$-factorization  approach, where the lower band is given by non-linear calculation and upper by the linear.
 The experimental data are from
LHCb measurements \cite{Aaij:2015bpa,Aaij:2013mga}.
}
\label{table:lhcb}
\end{table}

\begin{figure}[h]
\centering
\includegraphics[width=0.49\textwidth]{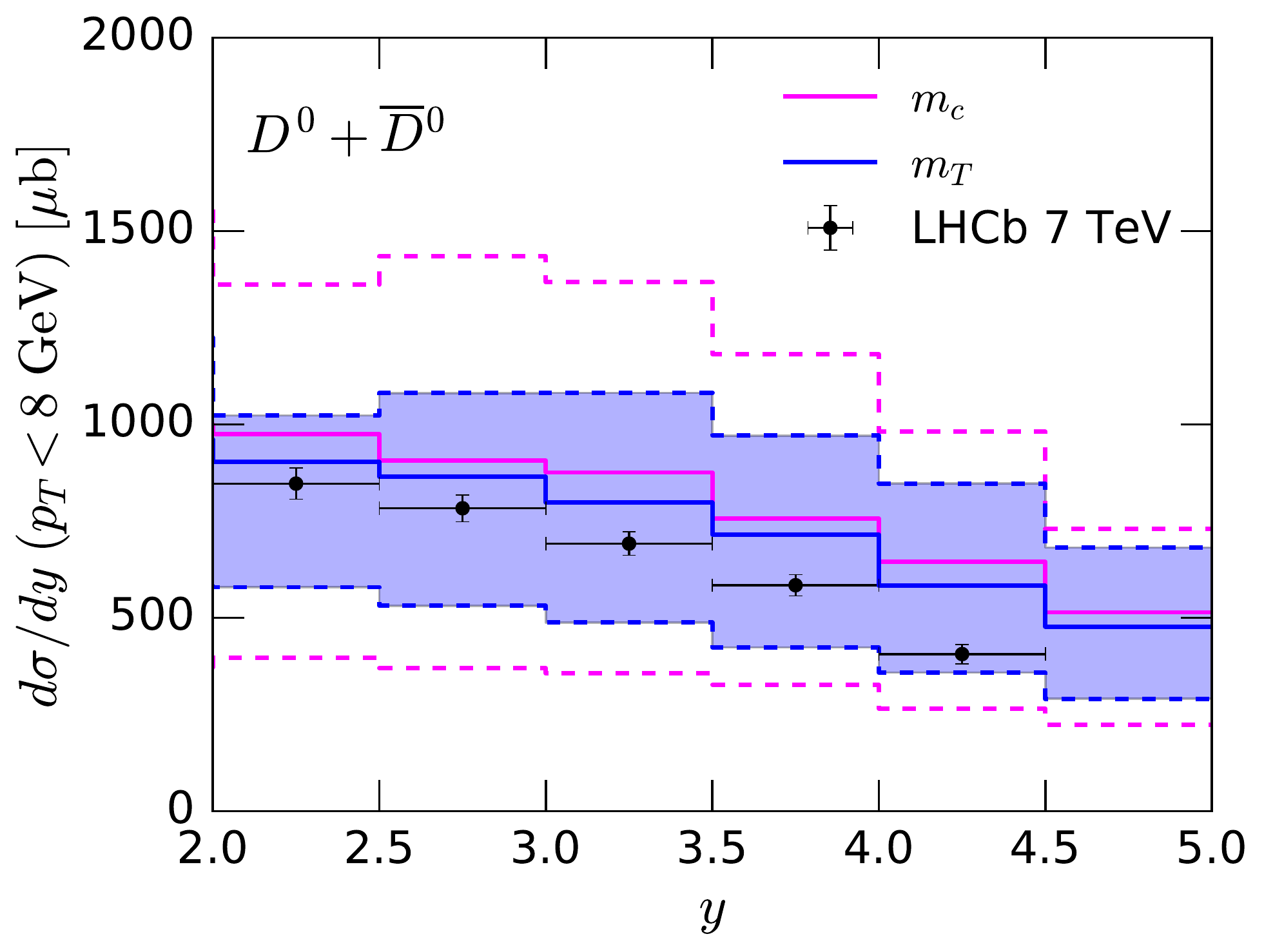}
\includegraphics[width=0.49\textwidth]{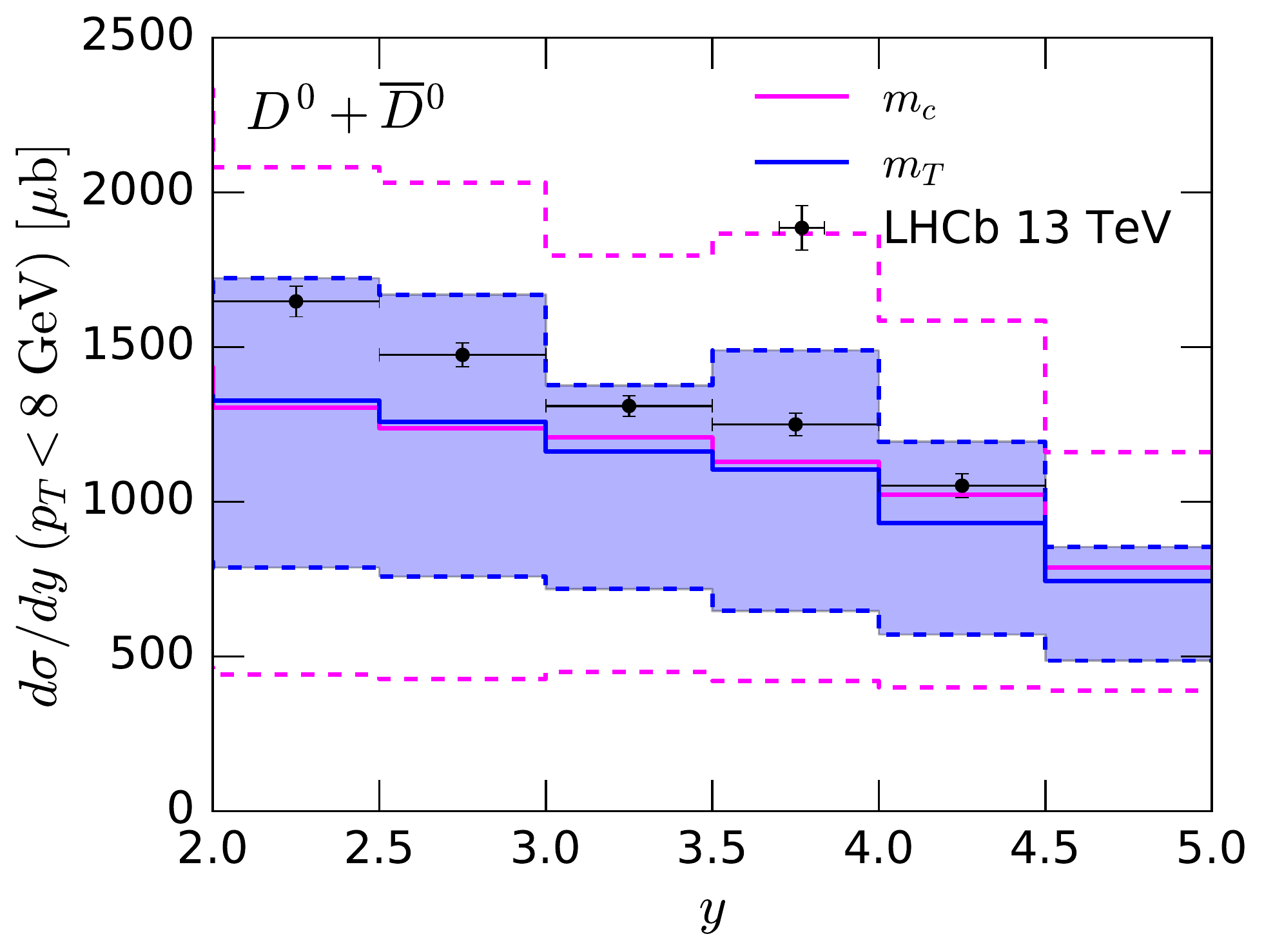}
\includegraphics[width=0.49\textwidth]{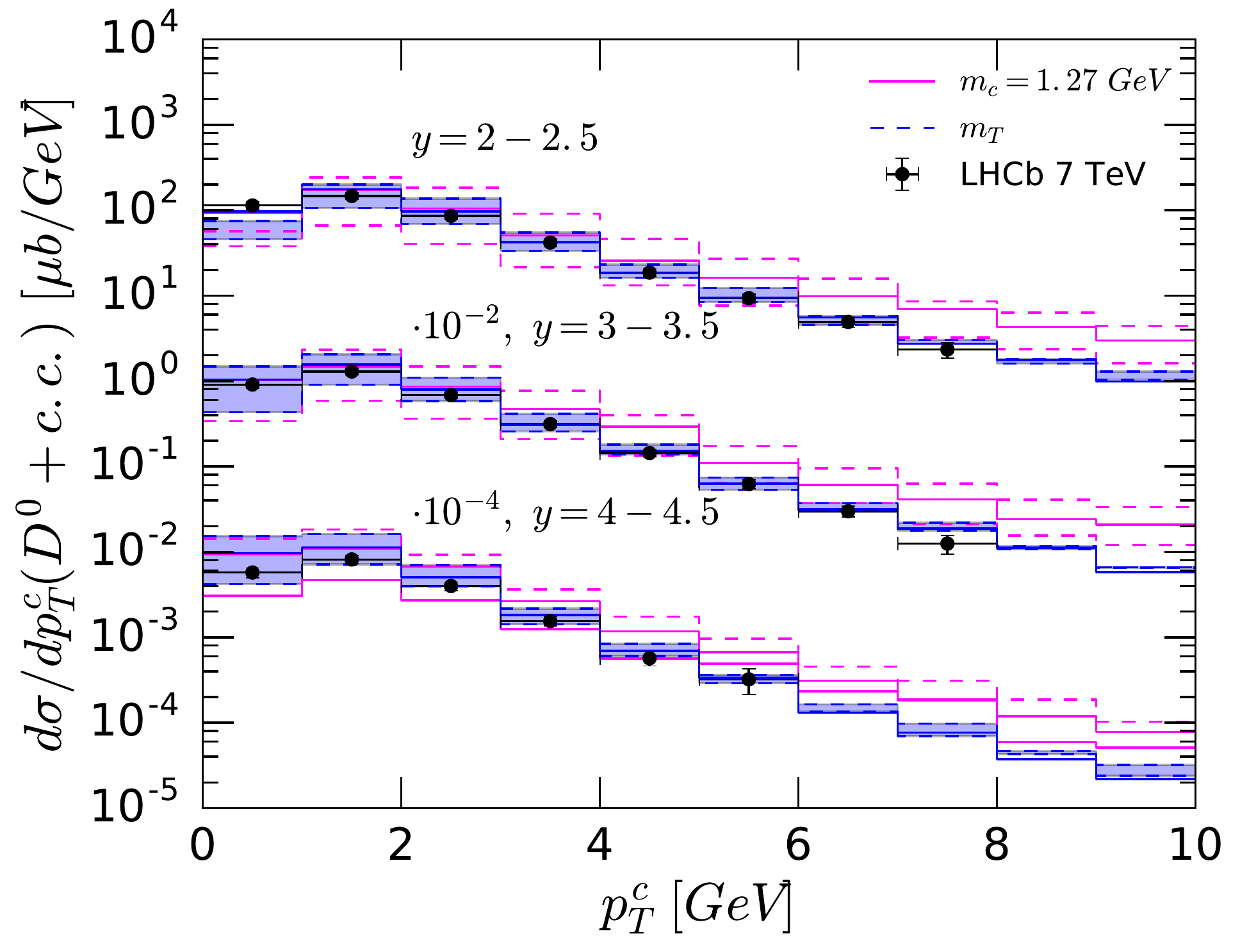}
\includegraphics[width=0.49\textwidth]{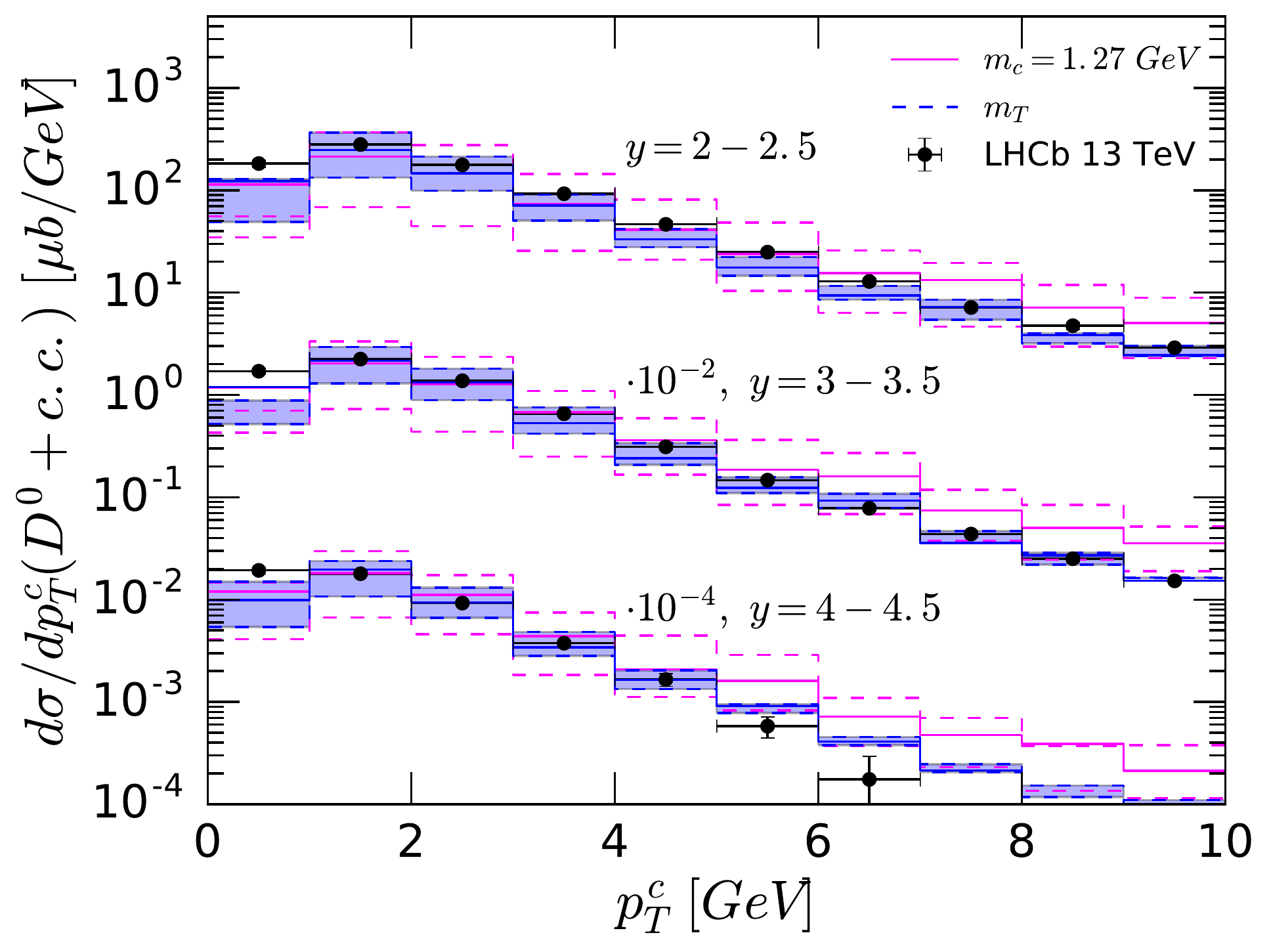}
\caption{Rapidity distributions for $pp\to D^0/\bar{D}^0\, X$
at $\sqrt{s}=7$ TeV (upper left)
  and at $ 13$ TeV (upper right)  for
 transverse momentum, $p_T<8$ GeV,
   and $p_T$ distributions in rapidity ranges with $\Delta y=0.5$, scaled by 1, 0.01
  and $10^{-4}$, for $\sqrt{s}=7$ TeV (lower left)
  and $ 13$ TeV (lower right), obtained with
nCTEQ15-01 PDFs \cite{Dulat:2015mca} 
  compared
with LHCb data
\cite{Aaij:2015bpa,Aaij:2013mga}.
The shaded blue region shows the range of scale dependence
proportional to  $m_T$, while the dashed magenta outer histograms show
the scale dependence proportional to $m_c=1.27$ GeV. { The range
of scales is the same as in fig.\ \ref{fig:sigpert}}}
\label{fig:dsdylhcbnlo}
\end{figure}
%
\begin{figure}[h]
\centering
\includegraphics[width=0.49\textwidth]{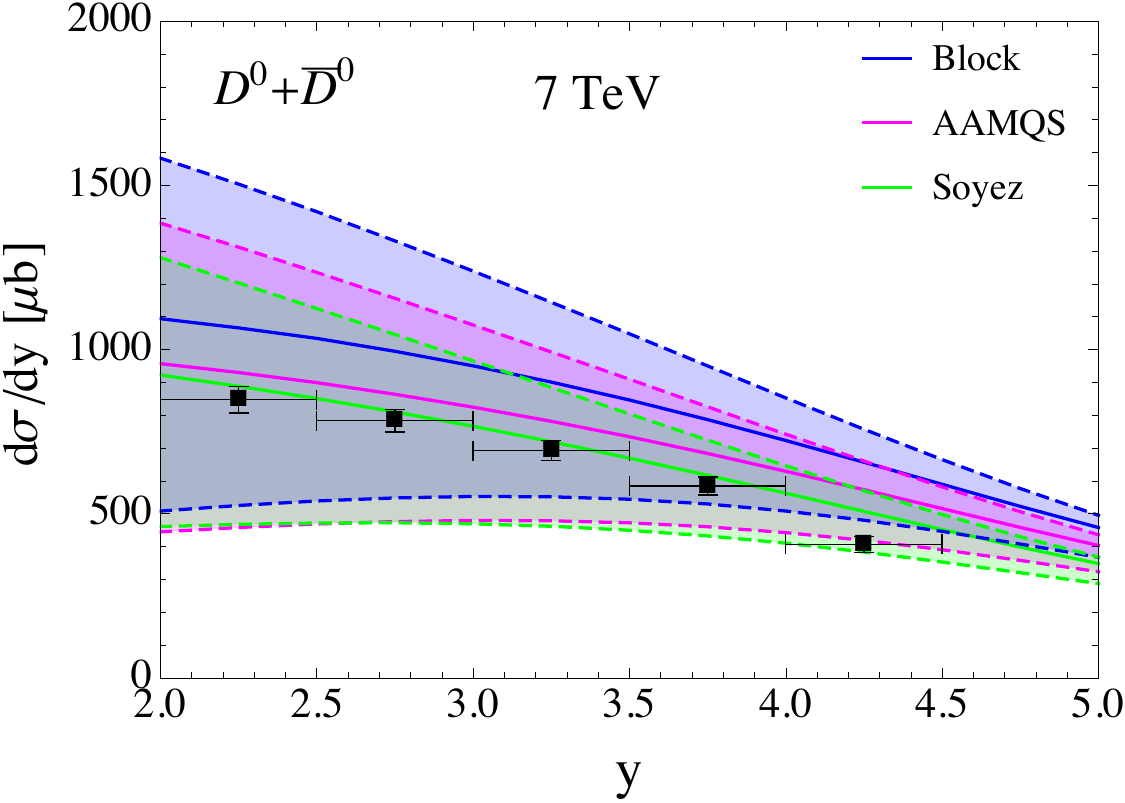}
\includegraphics[width=0.49\textwidth]{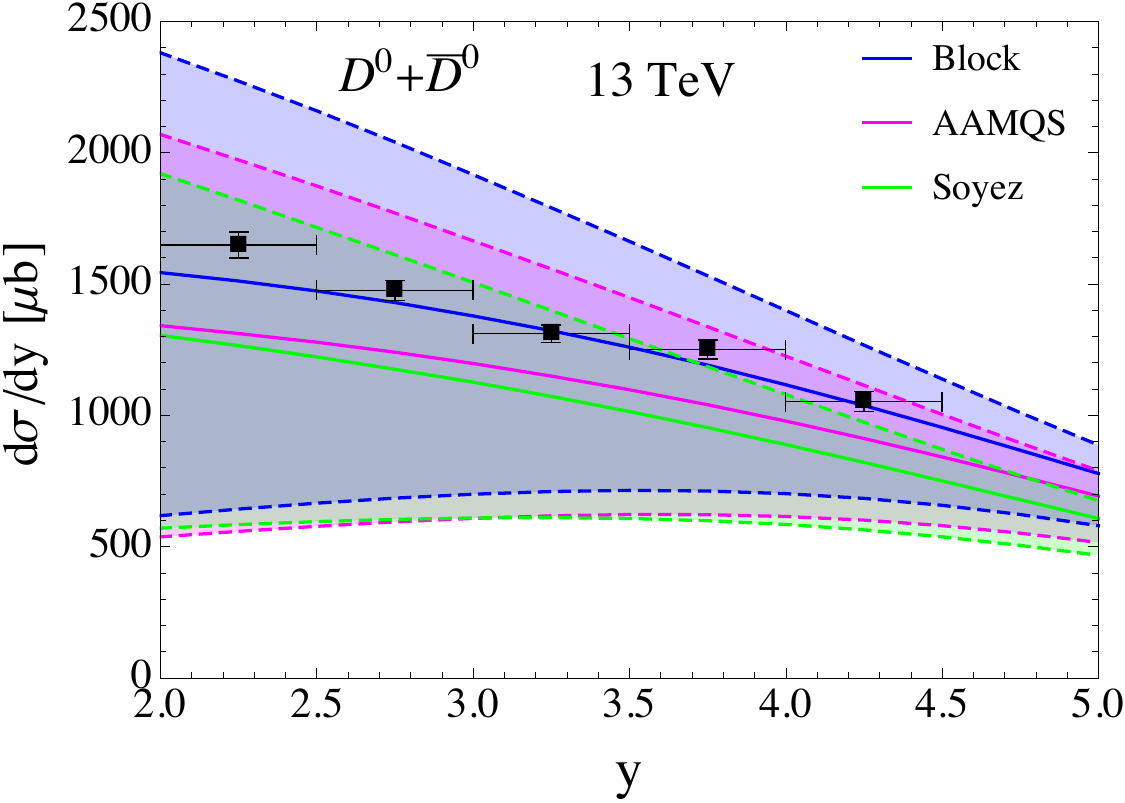}
\caption{As in fig.~\ref{fig:dsdylhcbnlo},
the rapidity distribution for $pp\to D^0/\bar{D}^0\, X$
at $\sqrt{s}=7$ TeV (left)
  and at $ 13$ TeV (right),
   calculated using dipole
  model
with Block, AAMQS and Soyez dipoles and the LO CT14 gluon PDF with factorization
scales ranging from $M_F=m_c$ to $4m_c$. The solid curves are the results with the $M_F=2m_c$.
The value of $\alpha_s$ is fixed to $\alpha_s=0.373$. The LHCb data are from refs.\ \cite{Aaij:2015bpa,Aaij:2013mga}.}
\label{fig:dsdylhcbdm}
\end{figure}
%
\begin{figure}[h]
\centering
\includegraphics[width=0.49\textwidth]{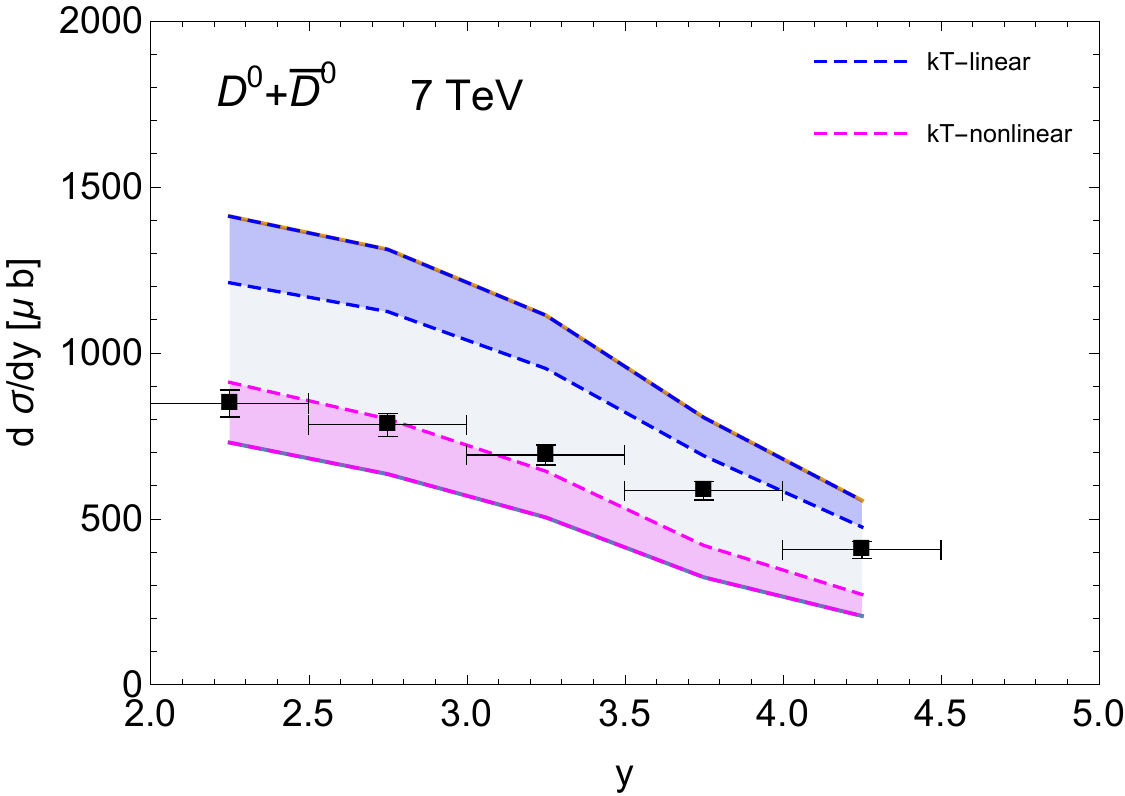}
\includegraphics[width=0.49\textwidth]{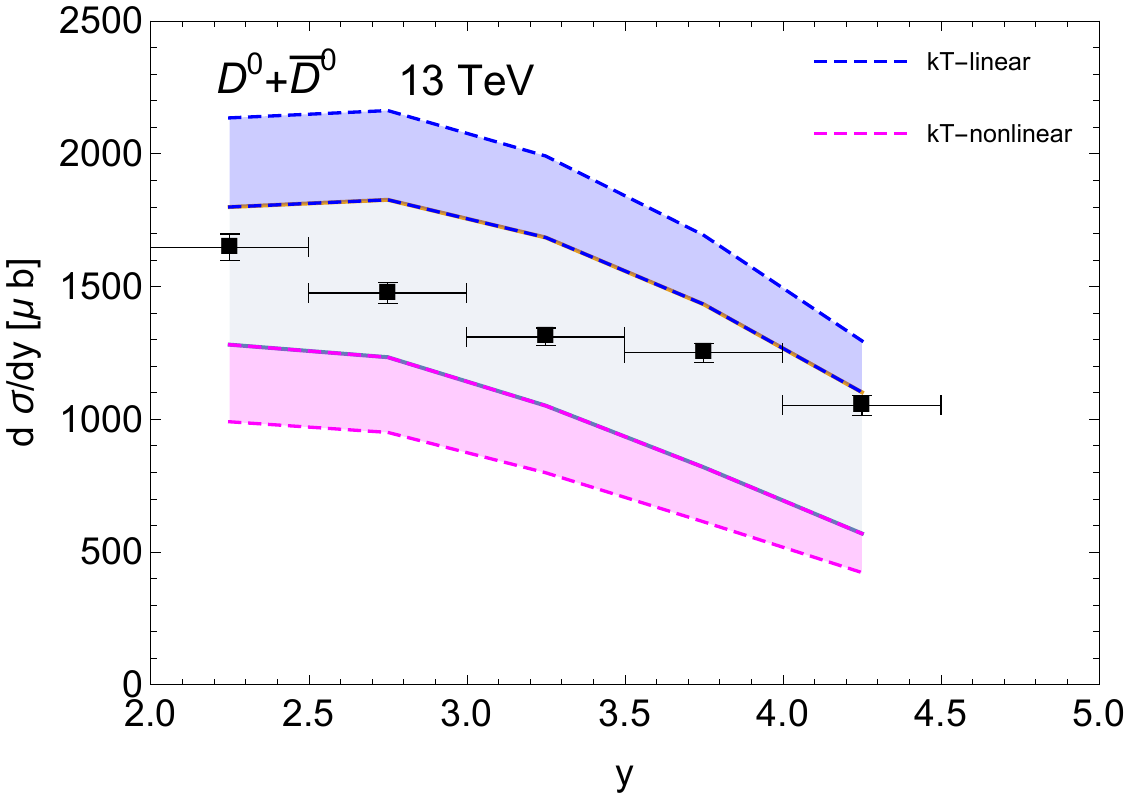}
\includegraphics[width=0.49\textwidth]{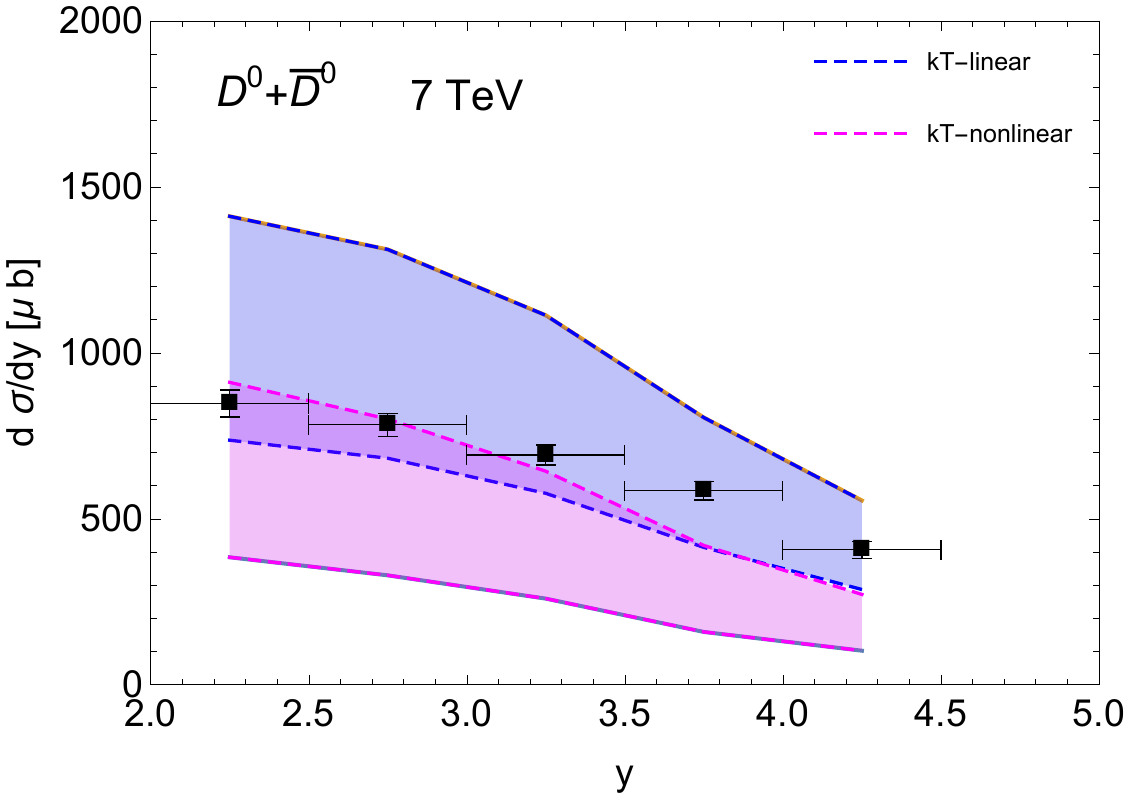}
\includegraphics[width=0.49\textwidth]{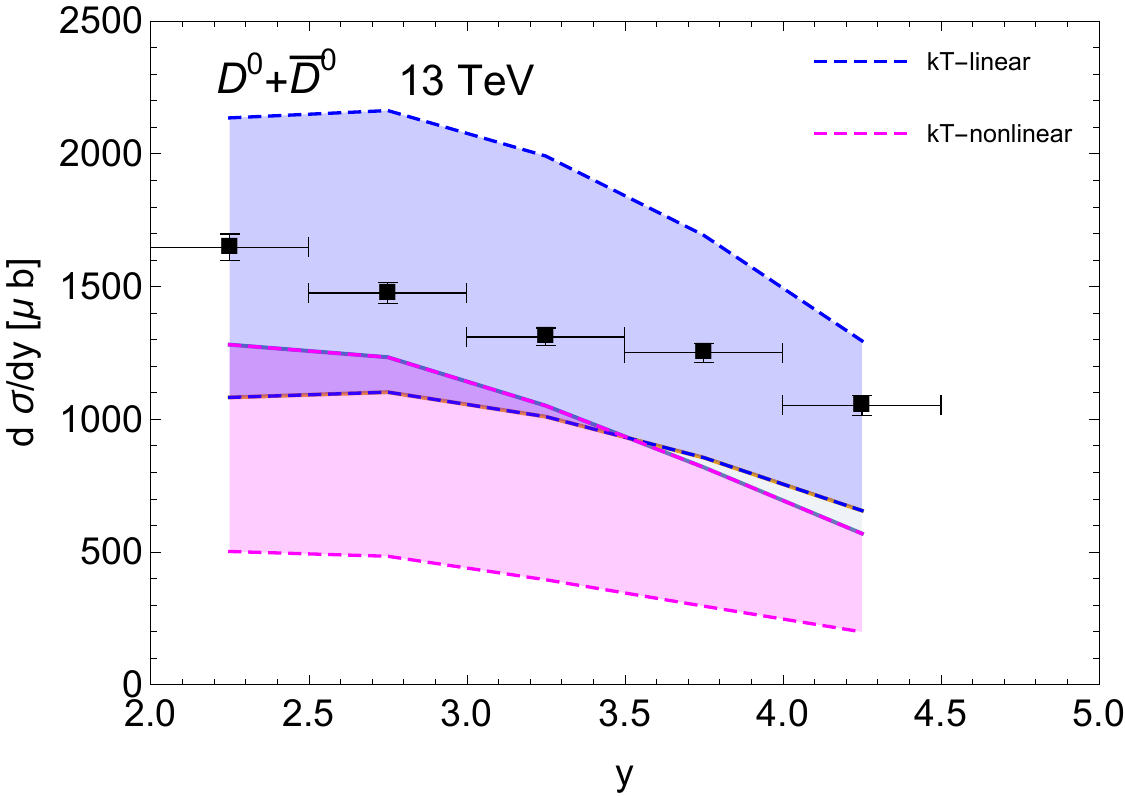}

\caption{Rapidity distribution for $pp\to D^0/\bar{D}^0\, X$
at $\sqrt{s}=7$  TeV (left)
  and at $13$ TeV (right) obtained
  using $k_T$ factorization formalism, and data from LHCb experiment \cite{Aaij:2015bpa,Aaij:2013mga}. Blue-bands correspond to the  gluon density without saturation, magenta to the calculation including saturation. Bands represent  the  variation in the $k_T$ integration upper limit corresponding to  $(m_T,k_{\rm max})$ (lower plots) and $(2.5 m_T,k_{\rm max})$ (upper plots). See text for more explanation. }
\label{fig:dsdylhcbkt}
\end{figure}

In refs.~\cite{Aaij:2015bpa,
  Aaij:2013mga},  data are presented for transverse momentum
and rapidity distributions.
Imposing a cut on transverse momentum,
$p_T<8$ GeV where possible (see below),
we show $d\sigma/dy$ for $2 \leq y \leq 4.5$
 evaluated using perturbative NLO, dipole model and
$k_T$
factorization. We also show the transverse momentum distributions in rapidity ranges $y=2-2.5$,
$y=3-3.5$ (scaled by $10^{-2}$) and $y=4-4.5$ (scaled by $10^{-4}$) where possible.
All the calculations were performed by computing the differential distribution of charm quarks,
multiplied by the fragmentation fraction for $c\to D^0$, and finally a factor of two  was included to account for antiparticles.
The results are shown in figs.~\ref{fig:dsdylhcbnlo},
~\ref{fig:dsdylhcbdm},~\ref{fig:dsdylhcbkt} respectively. The  highest rapidity bin from LHCb does not include
the $p_T$ to 8 GeV, but the distribution falls off rapidly. The dipole
model already includes the full $p_T$ range, but again, the steep
distribution in $p_T$ means the dipole result is a good approximation.

In fig.~\ref{fig:dsdylhcbnlo} we show NLO differential distributions of charm pairs evaluated
using the free proton nCTEQ15-01 PDFs. The blue
shaded band shows the prediction for the range of scales proportional to
$m_T$, while the dashed magenta lines show the predictions for the scale
dependence proportional to $m_c$, the scale range taken to be the same
as in fig.~\ref{fig:sigpert}.  The $m_T$ range used in our flux
evaluations is consistent with the LHCb results, as is the very large
range coming from  $m_c$ dependent scales. Based on this comparison,
in our discussion of the prompt flux we use our error band that 
corresponds to the range of scales proportional to $m_T$.

Fig.~\ref{fig:dsdylhcbdm} presents the differential cross sections for charm pair production from the dipole models for $2.0 < y < 4.5$.
In dipole model calculations, there is no explicit $p_T$ dependence so
we do not
make any $p_T$ cuts. As the LHCb data show, the
differential cross section decreases with $p_T$. For example, already
between $p_T=7-8$ GeV, the differential cross section in the $y=2-2.5$
bin is less than 3\% of the differential cross section in the
$p_T=1-2$ GeV bin for $\sqrt{s}=7$ TeV  \cite{Aaij:2013mga}, so the
lack
of a $p_T$ cut should not introduce a large error in the rapidity comparison.

The cross section in the dipole picture is naturally written in a mixed representation where the momentum dependence in the transverse plane is Fourier transformed to coordinate space, while the longitudinal momentum dependence is kept in momentum space (corresponding to the integration variables $\vec r$ and $z$ in eq.~(\ref{eq:dipolexsec}). It is possible to obtain a formula for the differential cross section
$d\sigma^{G p \to Q\bar Q X}/d^2k_T$, which, when integrated
over $d^2k_T$ gives eq.~(\ref{eq:dipolexsec})
\cite{Raufeisen:2002ka}, and in principle it
could be possible to obtain the $p_T$
dependence of the cross section in this way.
In the context of calculating the prompt neutrino flux,
this does not seem to be a fruitful approach, but
it has, however, been used to demonstrate that if the
dipole cross section is calculated in
LO QCD, then the LO pQCD
approach is exactly equivalent to the dipole approach \cite{Raufeisen:2002ka}.

In fig.~\ref{fig:dsdylhcbdm} the blue band is the range from the
different dipole models with the factorization scale $M_F= m_c$ to $M_F=4m_c$
for the Block dipole,
 the area shaded with magenta shows the differential cross sections
with the AAMQS dipole, and the shaded green band shows the Soyez
dipole.
The central value of each model with the scale $M_F=2m_c$ is also presented with the solid curve.
For the rapidity distributions of the charm pair produced cross
sections in the dipole models with the wide uncertainty due to
the factorization scales agree with the LHCb data for $\sqrt{s}$=7 TeV
and $13$ TeV, however this is possible only because of the
wide range of factorization scales.

The calculation using the $k_T$ factorization formalism is shown in fig.~\ref{fig:dsdylhcbkt}.  It can be seen that the differential distribution in rapidity is quite sensitive to the resummation of $\ln 1/x$ terms, in particular to the parton saturation effects. As expected, the calculation with saturation effects included is substantially below the calculation without it. To illustrate the sensitivity to the small $x$  effects we have varied the upper limit on the integral over the transverse momentum of the off-shell gluon. This illustrates how a significant contribution comes from the lack of transverse momentum ordering in this process, and illustrates the sensitivity to transverse momenta of partons larger than the typical transverse momentum of the produced charm quark.  We see that the calculation is quite sensitive to
the change of the upper limit in  integrals over the transverse momentum of the off-shell small $x$ gluon. The plots with wider bands (lower plots in fig.~\ref{fig:dsdylhcbkt})
were performed with the variation of the upper limit on this integral between $(m_T,k_{\rm max})$ where $k_{\rm max}$ is essentially
kinematical limit in the subprocess. The smaller bands  (upper plots in fig.~\ref{fig:dsdylhcbkt}) correspond to the variation of $(2.5 m_T,k_{\rm max})$.   We see
that apparently the LHCb data  exclude the most extreme limits of these calculations. For the evaluation of the neutrino flux, we have thus
included the linear and non-linear calculation with scale choices of $2.5m_T$ and  $k_{\rm max}$  correspondingly. They correspond  to the inner band (gray shaded area) in the upper
plots in fig.~\ref{fig:dsdylhcbkt}.

\begin{figure}[h]
\centering
\includegraphics[width=0.49\textwidth]{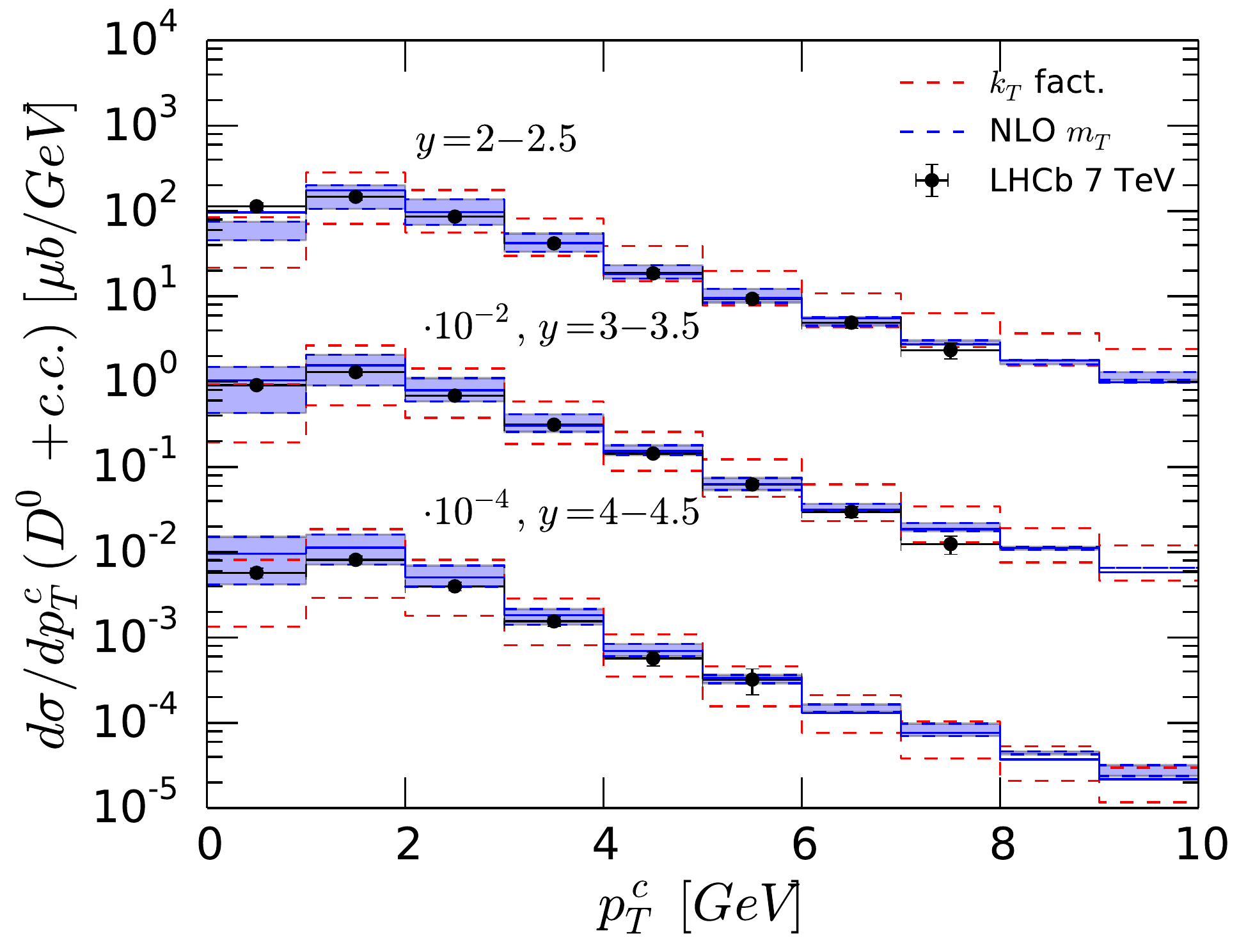}
\includegraphics[width=0.49\textwidth]{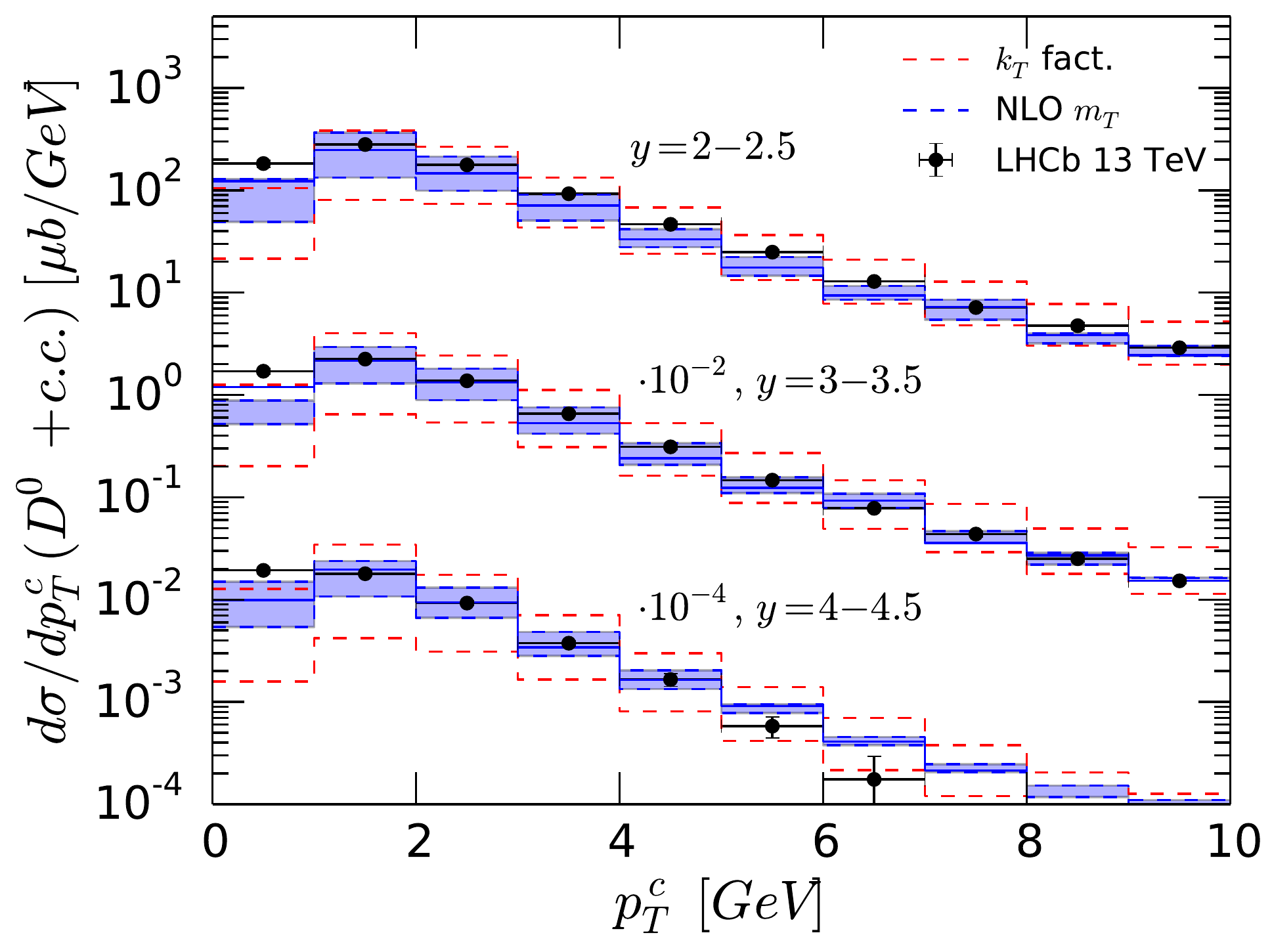}

\caption{Transverse momentum distribution for $pp\to D^0/\bar{D}^0\, X$
at $\sqrt{s}=7$  TeV (left)
  and at $13$ TeV (right) obtained
  using $k_T$ factorization formalism, with data from LHCb experiment
  \cite{Aaij:2015bpa,Aaij:2013mga}. Red-lines correspond to the
  calculation from $k_T$ factorization using gluon with saturation
  (lower lines) and without saturation (upper lines).  This
  calculation is compared with the NLO calculation with $m_T$ scale
  variation shown in fig.\ 
\ref{fig:dsdylhcbnlo}.}
\label{fig:dsdptlhcbkt}
\end{figure}

In Fig.~\ref{fig:dsdptlhcbkt} we show transverse momentum distributions obtained within the $k_T$ factorization formalism
as compared with the our pQCD  NLO results presented in fig.\
\ref{fig:dsdylhcbnlo}. 
 We see that both computations are in agreement with each other and the data for this distribution. 

Figs.~\ref{fig:dsdylhcbnlo}, \ref{fig:dsdylhcbdm} and \ref{fig:dsdylhcbkt} all show
common trends in the comparison of theory with experiment.
In each case, the ratio of the rapidity distributions at
$\sqrt{s}=13$ TeV and at
$\sqrt{s}=7$ TeV for fixed QCD parameters is smaller than the data.
However, including the theoretical uncertainty due to the choice of
factorization scale, one finds reasonable agreement with the experimental data.

\begin{figure}[h]
\centering
\includegraphics[width=0.7\textwidth]{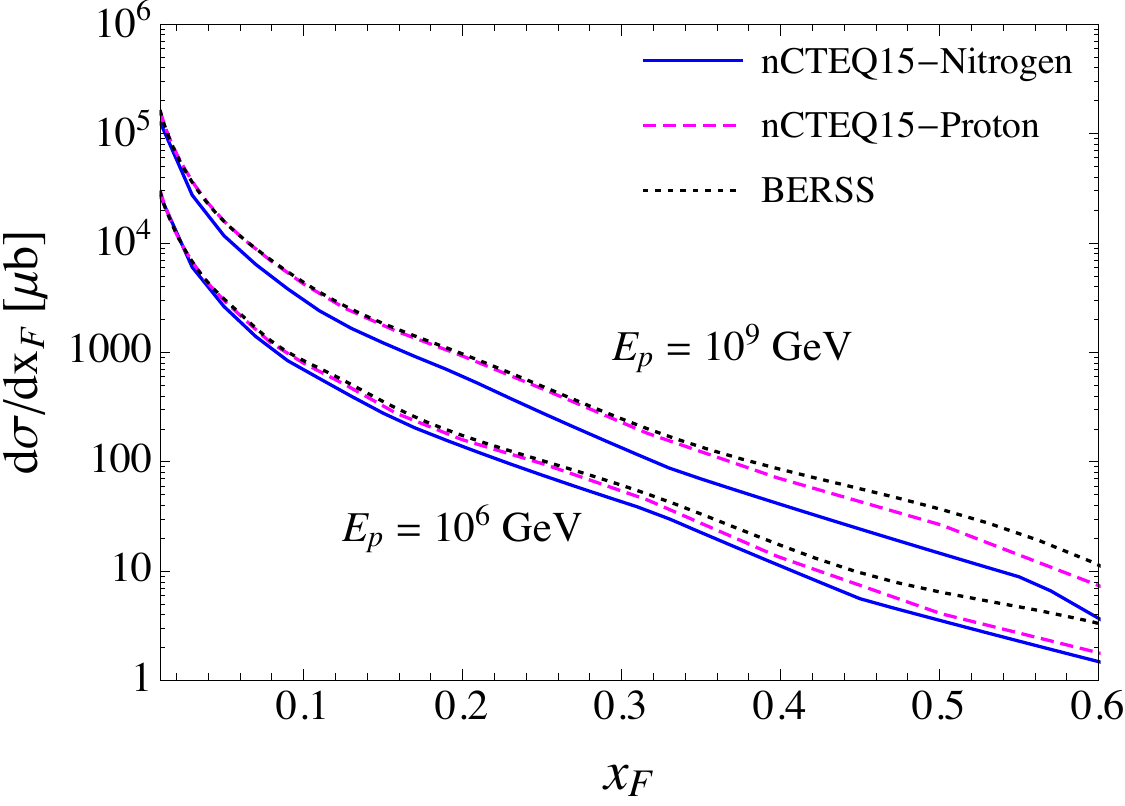}
\caption{ Charm quark differential cross section ${d\sigma}/{d x_E}$
obtained in NLO QCD at energies of $10^6$ GeV (left) and $10^9$ GeV
(right), compared with the central BERSS result (black dotted curve)
for free proton targets (magenta dashed) and bound nucleons (solid
blue curve). }
\label{fig:dsdxnlo}
\end{figure}

\begin{figure}[h]
\centering
\includegraphics[width=0.7\textwidth]{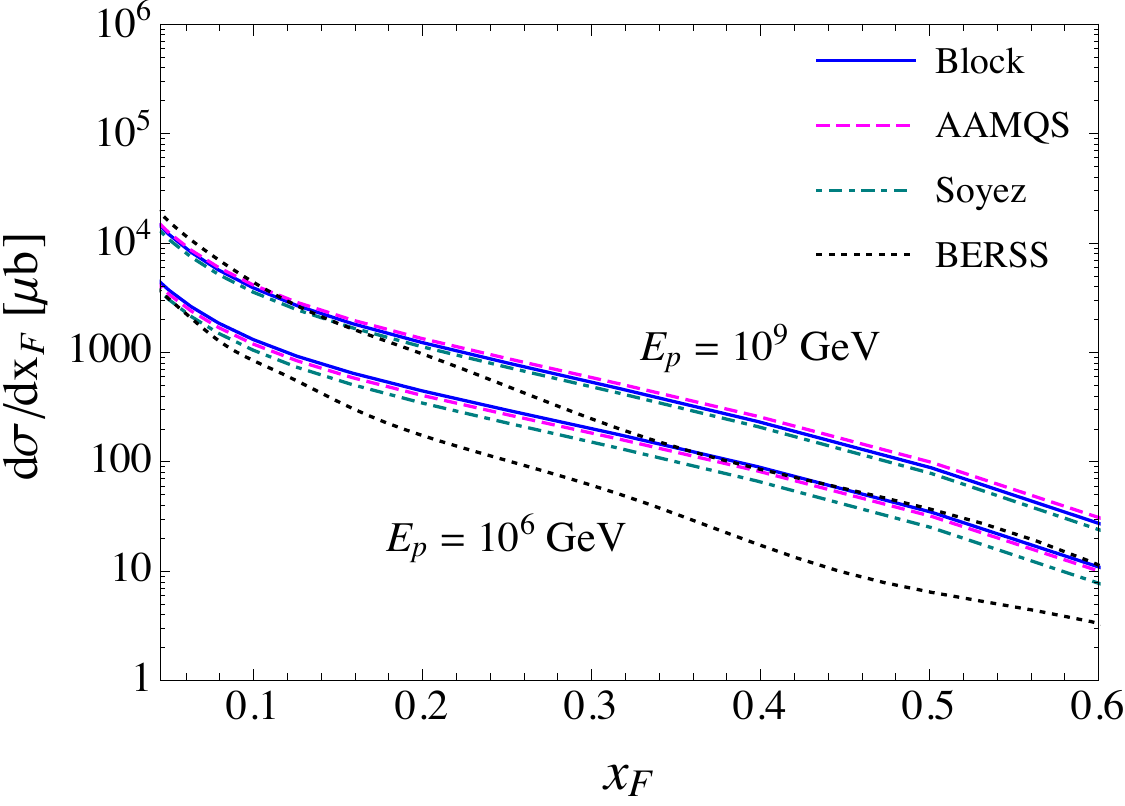}
\caption{The differential cross section  ${d\sigma}/{d x_F}$ as a function of $x_F$ from the dipole models for $c\bar{c}$ production,
evaluated with $\alpha_s=0.373$ and $\mu_F=2 m_c$ using the CT14 LO PDF set.
The charm mass is used 1.4 GeV for the Soyez dipole and 1.27 GeV for the AAMQS and the Block dipoles.
The differential cross section from ref.~\cite{Bhattacharya:2015jpa} is presented for comparison.}
\label{fig:dsdxdm}
\end{figure}

\begin{figure}
\centering
\includegraphics[width=0.45\textwidth]{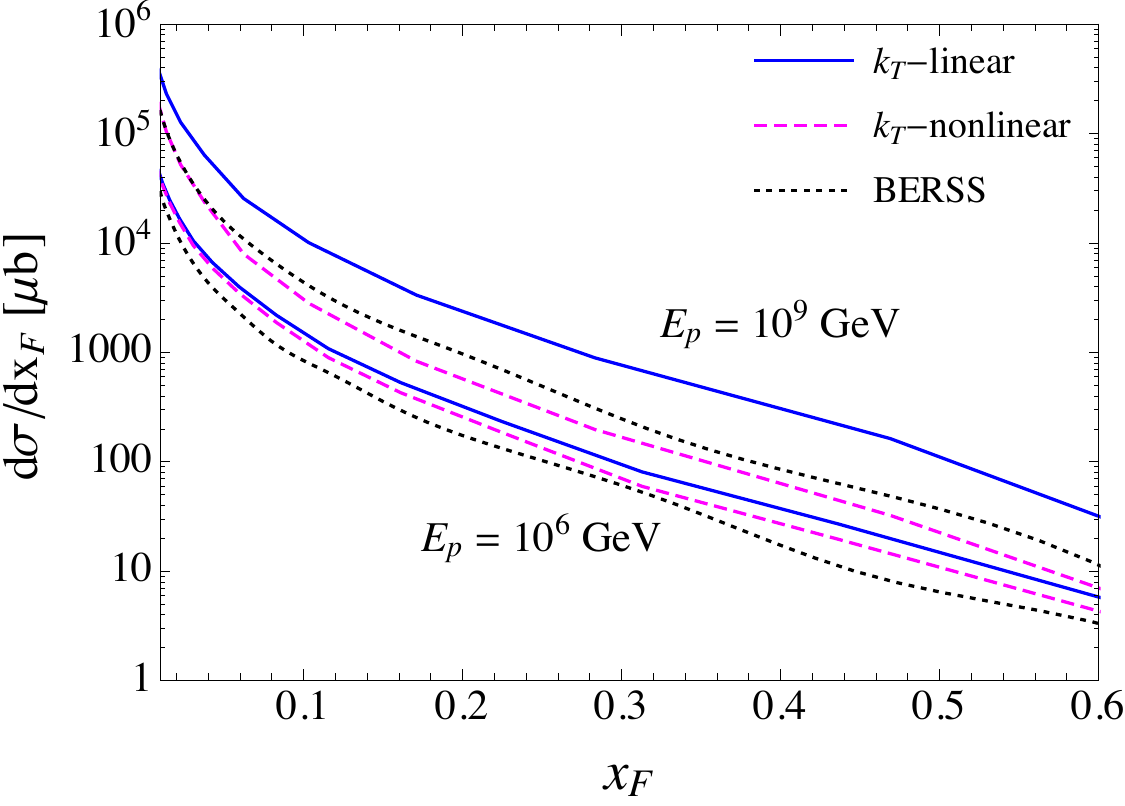}
\includegraphics[width=0.45\textwidth]{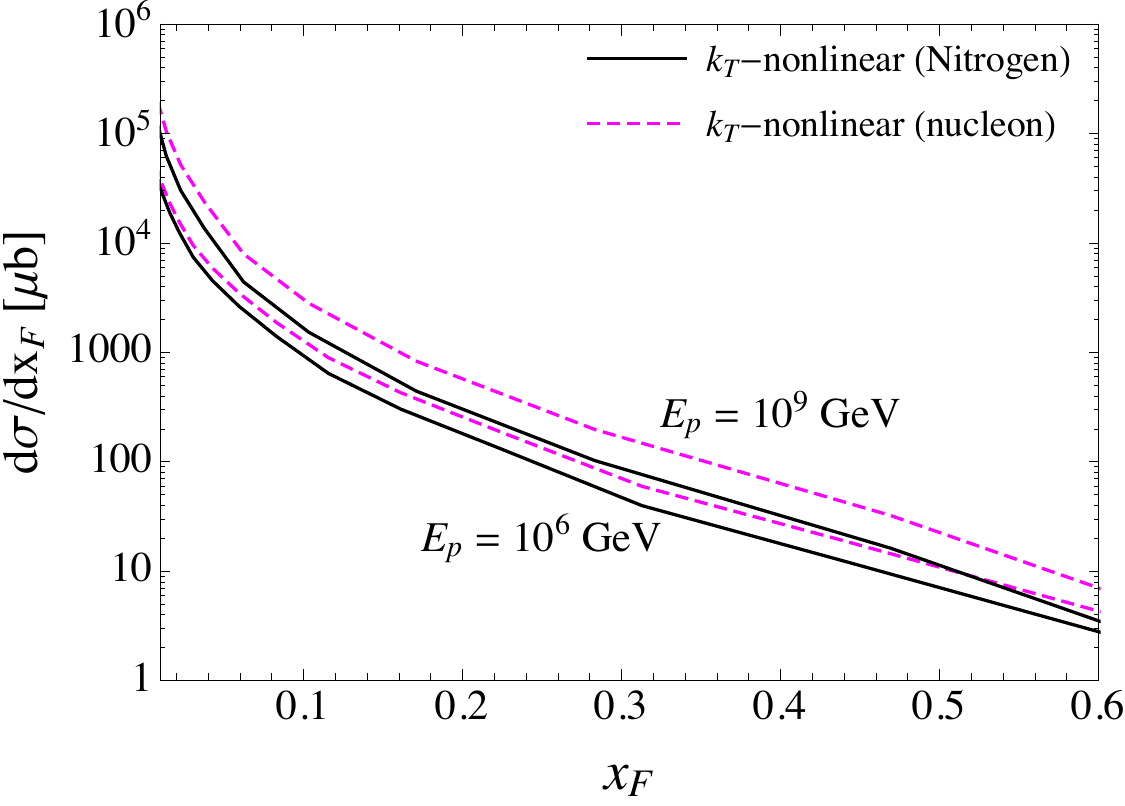}
\caption{Left: The differential cross section ${d\sigma}/{d x_F}$ as a function of $x_F$ for two energies $E=10^6$ GeV and $E=10^9$ GeV from $k_T$ factorization, with linear evolution (solid upper blue),
and non-linear evolution (lower dashed magenta). Shown for comparison is the perturbative
differential cross section from ref.~\cite{Bhattacharya:2015jpa}. Right: Comparison of the  $k_T$ factorization with nonlinear evolution for the proton case (dashed magenta) and the nitrogen (solid black).}
\label{fig:dsdxkt}
\end{figure}

In figs.~\ref{fig:dsdxnlo},~\ref{fig:dsdxdm},~\ref{fig:dsdxkt},
differential  distributions ${d\sigma}/{dx_c}$  are shown for two
energies $E=10^6$
and $E=10^9$ GeV, for  perturbative NLO ($x_c=x_E\simeq x_F$),  for the dipole model ($x_c=x_F$)
and for the calculation using $k_T$ factorization ($x_c=x_F$), respectively.

Additionally, each of the calculations is compared with BERSS calculation
\cite{Bhattacharya:2015jpa}.  The difference in fig.~\ref{fig:dsdxnlo}
between the BERSS results (black dotted curves) and the current
central NLO calculation (magenta dashed curves) can be attributed to
 the different choice of the PDFs (CTEQ10 vs nCTEQ15).
As we show below, this PDF choice reduces the flux by about $1\% - 6\%$ 
between $10^3 - 10^8$ GeV relative to the CTEQ10 choice.

\begin{figure}
\centering
\includegraphics[width=0.45\textwidth]{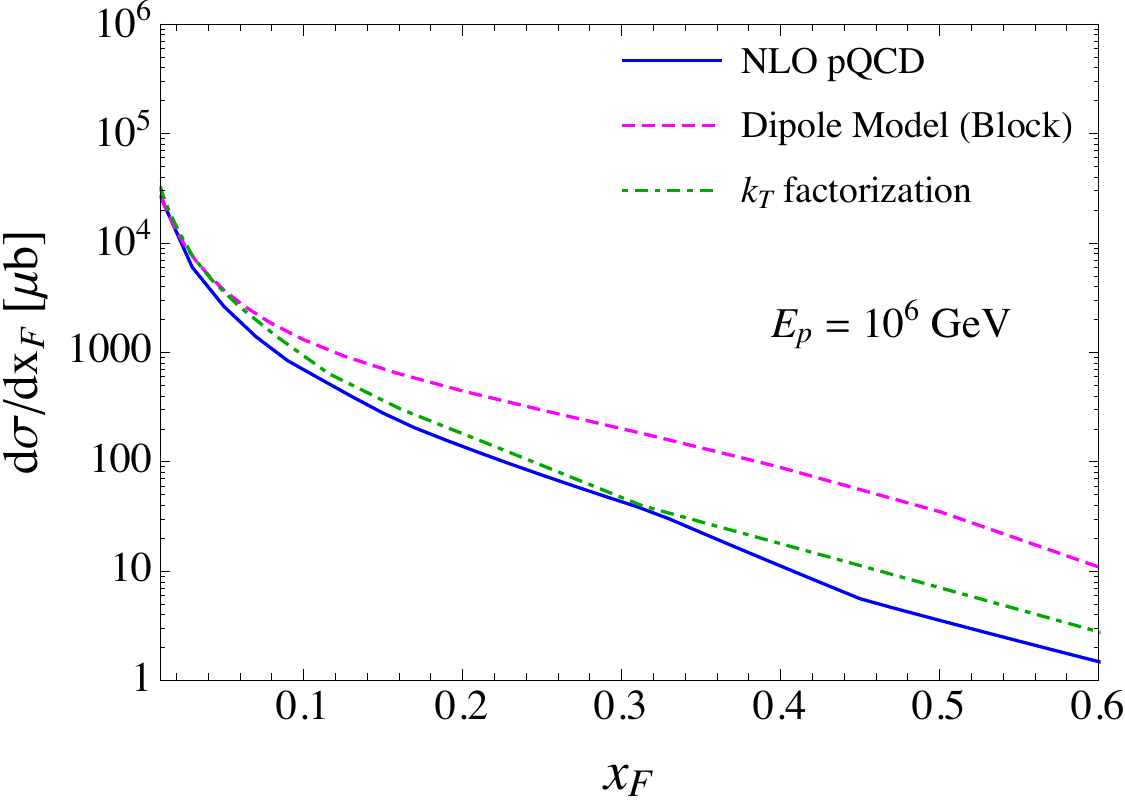}
\includegraphics[width=0.45\textwidth]{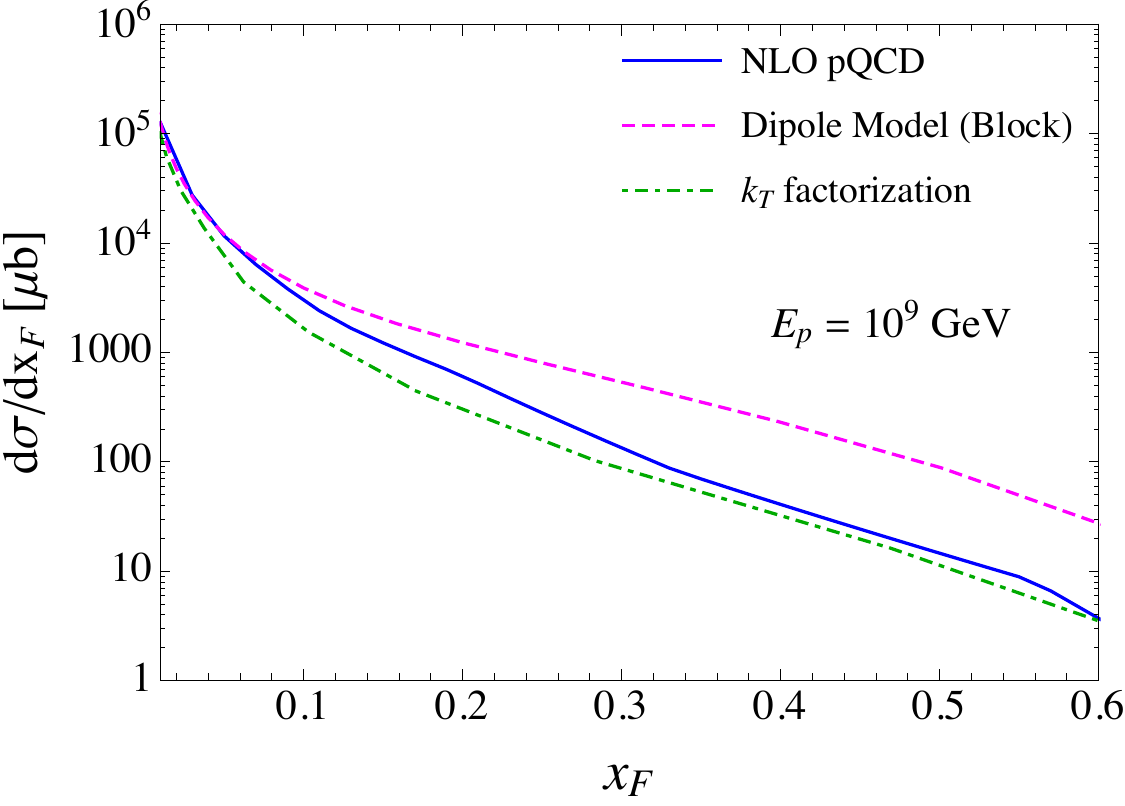}
\caption{The comparison of the differential cross section ${d\sigma}/{d x_F}$ as a function of $x_F$ 
from NLO pQCD (Blue), the dipole model (Magenta) and the $k_T$ factorization with  non-linear evolution (Green) 
at energies of $E=10^6$ GeV and $E=10^9$ GeV. All calculations contain nuclear corrections.
}
\label{fig:dsdxall}
\end{figure}

In fig.~\ref{fig:dsdxnlo} we also show nuclear effects on the differential distribution (blue solid
curve) in the perturbative NLO calculation.  We observe that the nuclear corrections evaluated here are non-negligible for higher energies.
LHCb data on charm production for proton-lead collisions will be able
to constrain nuclear effects for heavy nuclei
in the future, as noted in ref. \cite{Gauld:2015lxa}. 
 Fig.\ \ref{fig:dsdxdm} shows that the $x_F$ distributions in
the dipole model are harder than the BERSS pQCD distributions.
 In fig.~\ref{fig:dsdxkt}, which is from the $k_T$ factorization approach, 
one can clearly see the effect of resummation of $\ln 1/x$ effects, as well as the impact of gluon saturation. The linear BFKL-DGLAP evolution
tends to give large contribution at large $x_F$, which increases at larger energies. On the other hand the calculation with the non-linear evolution
tends to give lower values due to the suppression of the gluon density at low $x$.   We see that the effect of the nuclear corrections are also non-negligible in this approach, and further  reduce the cross section for higher energies and large values of $x_F$.
Finally, in Fig.~\ref{fig:dsdxall} we compared calculations from all approaches which include the nuclear corrections. The NLO perturbative and $k_T$ factorization seem consistent with each other, on the other hand the dipole calculation is somewhat higher than the other two. 

%
%
%

\section{Prompt fluxes}

\subsection{Overview}

The prompt fluxes are evaluated using the semi-analytic $Z$-moment method. This procedure is described in detail
in, e.g.,  refs.~\cite{Gaisser:1990vg} and \cite{Lipari:1993hd}.
This one-dimensional method consists of using spectrum weighted
differential cross section for the production of hadrons, and for decays of hadrons to neutrinos, as inputs to approximate low energy and high energy solutions to the coupled cascade
equations for $p,N,h,\nu$. The prompt flux contributions come from charmed hadrons
$h=h_c=D^0,\ D^+,\ D_s,\ \Lambda_c$ and $b$ hadrons
$h=h_b=B^0,\ B^+,\ B_s,\ \Lambda_b $ and their antiparticles.
The general form of the cascade equations for particle $j$ and column depth $X$ are
\begin{eqnarray}
\frac{d\phi_j(E,X)}{dX}&=&-\frac{\phi_j(E,X)}{\lambda_j(E)} - \frac{\phi_j(E,X)}{\lambda_j^{\rm dec}(E)}
+ \sum S(k\to j)\, ,\\
S(k\to j) &=& \int_E^{\infty}dE ' \frac{\phi_k(E',X)}{\lambda_k(E')}
\frac{dn(k\to j;E',E)}{dE}\, ,
\end{eqnarray}
\begin{eqnarray}
\frac{dn(k\to j;E',E)}{dE}&=& \frac{1}{\sigma_{kA}(E')}\frac{d\sigma(kA\to jY;E',E)}{dE }
\quad {\rm (interaction)\, ,}\\
\frac{dn(k\to j;E',E)}{dE}&=& \frac{1}{\Gamma_k(E')}\frac{d\Gamma (k\to jY;E',E)}{dE}\quad {\rm (decay)}
\ .
\end{eqnarray}
The $Z$-moment method approximates the source term for $k\to j$
with interaction
length $\lambda_k$
\begin{eqnarray}
S(k\to j) &\simeq & Z_{kj}(E)\frac{\phi_k(E,X)}{\lambda_k(E)}\, ,\\
Z_{kj}(E) &=& \int _E^{\infty}dE ' \frac{\phi_k^0(E')}{\phi_k^0(E)}
\frac{\lambda_k(E)}{\lambda_k(E')}
\frac{dn(k\to j;E',E)}{dE} \, ,
\label{eq:zmomdef}
\end{eqnarray}
for
$\phi_k (E,X)=\phi_k^0(E)f(X)$. The factorization of the $X$ dependence in the flux is
a good approximation for the Earth's atmosphere, where we approximate the target nucleon density
with an exponential atmosphere
\begin{equation}
\rho = \rho_0\exp(-h/h_0)\ ,
\end{equation}
where $h_0=6.4$ km and $\rho_0 h_0=1300$ g/cm$^2$. The column depth is then given by $X(\ell,\theta) = \int_\ell^\infty d\ell' \rho(h(\ell',\theta))$,
where $h(\ell,\theta)$ is the height at distance from the ground $\ell$ and
zenith angle $\theta$. We shall be focusing on vertical fluxes, $\theta=0$.

Using the assumption of the  exponential dependence of density on height in the  atmosphere, the approximate solutions can be conveniently written in terms of the interaction
lengths $\Lambda_k=\lambda_k/(1-Z_{kk})$, giving
$f(X) = \exp{(-X/\Lambda_k)}$. For particle $k$ decays in the relativistic limit,
$\lambda_k^{dec} = E_k \tau_k\rho/m_k$.
 As a result, the high and low energy
lepton fluxes at Earth scale with the cosmic ray flux and
can be expressed as
\begin{eqnarray}
\phi_{h\to \nu}^{\rm low} &= &\sum_h \frac{Z_{Nh}Z_{h\nu}}{1-Z_{NN}}\phi_N^0\, ,\\
\phi_{h\to \nu}^{\rm high} &= &\sum_h \frac{Z_{Nh}Z_{h\nu}}{1-Z_{NN}}\frac{\ln(\Lambda_h/\Lambda_N)}
{1-\Lambda_N/\Lambda_h}\frac{\epsilon_h}{E}
\phi_N^0 \, ,
\end{eqnarray}
where each $Z$-moment and effective interaction length $\Lambda_k$
depends on the energy of the prompt lepton. We have here defined the critical energy
$\epsilon_k=(m_k c^2h_0/c\tau_k) g(\theta)$ for hadron $k$, which separates the high- and low-energy regimes. The angular dependence of the flux enters through the critical energy, with the function
$g(\theta)\simeq 1/\cos \theta$ for small angles close to vertical, but for angles near
horizontal, it is more complicated due to the geometry of the Earth and the atmosphere. We will not need the details here, but more information can be found in  e.g.~\cite{Lipari:1993hd}.

For both $h_c$ and $h_b$, the transition between low energy and high energy fluxes is at
energies $E_\nu \sim 10^8$ GeV. We evaluate the flux by taking the sum over hadron
contributions:
\begin{equation}
\phi_\nu = \sum_h \frac{\phi_{h\to \nu}^{\rm low}\phi_{h\to \nu}^{\rm high}}{(
\phi_{h\to \nu}^{\rm low}+\phi_{h\to \nu}^{\rm high})}\ .
\end{equation}

This approach is applicable to $\nu=\nu_\mu, \nu_e$ and $\nu_\tau$
plus their antiparticles (and to $\mu^++ \mu^-$, which are stable
nearly massless leptons to first approximation).
The prompt electron neutrino flux and prompt muon flux are essentially equal to the prompt
muon neutrino flux. This comes from the nearly identical kinematics in the semileptonic
charmed hadron and $b$-hadron decays for the $e\nu_e$ and $\mu\nu_\mu$ final states.
Tau neutrinos, however, are produced at a lower rate, with the
dominant source being $h=D_s$
\cite{Pasquali:1998xf,Martin:2003us}
where
$D_s\to \nu_\tau \tau$ comes with a branching fraction of $5.54\pm
0.24\% $ \cite{Agashe:2014kda}.

Tau decays are also prompt, with $c\tau_\tau=87.03\ \mu$m
(as compared to $c\tau_{D_s}=149.9\ \mu$m  \cite{Agashe:2014kda}.). Tau energy
loss in the atmosphere is negligible, so we include the chain decay
$D_s\to \tau\to \nu_\tau$ as well as the direct decay $D_s\to
\nu_\tau$.
Since most of the $D_s$ energy goes to the tau lepton, and approximately 1/3 of the tau energy goes to the tau neutrino, the chain decay
contribution is larger than the direct contribution of $D_s$ decays to the prompt tau neutrino flux.
For the chain decay $D_s\to \tau\to \nu_\tau$,
\begin{equation}
Z_{D_s\nu_\tau}^{(chain)} (E)=\int_E^\infty dE_D \int_E^{E_D} dE_\tau \frac{\phi_D(E_D)}{\phi_D(E)}\frac{E}{E_D}
\frac{dn_{D_s\to \tau}}{dE_\tau}(E_D,E_\tau)\frac{dn_{\tau \to \nu_\tau}}{dE}(E_\tau,E) \ .
\end{equation}
In the results below, we approximate all $\tau$ decays as prompt. This
overestimates the tau neutrino flux above $E_\nu\sim 10^7$ GeV \cite{Martin:2003us},
however, at these high energies, the tau neutrino flux is quite low anyway.

\subsection{Cosmic ray flux, fragmentation and decays}

A broken power law (BPL) approximation of the cosmic ray nucleon flux in terms of
the energy per nucleon $E$ is of the form
\begin{eqnarray}
\nonumber
\phi_N^0(E) \Biggl[\frac{\rm nucleons}{\rm cm^2\, s\, sr\,
  GeV}\Biggr]
&=& 1.7\ (E/{\rm GeV})^{-2.7}\quad E< 5\cdot 10^6\ {\rm GeV}\\
&=& 174 \ (E/{\rm GeV})^{-3}\quad E>5\cdot 10^6 \ {\rm GeV}\ .
\end{eqnarray}
The BPL has been used to evaluate the prompt flux in many earlier
references. While the all particle spectrum resembles a broken power
law, recent analyses have shown that it poorly represents the nucleon spectrum, even though the composition of
the cosmic rays is not completely known
\cite{Gaisser:2011cc,Gaisser:2013bla,Stanev:2014mla}.
Recent three component models \cite{Gaisser:2011cc} with extragalactic
protons (called H3p here) or an extragalactic mixed composition (H3a)
come from an analysis of cosmic ray measurements paired with fits to
functional forms for the spectrum by composition. The four component
model by Gaisser, Stanev and Tilav (GST*)\cite{Gaisser:2013bla} is labeled
here by GST4. To compare with other flux calculations, we use the BPL
for reference. The H3a flux dips more precipitously at high energies
than the H3p flux. We show below that the H3p and GST4 inputs to
the prompt flux lead to similar predictions.

Our three approaches to charm production provide the $x_E$ or $x_F$
distribution of the charmed quark, and similarly for the $b$ quark. We
use the Kniehl and Kramer fragmentation functions for charm
\cite{Kniehl:2006mw}, using the LO
parameters with the overall normalization scaled to account
for updated fragmentation fractions determined from a recent review of
charm
production data in ref.~\cite{Lisovyi:2015uqa}. The original
normalizations in  ref.~\cite{Kniehl:2006mw} had the fragmentation fractions add to 1.22 rather than
1.00.  Using the updated fragmentation fractions to rescale the
fragmentation functions, the sum of fractions is 0.99.
For the $B$ meson fragmentation, we use the power law form of Kniehl
et al.\ from ref.~\cite{Kniehl:2008zza},
rescaled to match the fragmentation fractions of
ref.~\cite{Aaij:2011jp}.  The details of
the fragmentation functions and parameters are shown in
Appendix \ref{ssec:appfrag}.
Decay distributions and parameters for the branching fractions and
effective hadronic masses for semi-lepton heavy meson decays are
listed in Appendix \ref{ssec:appdecay}.

The Kniehl and Kramer (KK) provide LO and NLO
fragmentation functions \cite{Kniehl:2006mw},
the later, in principle, being more suitable for our NLO pQCD calculation.
However, the dipole model approach is a LO calculation,
so  we use the LO KK fragmentation, consistent with
the way fragmentation was used in ERS \cite{ERS}.
We note that in our previous NLO pQCD work, BERSS
\cite{Bhattacharya:2015jpa}, we also used the LO KK
fragmentation functions.
Therefore, in order to compare our new  NLO pQCD  results and
 results from different dipole models and $k_T$ factorization
 with the previous work in ERS and BERSS,
we use the LO parametrization of the KK fragmentation
function even for the NLO pQCD calculation. Since the dominant
contribution is at threshold, the fragmentation functions are not evolved.
We find that if we use NLO KK fragmentation functions,
it results in only $2\%$ lower fluxes than obtained
with the LO KK fragmentation functions when both are rescaled to match
the fragmentation fractions in \cite{Lisovyi:2015uqa}.

The results presented below use $z=E_h/E_Q$ as the
fractional variable in the fragmentation functions.
There are other choices possible, however, this
definition can be used for all three approaches.
The flux has some sensitivity to the charmed hadron fragmentation
functions.
We have also used
the  Braaten et al.~\cite{Braaten:1994bz} (BCFY) fragmentation
functions for
quark fragmentation to pseudoscalar and vector states,
updating the discussion of Cacciari and Nason in
ref.~\cite{Cacciari:2003zu},  for $D^0$ and $D^\pm$, and approximate
forms for $D_s$ and $\Lambda_c$. In comparison to the Kniehl-Kramer fragmentation results, the flux evaluated with
the BCFY fragmentation
functions may be as much as  30-50\%  larger depending on parameter choices.

The fragmentation process is thus somewhat uncertain. An alternative to using fragmentation functions is to use Monte Carlo event generators such as Pythia \cite{Sjostrand:2006za} to hadronize the partonic state. This typically leads to a harder spectrum of $D$-mesons than with fragmentation functions, see e.g.\ \cite{Norrbin:1998bw,Norrbin:2000zc}. The reason is that the hadron that contains the  charm (or anticharm) quark can pick up some momentum from the beam remnant of the interaction, and the charmed hadron can in fact have a larger longitudinal momentum than the original charm quark. In Pythia this is modeled as a drag effect from the string. This effect is not possible in $e^+e^-$ collisions, and in hadron-hadron collisions it is larger for more forward production. Fragmentation functions, however, are fitted from $e^+e^-$ data and more central hadron-hadron data, so if such an effect is real, it may not show up in the fit.

Garzelli et al.~\cite{Garzelli:2015psa} computed the charm cross section at NLO using the NLO Monte Carlo POWHEG BOX \cite{Nason:2004rx,Frixione:2007vw,Alioli:2010xd,Frixione:2007nw}, and used Pythia for fragmentation. The resulting prompt flux is compared with the BERSS flux~\cite{Bhattacharya:2015jpa}, which is computed using fragmentation functions, and is found to be around 30\% larger than BERSS. We believe that one reason for this is the different fragmentation method used. We have checked, using POWHEG BOX, that the produced spectrum of $D$-mesons is indeed harder when the Pythia fragmentation is used rather than the KK fragmentation functions applied to the same spectrum of charm quarks. The difference becomes larger for larger $x_F$, but is small for small $x_F$.
The situation is therefore somewhat unclear, and ideally more forward data would be needed to resolve the question. The most forward data that we have considered is the LHCb data discussed above, but this is not forward enough to be sensitive to differences in fragmentation. The prompt flux, however, could be sensitive to such effects.

Finally, to evaluate the prompt atmospheric lepton fluxes, we need the $Z$-moments
$Z_{pp}$, $Z_{DD}$, $Z_{BB}$ and interaction lengths.
We make the same assumptions as in ref.~\cite{Bhattacharya:2015jpa}
(BERSS). We use expressions for $d\sigma/dx_E$ that factorize into
energy and $x_E$ dependent functions. The energy dependence ensures
that the differential cross section grows with energy following the
relevant cross section growth with energy. For $pA$ scattering, the
cross section parametrization of EPOS 1.99 is used \cite{Pierog:2009zt}, with
the $x_E$ dependent function proportional to $(1-x_E)^n$ with $n=0.51$. For
$DA$ and $BA$ scattering, we approximate the energy dependence with
the cross section for kaon-nucleon scattering determined by the COMPAS
group (see ref.~\cite{Agashe:2014kda}), scaled by $A^{0.75}$ for nuclear
corrections, and use the same $(1-x_E)^n$ behavior for both $B$ and
$D$ scattering with air nuclei, with $n=1$.

\subsection{Prompt muon neutrino flux}

We begin with the muon neutrino fluxes predicted by each of the three
approaches. All of the plots shown are for $\nu+\bar{\nu}$ for a
single flavor. The prompt fluxes for $\nu_\mu+\bar{\nu}_\mu$ are equal
to the $\nu_e+\bar{\nu}_e$ and $\mu^+ + \mu^-$ fluxes. We show the
vertical fluxes. Angular correlations are relevant only for $E\gsim
10^7$ GeV.

\begin{figure} [t]
\centering
\includegraphics[width=0.49\textwidth]{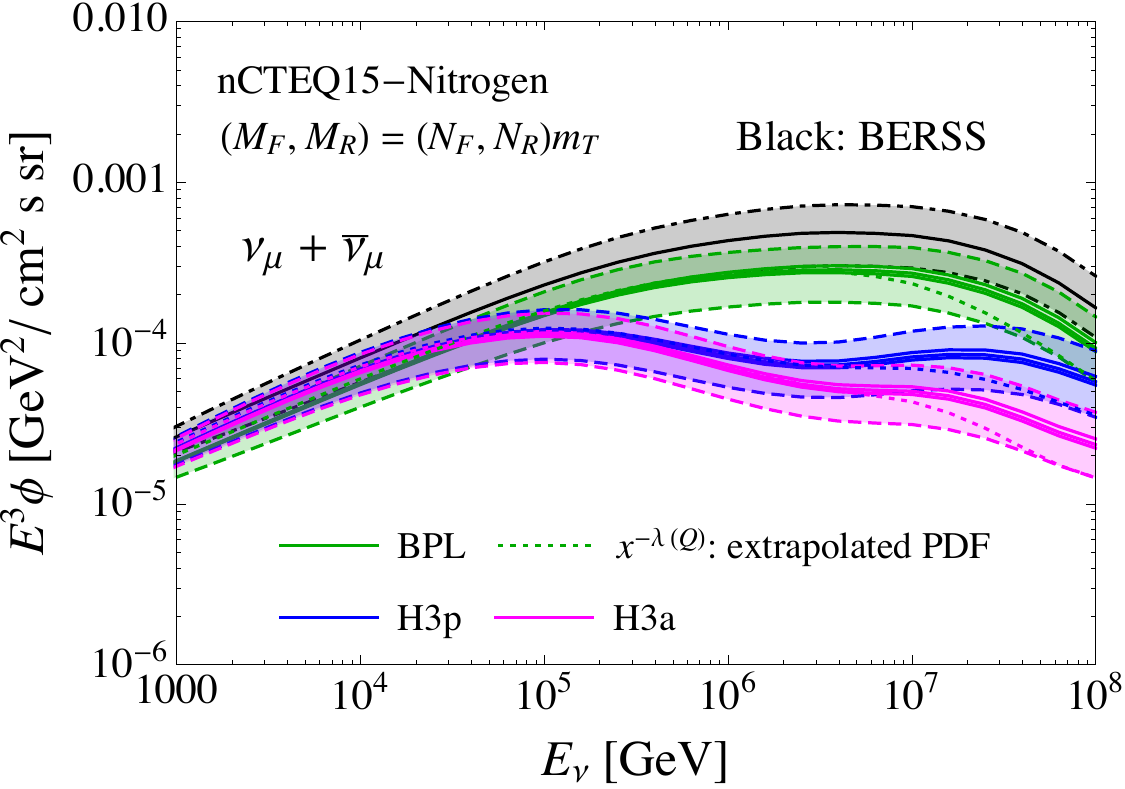}
\includegraphics[width=0.49\textwidth]{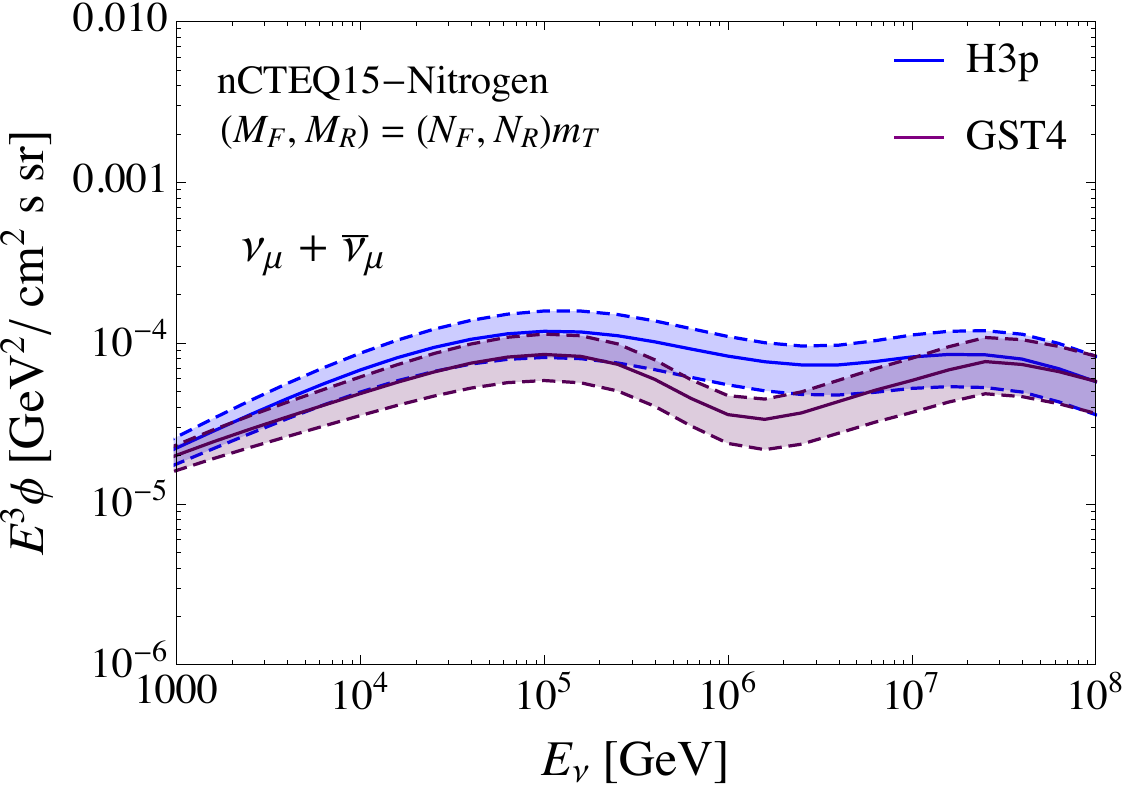}
   \caption{The NLO pQCD flux predictions from decay of charm and bottom
     hadrons.
The BERSS flux was evaluated for charm contributions
using the CT10 PDFs without
     nuclear corrections. The other curves shown the sum of the charm
     and bottom contributions evaluated using the nCTEQ15-14 PDFs.
   }
\label{fig:flux-nloqcd}
\end{figure}

The NLO pQCD evaluation of the fluxes with nuclear corrections are
shown for the broken power law, H3p and H3a cosmic ray fluxes in
fig.~\ref{fig:flux-nloqcd} in the left panel.
The flux is conventionally shown scaled by $E_\nu^3$ to display the
features of the flux more clearly.
For comparison, we show
the BERSS \cite{Bhattacharya:2015jpa} perturbative result. The right
panel shows the flux result for H3p and GST4 cosmic ray inputs, where
the GST4 curve shows a dip at $E_\nu\sim 10^6$ GeV relative to the H3p curve.
In each of the curves, the central solid line is from central scale
choice $(M_F,M_R)=(2.1,1.6)m_T$, and the error band reflects the uncertainty in the nuclear PDFs and 
the range of scales $(M_F,M_R)=(1.25-4.65,1.48-1.71)m_T$, scale factors
representing the best fit range discussed in ref.~\cite{Nelson:2012bc}
and used in our earlier flux evaluation \cite{Bhattacharya:2015jpa}.

\begin{figure} [t]
\centering
\includegraphics[width=0.495\textwidth]{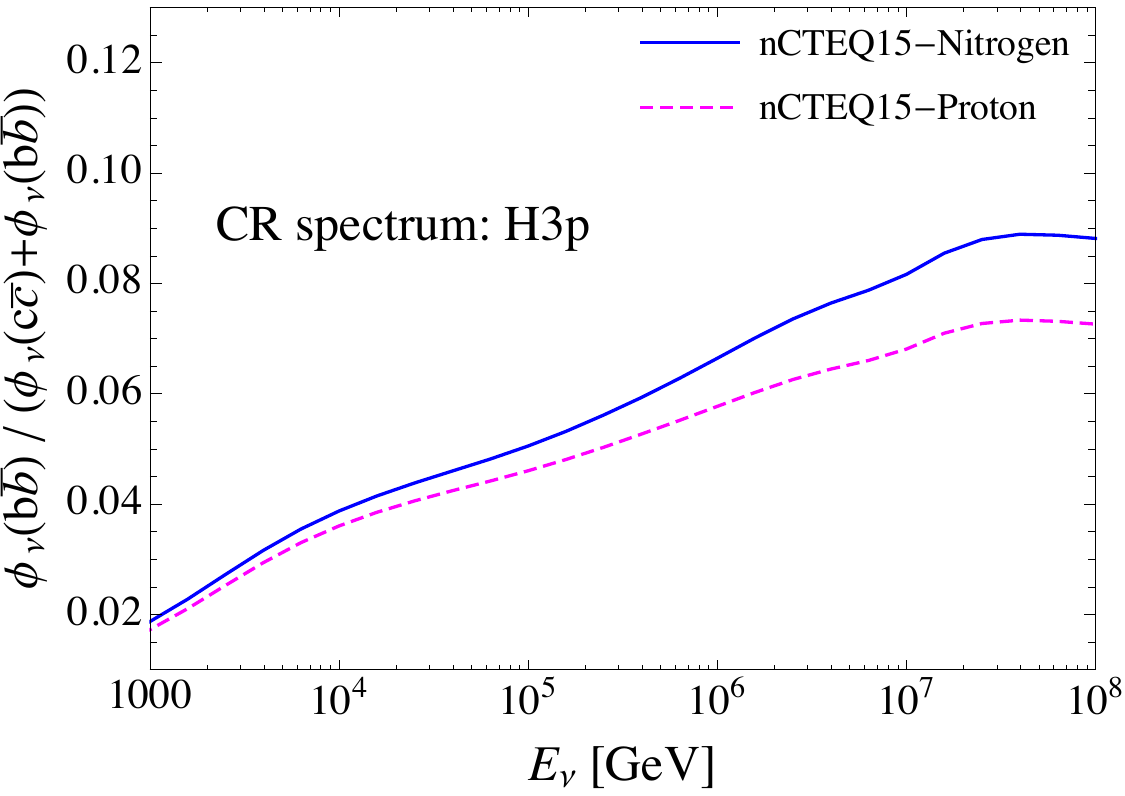}
\includegraphics[width=0.485\textwidth]{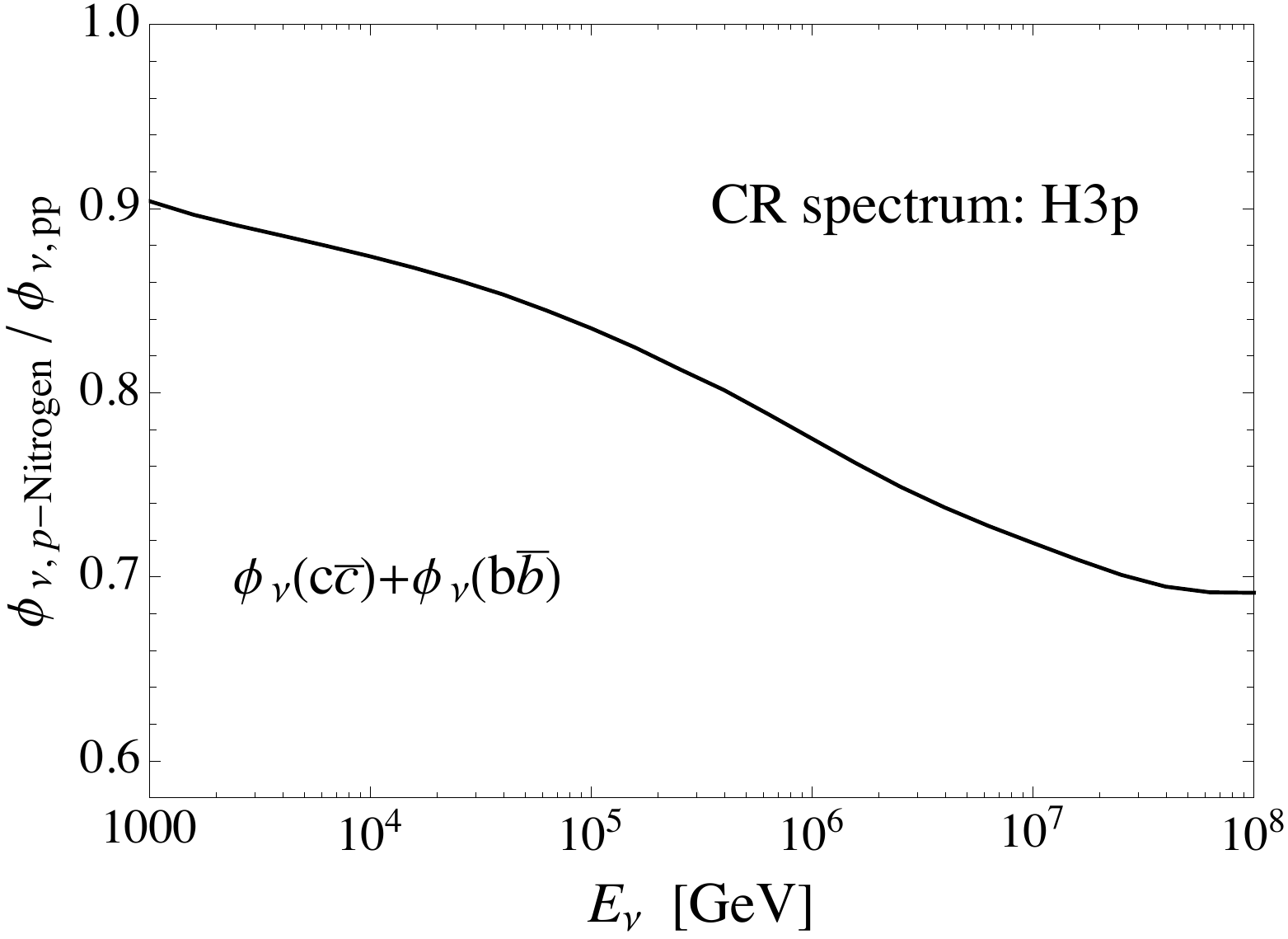}
   \caption{ Left: The NLO pQCD flux prediction from bottom hadrons.
   The fluxes from B hadrons have a ratio of about 2 \%,  7 (6)\% and 9 (7) \% to those from charm at $10^3$, $10^6$ and $10^8$\gev, respectively, for nitrogen (Proton) PDF.
  Right: Nuclear effect in the prompt neutrino flux evaluated in the
NLO pQCD approach.
    }
\label{fig:flux-nloqcd-bb}
\end{figure}

The new evaluation of the prompt flux has several differences relative
to the BERSS result shown in black.
Here, with the updated charm
fragmentation fractions \cite{Lisovyi:2015uqa},
the flux is reduced by about
 20\%.
 The
$b\bar{b}$ contribution was not included in BERSS. It contributes
5-10\%
of
 the flux at $E_\nu\sim 10^5-10^8$ GeV, as shown in the left panel
of fig.~\ref{fig:flux-nloqcd-bb}.
Finally, in this work we have included nuclear effects on the prompt neutrino
flux, which reduces the flux by  20\%-30\%
for energies between $E=10^5-10^8$ GeV for the nCTEQ15-14 PDFs, with the
largest effect at the highest energy.
We note that the nuclear effect is more pronounced on the prompt
neutrino flux than it is for the total charm cross section, due to
the fact that the flux calculation probes forward charm production (small
$x$ of the parton from the target air nucleus)
 where nuclear suppression is larger.
For reference, the total charm cross section section
suppression due to nuclear effects is between  4\% and 13\% at $10^5-10^8$ GeV
with the nCTEQ15-14 PDFs.
 The gluon PDF in
fig.~\ref{fig:gluonpdfncteq} helps illustrate this point. The cross
section is dominated by the small $x_F$ region, where the parton
momentum fractions are nearly equal, so probing less the shadowing
region. The ratio of the flux with nuclear effects to the flux using free
protons (nCTEQ15-14 PDFs compared to nCTEQ15-01 PDFs) is shown as a
function of energy in the right panel of
fig.~\ref{fig:flux-nloqcd-bb}.

The combination of all these effects results in our
NLO pQCD prompt flux estimate being $30\%$ lower than BERSS at
$10^3$ GeV, about $40\%$ lower at $10^6$ GeV and almost
$45\%$ lower at $10^8$ GeV, when we use nCTEQ15-14 PDF as parton
PDFs in the air.

%
\begin{figure} [t]
\centering
\includegraphics[width=0.49\textwidth]{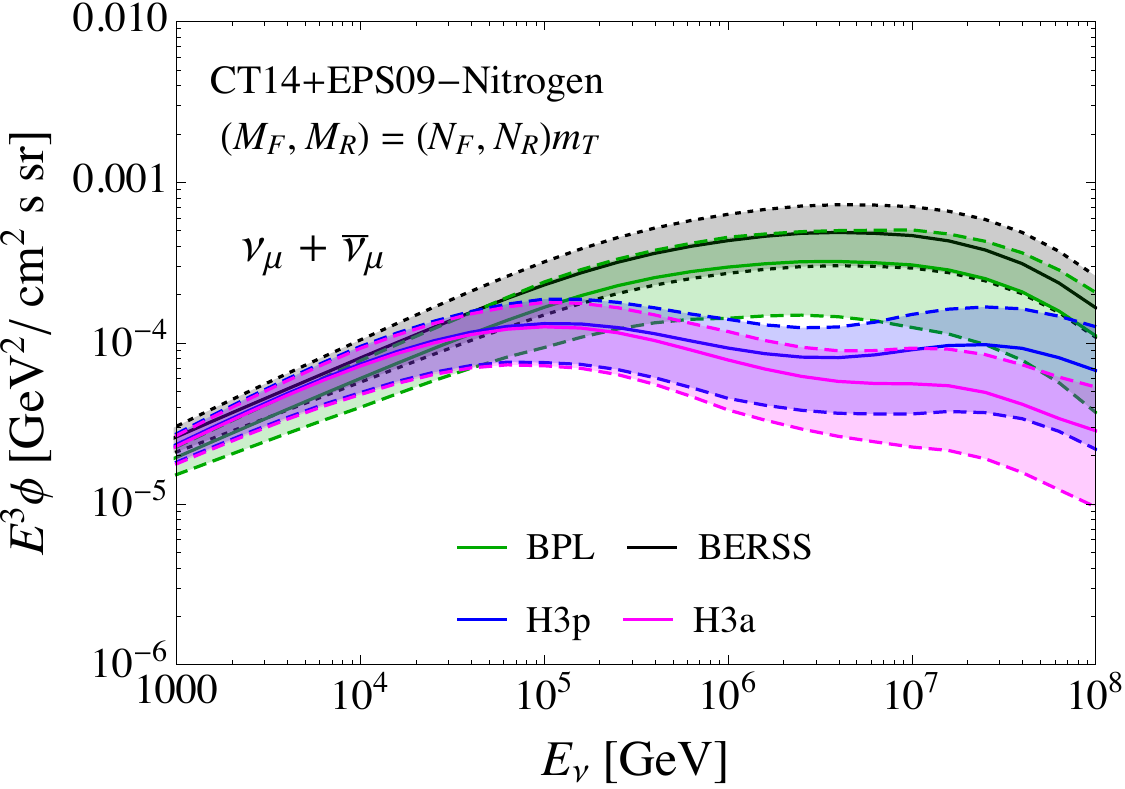}
\includegraphics[width=0.49\textwidth]{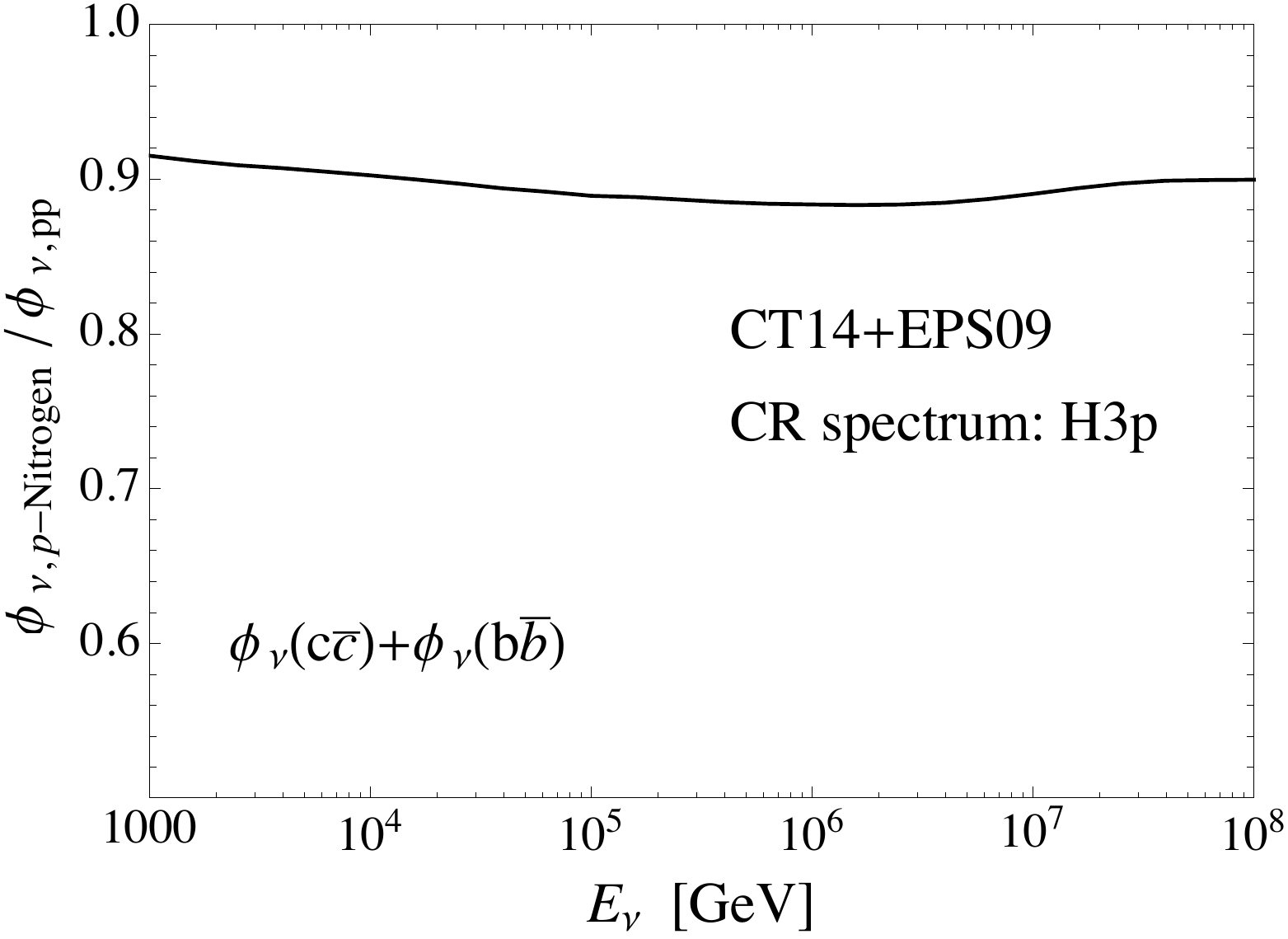}
   \caption
{The central prompt neutrino
flux prediction using the CT14 PDFs with EPS nuclear corrections
  (left), and the ratio of the fluxes with and without the nuclear
  corrections (right), as a function of neutrino energy.
 As in fig.~\ref{fig:flux-nloqcd}, the upper and lower limits 
correspond to 
 variation in the QCD scales and the uncertainty from the different PDF sets.
}
\label{fig:flux-nloqcd-eps}
\end{figure}

When we use CT14 PDFs plus EPS09
for nuclear effects, our results are only moderately affected by
nuclear corrections.
  In the left panel of
fig.~\ref{fig:flux-nloqcd-eps}, we show the fluxes, and in the right
panel, the ratio of the flux with nuclear effects to the one without.
At very high energies, the CT14 PDFs predict a
similar flux to the one obtained with nCTEQ15 PDFs, with 
 the nuclear correction
being somewhat smaller than for nCTEQ15 case.  
Nuclear
corrections are uncertain for a larger range of $x$. The EPS09
suppression
factors are frozen at $R_i^A(x_{min},Q)$ for $x<x_{min}=10^{-6}$,
halting a decline in energy for the ratio of fluxes with nuclear corrected and free
nucleon targets.

\begin{figure} [h]
\centering
\includegraphics[width=0.7\textwidth]{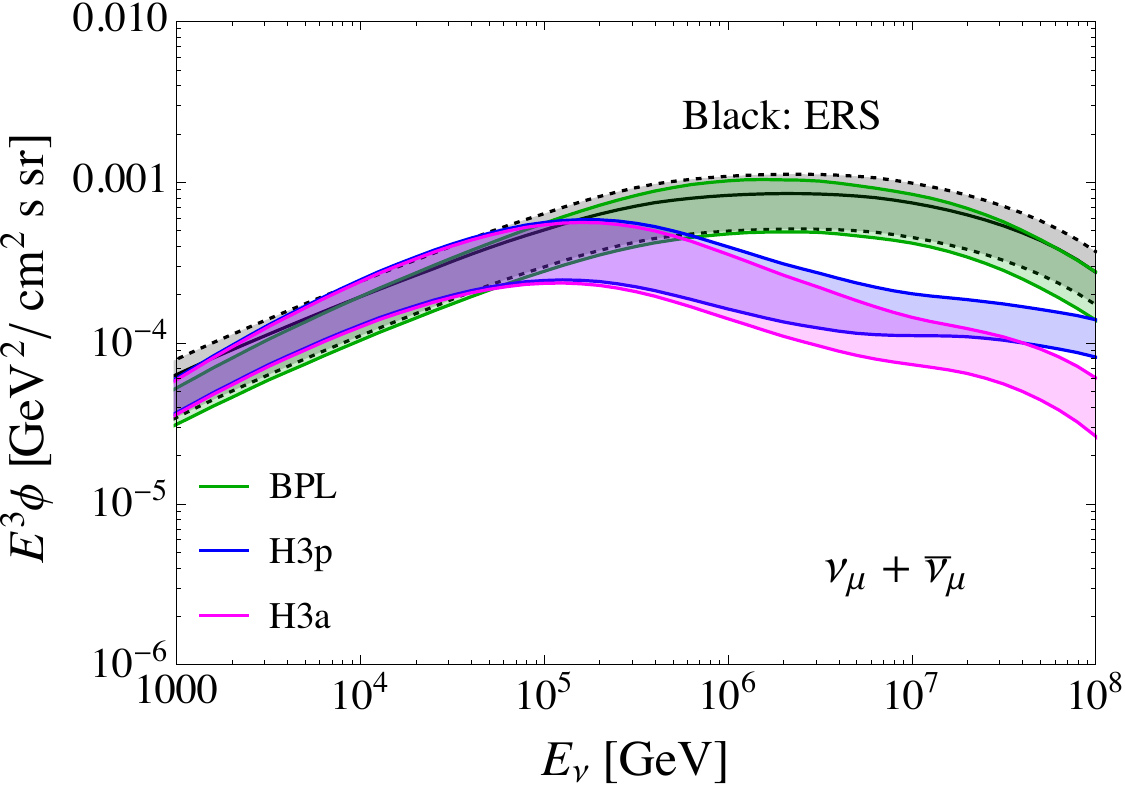}
   \caption{Dipole model flux predictions for $\nu_\mu+\bar{\nu}_\mu$ as a function of energy, compared to the ERS dipole results
   \cite{ERS}. The error bands come from the three dipole models, and
   varying the factorization scale
dependence from
   $M_F=m_c$ to $M_F=4 m_c$. }
\label{fig:flux-dipole}
\end{figure}

The dipole model results are shown in fig.~\ref{fig:flux-dipole},
together with our ERS dipole result from ref.~\cite{ERS} for the
broken power law. Compared to
the ERS result, we have used updated PDFs (LO CT14)
and included the $b\bar{b}$ contribution,
and we have considered two other dipole
models beyond the Soyez model used in ref.~\cite{ERS}. In comparing
the Soyez dipole calculations, the updated fragmentation fraction
 reduces
the overall flux by approximately 20\%. Relative to the ERS
calculation, we have updated the $Z_{pp}$ and $Z_{DD}$ moments, as
discussed in detail in ref.~\cite{Bhattacharya:2015jpa}, which gives a
further reduction (about $35\%$) in the flux prediction.
  The AAMQS dipole and
phenomenological Block dipole give very similar results and are the
upper part
of the uncertainty bands. Nuclear corrections to the dipole model flux
predictions reduce the flux by about 10\%  at
$E_\nu \sim 10^5\gev$ and reduce by
about 20\% at
$E_\nu \sim 10^8\gev$.

\begin{figure} [h]
\centering
\includegraphics[width=0.49\textwidth]{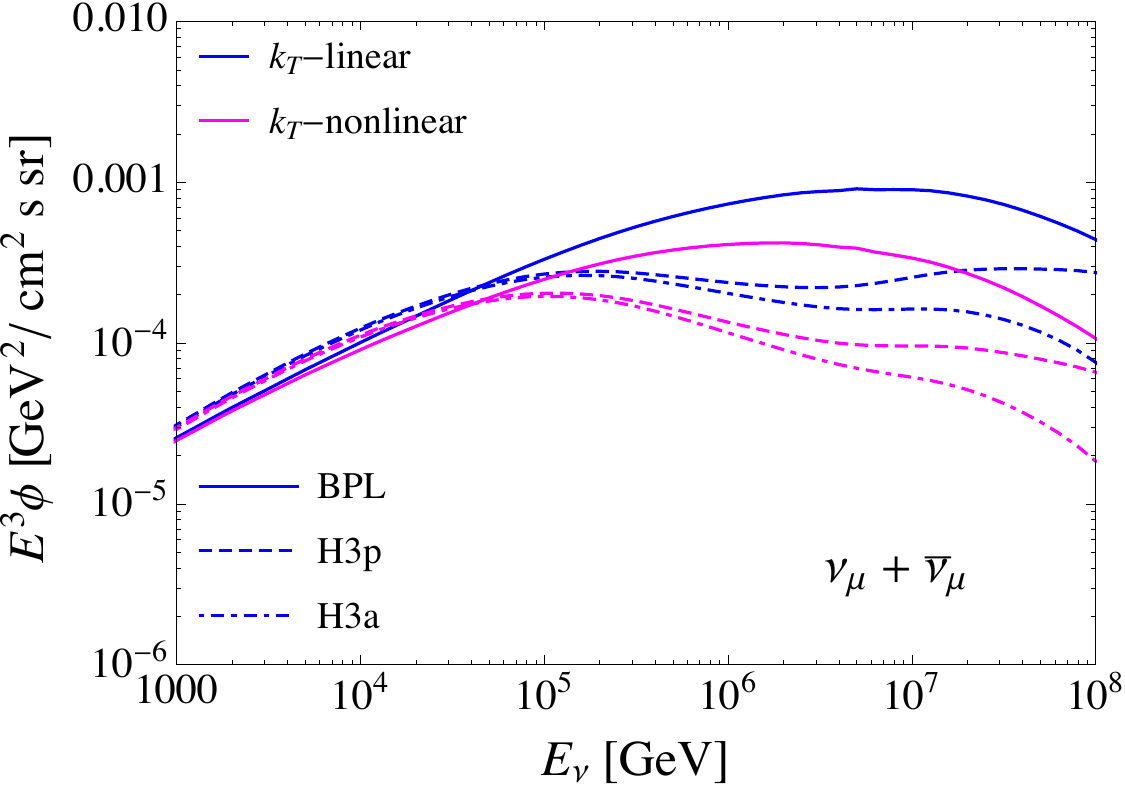}
\includegraphics[width=0.49\textwidth]{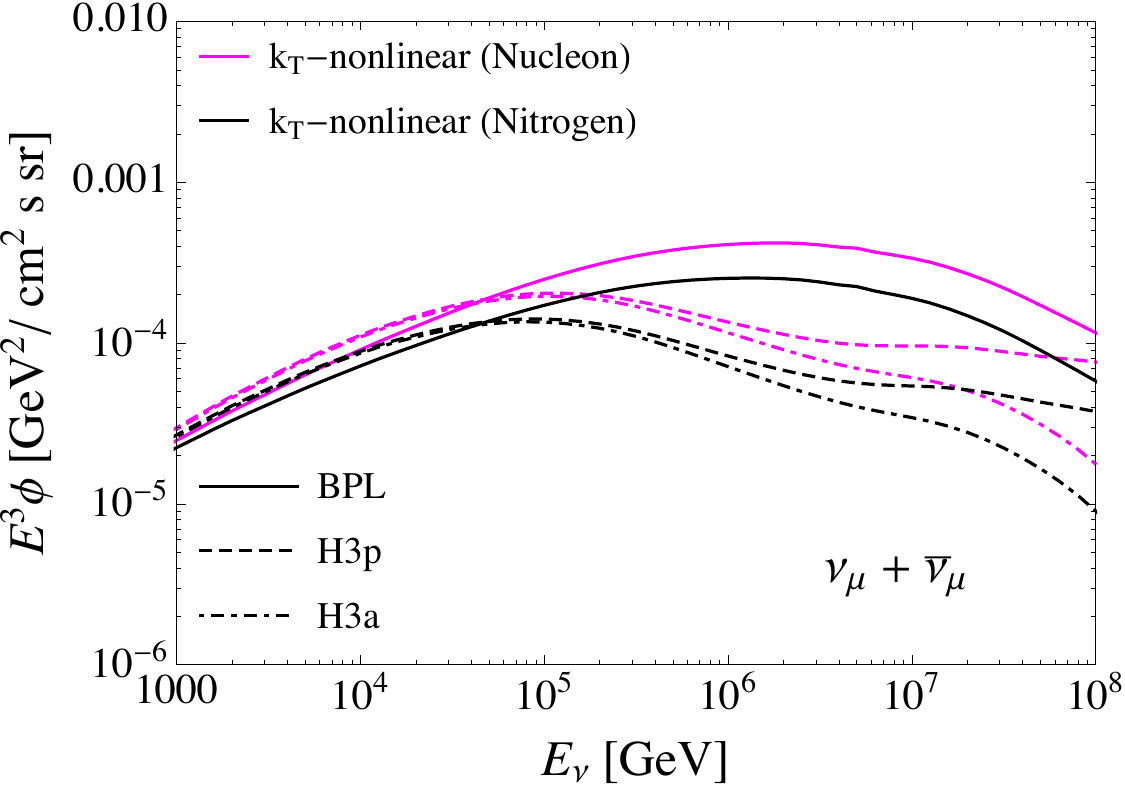}
   \caption{Muon neutrino plus antineutrino prompt flux
     predictions from the $k_T$ factorization model. 
     Left: comparison of the calculations based on linear
     evolution (without saturation) and with nonlinear evolution (with saturation) in the proton case. Right: comparison of the calculations which include saturation effects in proton and nitrogen.}
\label{fig:flux-kt}
\end{figure}

Fig.~\ref{fig:flux-kt} shows the $\nu_\mu+\bar{\nu}_\mu$ flux
predictions in the $k_T$ formalism with linear and nonlinear evolution for the unintegrated gluon density. Nuclear
corrections for nitrogen are included in this
approach.  The predicted high energy flux
in the $k_T$ factorization formalism
is consistent the other approaches, with the exception of the  low energy where the  flux is
somewhat smaller. The low energy deficit reflects the same deficit of the cross
section shown in fig.~\ref{fig:ktcrosssection} since the $k_T$
factorization model applies to small $x$ physics and therefore applies
to high energies. At the  high
energies shown,
the linear $k_T$ approach is about 7 times larger than the non-linear $k_T$ flux
prediction, reflecting the range of impact that small-$x$ effects
can have on the high energy prompt flux.

\begin{figure} [h]
\centering
\includegraphics[width=0.49\textwidth]{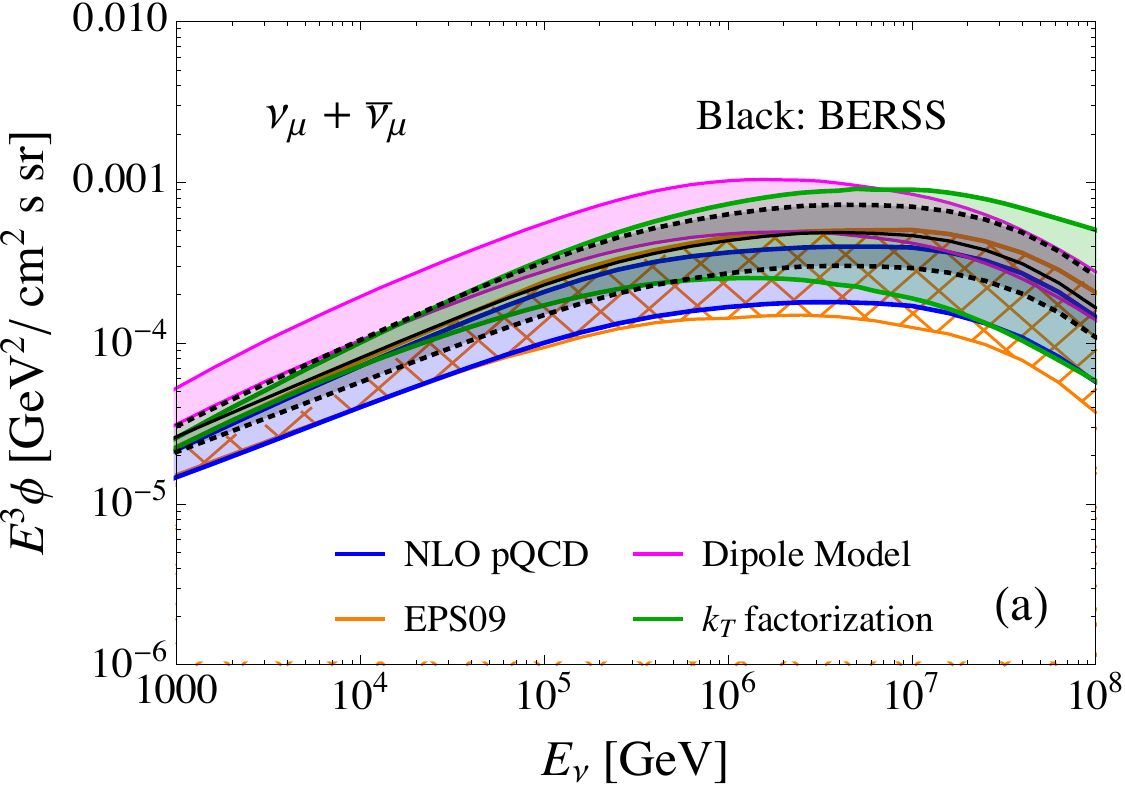}
\includegraphics[width=0.49\textwidth]{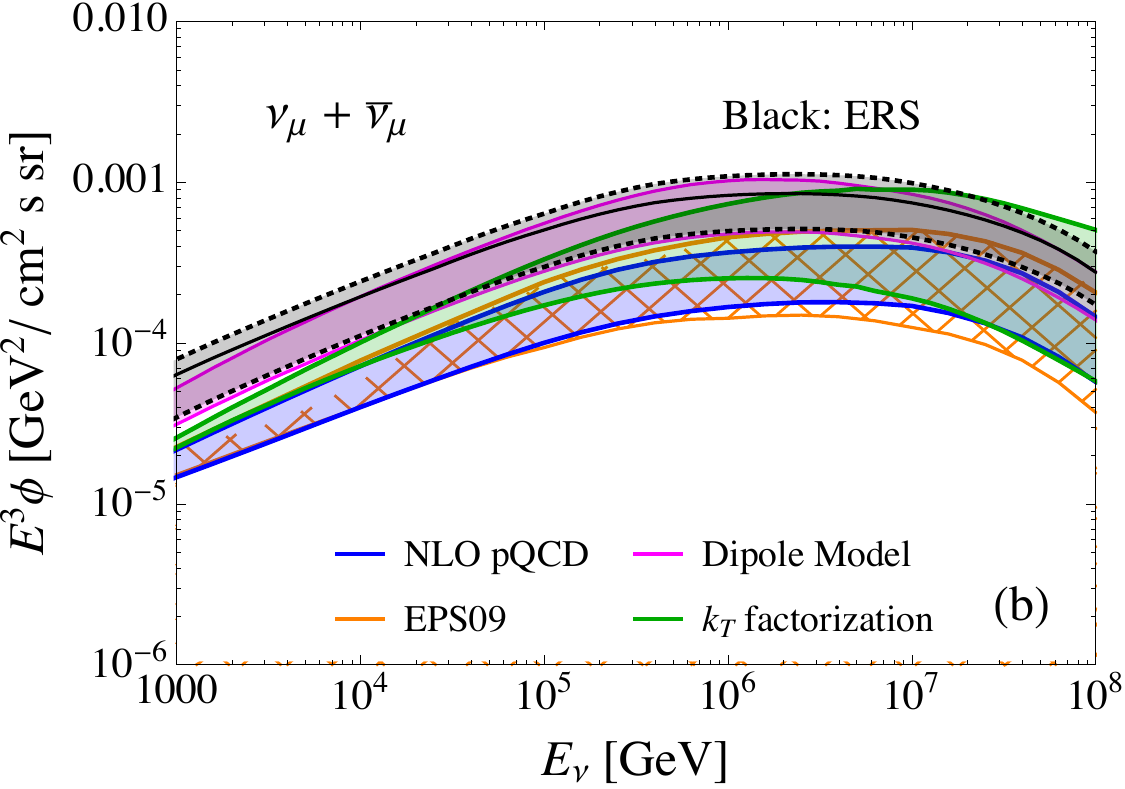}
\includegraphics[width=0.49\textwidth]{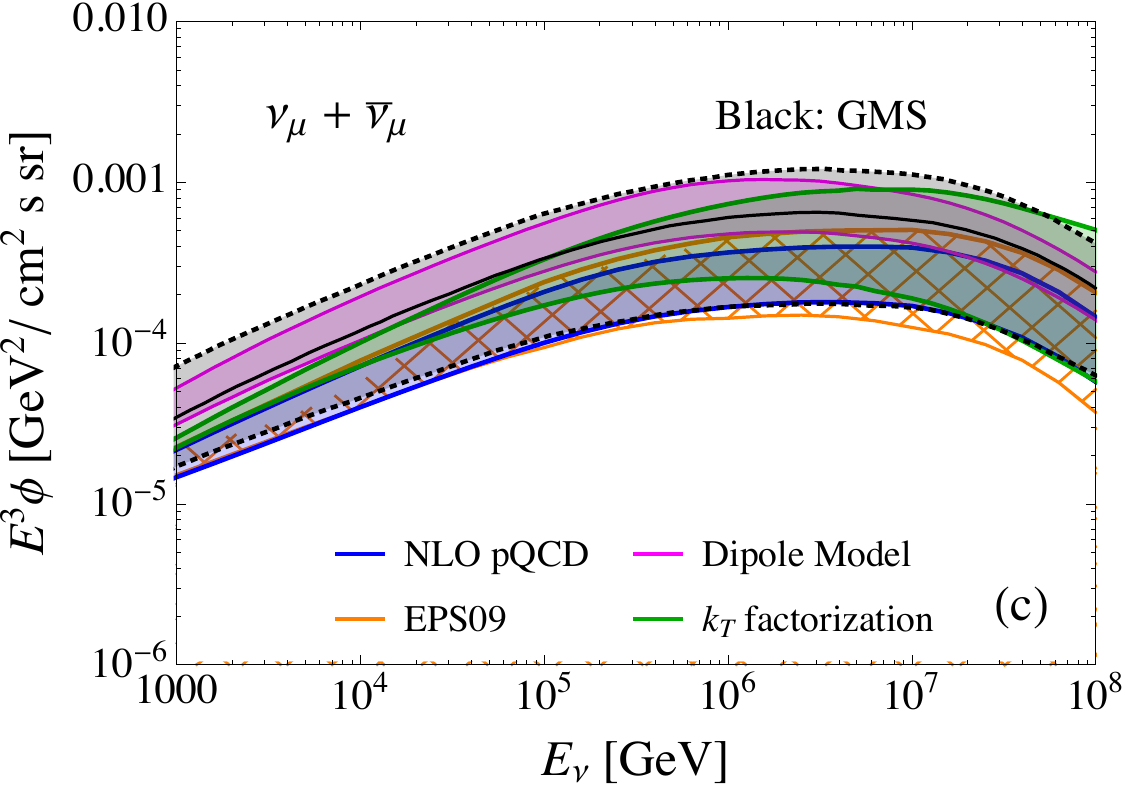}
\includegraphics[width=0.49\textwidth]{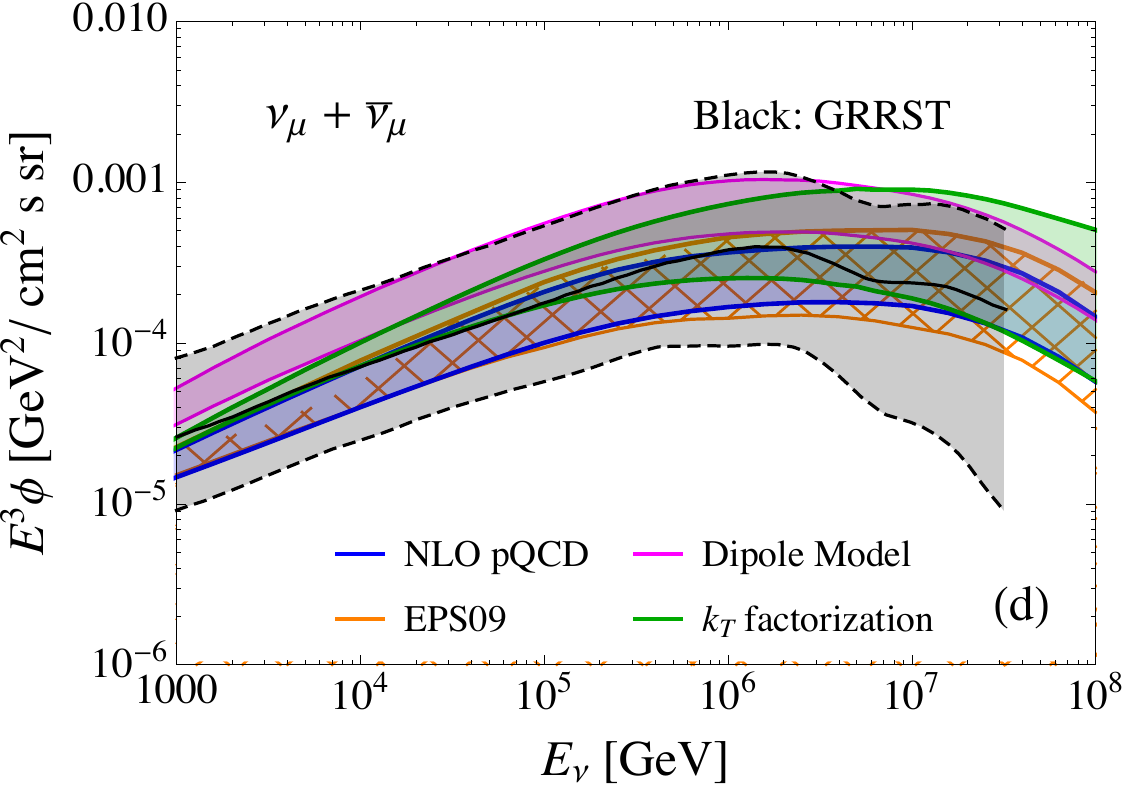}
   \caption{Comparison of the muon neutrino plus antineutrino fluxes using all the approaches: NLO perturbative QCD with nCTEQ15 (blue) and EPS09 (orange),  dipole model (magenta), $k_T$ factorization (green) with the other calculations (black): BERSS \cite{Bhattacharya:2015jpa}, ERS \cite{ERS}, GMS \cite{Garzelli:2015psa} and GRRST \cite{Gauld:2015kvh}.}
\label{fig:flux-comp}
\end{figure}

Finally, in fig.~\ref{fig:flux-comp}, we compare the three approaches
using the broken power law with the BERSS \cite{Bhattacharya:2015jpa}, ERS \cite{ERS}, GMS \cite{Garzelli:2015psa} and GRRST \cite{Gauld:2015kvh}
results. Relative to the BERSS flux, the dipole model predicts a
larger low energy flux, while the $k_T$ factorization model based on the linear evolution predicts a
larger high energy flux. On the other hand the flux based on the $k_T$ 
factorization with nuclear corrections is consistent with the lower 
 end of the NLO pQCD  calculation. Our new perturbative result lies below the
BERSS band for most of the energy range, due to a combination of the
nuclear shadowing and the rescaling of the fragmentation fractions
to sum to unity.
The total uncertainty range of our predictions from the different
approaches is compatible with the other recent evaluations of GMS and
GRRST, with the latter giving a larger error band.

\subsection{Prompt tau neutrino flux}
\label{sec:nutau}

\begin{figure} [h]
\centering
\includegraphics[width=0.49\textwidth]{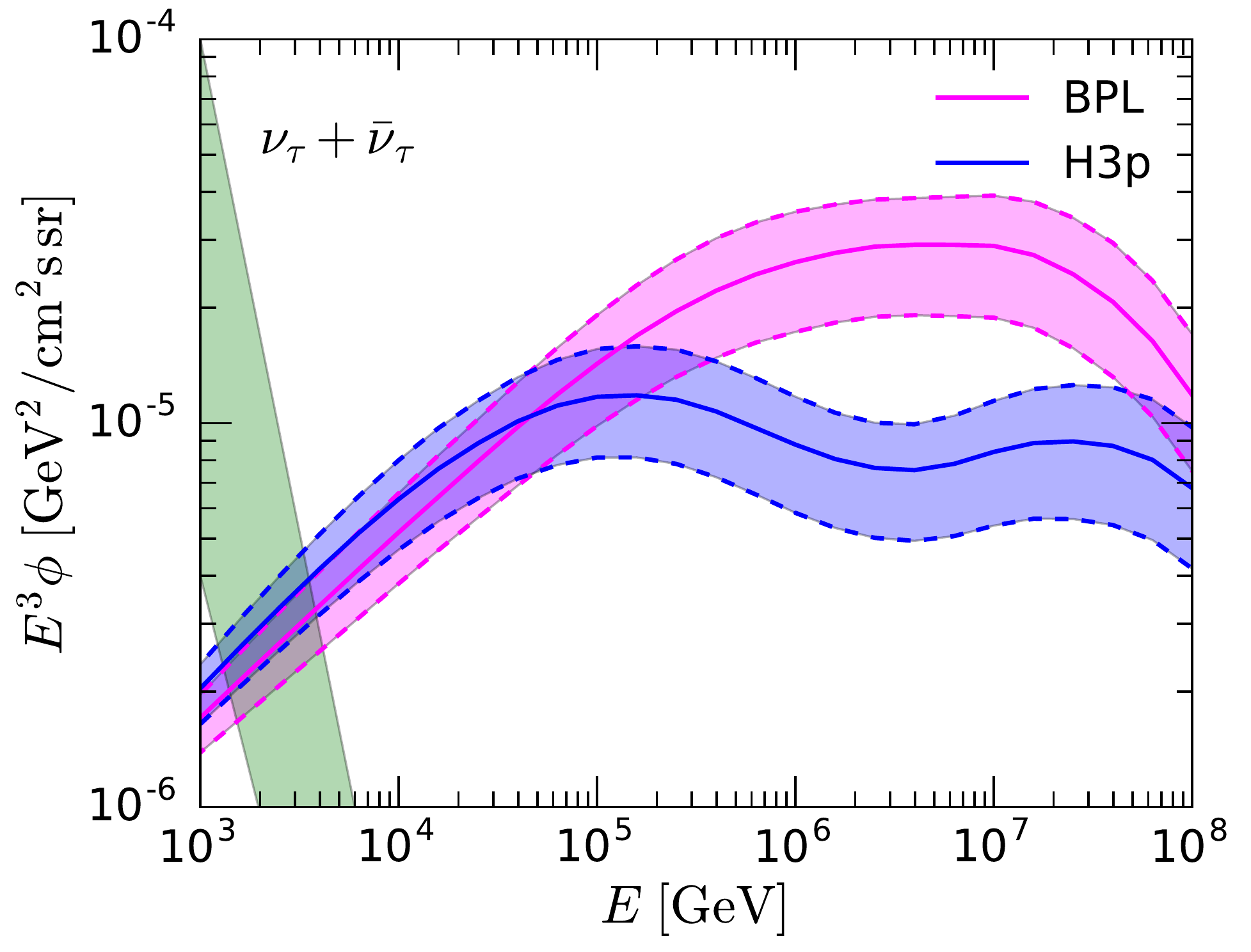}
\includegraphics[width=0.49\textwidth]{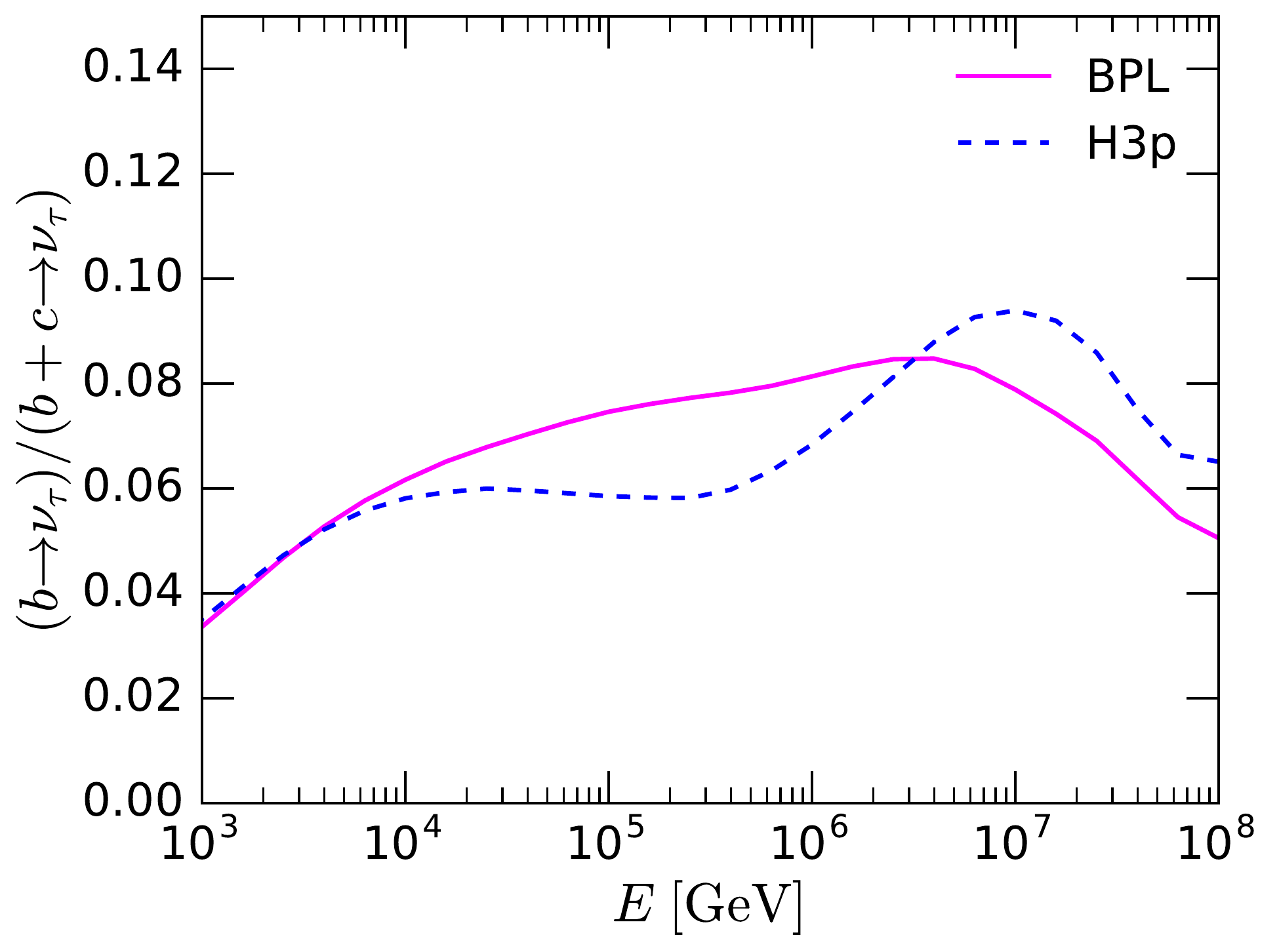}
	   \caption{Left: The prompt atmospheric tau neutrino flux $E_\nu^3
\phi_{\nu_\tau+\bar{\nu}_\tau}$ as a function of neutrino energy using
NLO pQCD for the broken power law and the H3p cosmic ray fluxes. The
vertical green band shows the oscillated conventional muon neutrino
plus antineutrino flux.
Right: Fraction of the flux from $B^0+B^+$  and charge conjugate mesons.}
  \label{fig:phinutau}
\end{figure}

To finish our discussion of prompt neutrino fluxes, for completeness
we show the tau neutrino plus antineutrino flux in
fig.\ \ref{fig:phinutau} in the NLO pQCD framework.
Using the central scale choice in the NLO pQCD calculation of charm pair
production with nCTEQ15-14 PDFs, together with scale variations to obtain the uncertainty
band,
and the KK fragmentation functions, the sum of the direct
and chain contributions from $D_s$ and the semileptonic $B^0$ and
$B^+$ (and charge conjugate) decays to
$\nu_\tau$,  the prompt tau neutrino flux is shown in fig.\ \ref{fig:phinutau} for the broken
power law and H3p cosmic ray fluxes \cite{Gaisser:2012zz}.
As above, the KK fragmentation fraction of 12.3\%\ of charm to $D_s$ has been updated to 8.1\%\ following ref.~\cite{Lisovyi:2015uqa}.

The
flux is about 10\% of  the prompt muon neutrino plus
antineutrino flux. As noted above, the very high energy flux shown here
overestimates the tau neutrino flux because we have approximated the
tau decays as all prompt. The steeply falling shaded green band shows
the range of the tau neutrino flux coming from $\nu_\mu\to \nu_\tau$
oscillations for $\nu_\mu$'s from pion and kaon decays, with the upper edge
coming from $\nu_\mu\to\nu_\tau$ conversion through the diameter of the Earth.
Secondary production of tau neutrinos from interactions of atmospheric leptons in the Earth are quite small
\cite{Iyer:1999wu,Becattini:2000fj,Bulmahn:2010pg}.

The chain decay
$D_s\to \tau\to \nu_\tau$ dominates the flux. The tau neutrino flux from the chain decay is approximately a factor of 4 larger at low energies than from the direct decays, and becoming
larger at higher energies.
The fraction of the tau neutrino flux
from $B$'s is shown in the right panel of fig.~\ref{fig:phinutau}. We
have not included the $B_s$ and $\Lambda_b$ decays as the branching fractions to $\nu_\tau$ are not
measured. They may contribute as much to the prompt tau neutrino flux as
the $B^0+B^+$ contributions.

\subsection{Comparison with IceCube limit}

\begin{figure} [h]
\centering
\includegraphics[width=0.49\textwidth]{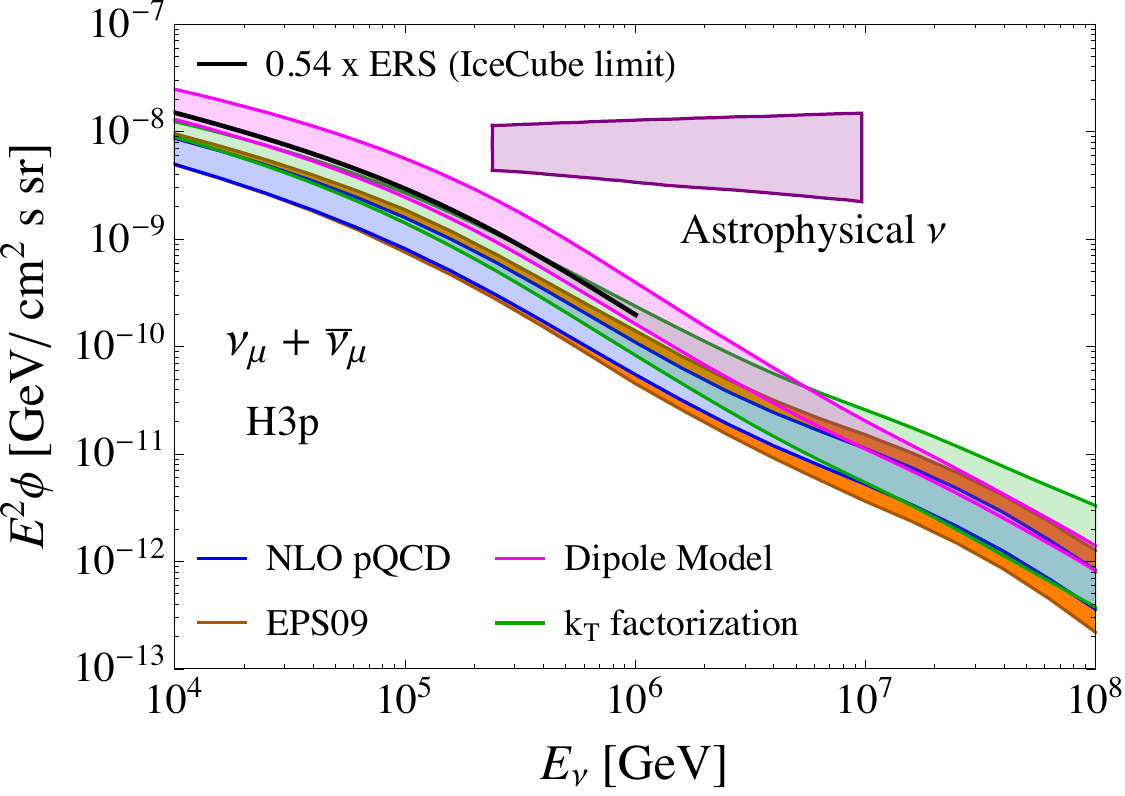}
\includegraphics[width=0.49\textwidth]{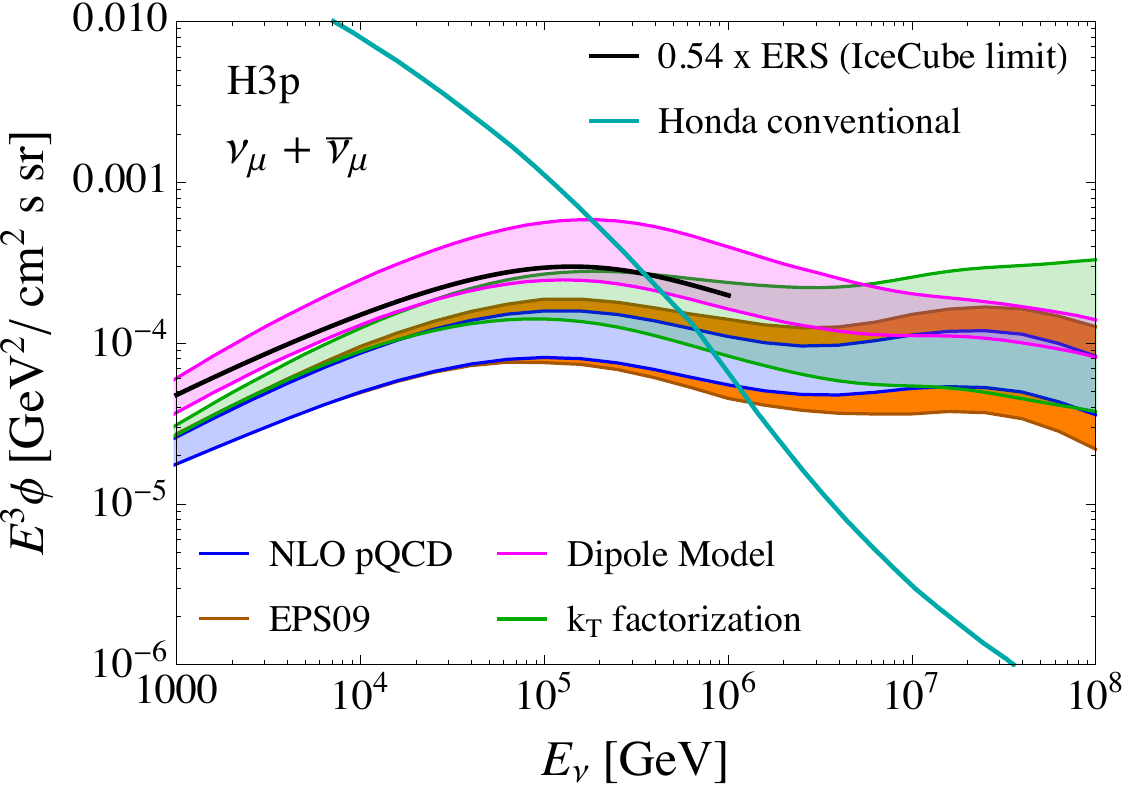}
   \caption{A comparison of our neutrino fluxes from different QCD
models with the IceCube upper limit from ref.~\cite{Radel:2015rsz}.
   The results are for the H3p cosmic ray spectrum.  We also show the
conventional vertical neutrino flux. }
\label{fig:flux-icecube}
\end{figure}

In fig.~\ref{fig:flux-icecube} we show our results for the prompt neutrino flux obtained in the
three different QCD approaches, compared with the conventional neutrino flux and the
IceCube limit on the prompt neutrino flux using 3 years of data \cite{Radel:2015rsz}.
In the left figure, we scale the flux by $E^2$ and in the right
figure, by $E^3$.
The IceCube result is an upper limit at $90\%$ C.L.\ that places a bound on the normalization
of a flux with the same spectral shape as the ERS model~\cite{ERS}, and is thus
not completely model independent. The upper limit corresponds to $0.54$ times the ERS flux,
rescaled from the broken power law CR flux used by ERS to the H3p CR flux, using the method
proposed in~\cite{Gaisser:2011cc}. For comparison, the red band in fig.~\ref{fig:flux-icecube}
represents the $1\sigma$ error on the measured astrophysical neutrino flux.
It is important to note that the evaluation of this IceCube limit is not independent of the
modeling of the astrophysical neutrino flux, which in this case is taken as an unbroken power law,
and the normalization of the ERS flux is taken as a free parameter in a likelihood fit to the data,
yielding the displayed upper limit.

From fig.~\ref{fig:flux-icecube} we note that the IceCube limit is in tension with all dipole model
predictions, and very close or at the border of the upper limit of the $k_T$ factorization approach.  On the other hand both the  NLO pQCD prediction   which includes nuclear effects via the
nuclear parton distributions, and the nonlinear $k_T$ calculation, are below the IceCube limit.  We  note, however, that the nuclear effects
in the dipole model and with the EPS09 pQCD approach are smaller than
in the nCTEQ15-14 pQCD
approach. IceCube data may help distinguish between nuclear
suppression models at small-$x$.

\section{Discussion and conclusions}

\subsection{LHC and IceCube}

As figs.\ \ref{fig:dsdylhcbnlo}, \ref{fig:dsdylhcbdm}, \ref{fig:dsdylhcbkt}
show, rapidity distributions measured at
$7$ and $13$ TeV \cite{Aaij:2015bpa,Aaij:2013mga} seem to be somewhat
in tension within all three approaches  if one considers a fixed
prescription for the scales independent of energy. The theoretical error bands, however, do accommodate the data as noted in 
ref. \cite{Gauld:2015kvh}.  Figs. \ref{fig:dsdylhcbnlo}, \ref{fig:dsdylhcbdm}, \ref{fig:dsdylhcbkt}
compare the distributions of charm quarks with the measured $D^0$
distributions. 
In the case of the $k_T$ factorization approach the $7$ TeV data seem
to be more consistent with the calculation with the nonlinear gluon
density, or the lower band of the calculation with linear gluon
density, whereas the data at $13$ TeV are more in line with the
evaluation with the linear evolution. This is rather counterintuitive
and perhaps could suggest that the calculation with nonlinear
evolution is disfavored by the data. However, given the spread of the
uncertainty of the calculation it is not possible to make decisive
statement at this time and more studies are necessary.
The dipole model evaluation favors the Soyez form for $\sqrt{s}=7$ TeV
and the AAMQS or Block form for $\sqrt{s}=13$ TeV for our fixed value
of $\alpha_s$. The central pQCD predictions seem to indicate that the
distributions don't rise quickly enough with increasing $\sqrt{s}$. In
ref.~\cite{Gauld:2015yia}, the NLO pQCD prediction of the ratio of
$d\sigma/dy$ in the forward region for LHCb for $\sqrt{s}=13$ TeV to
$\sqrt{s}=7$ TeV was predicted to be on the order of 1.3-1.5, which we
also see in fig.~\ref{fig:dsdylhcbnlo}. The data
show the ratio to be closer to a factor of 2. Nevertheless, for all three
approaches, the LHCb data fall within the theoretical uncertainty bands.

We have calculated the rapidity and $p_T$ distribution 
using our theoretical QCD parameters, i.e. the range of factorization scales 
for a 
given charm mass which was determined from the energy dependence of the total 
charm cross section.  We have found our results to be consistent with 
 LHCb data.
The range of
$m_T$ dependent
factorization scales in the  pQCD evaluation
adequately cover the range of LHCb data, while
the range of $m_c$ dependent factorization scales overestimate the
uncertainty.
In the case of the dipole models, in
which there is no explicit $p_T$ dependence, we have
only made comparison with the LHCb rapidity distributions.

\begin{figure} [t]
\centering
\includegraphics[width=0.5\textwidth]{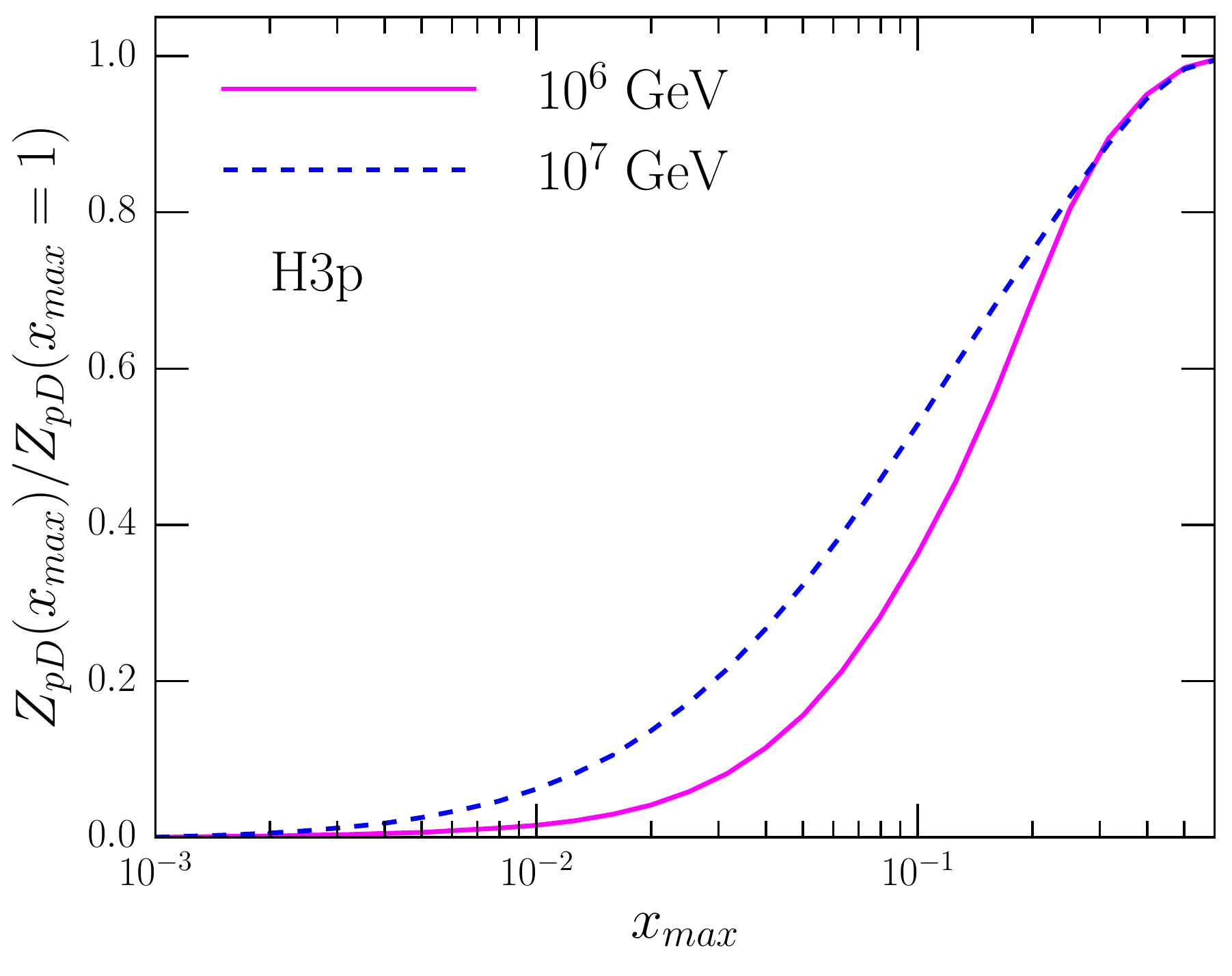}
   \caption{$Z_{pD^0}(x_{max})/Z_{pD^0}(x_{max}=1)$ for the H3p flux
     and
$E=10^6$ and $10^7$ GeV.}
\label{fig:xmax}
\end{figure}

The IceCube Neutrino Observatory and other high energy neutrino
detectors may be useful in getting a handle on forward charm
production. Indeed, the high energy prompt lepton flux depends on charm
production at even higher rapidity than measured by LHCb, as can be
seen by the following argument.
In both the high and low energy forms of the prompt lepton fluxes, the
$Z$-moments for cosmic ray production of charm, e.g., $Z_{pD^0}(E)$,
depend on the lepton energy $E$.
To evaluate the $Z$-moment for charm production, the energy integral over $E'$ in
eq.~(\ref{eq:zmomdef})
can be cast in the form of an integral over $x_E=E/E'$ that runs from
$0\to 1$, account for incident cosmic rays ($p$) with energy $E'$
producing, in this case, $D^0$ with energy $E$.
Fig.~\ref{fig:xmax} shows the fraction of the $Z$-moment
integral in eq.~(\ref{eq:zmomdef}) for $x_E=0\to x_{max}$ for two different
energies using NLO pQCD with the central scale choice and the H3p
cosmic ray flux.
For $E=10^6$ GeV, about $10\%$ of the $Z$-moment comes from
$x_E<x_c=3.6\times 10^{-2}$, while for $E=10^7$ GeV, this same percentage
comes from $x_E<x_c=1.5\times 10^{-2}$.
We can use the value of $x_E>x_c$ that gives 90\% of the $Z$-moment as a guide
to what are the useful kinematic ranges in high energy $pp$ collider experiments.

We approximate
\begin{equation}
x_E\simeq x_F\simeq \frac{m_T}{\sqrt{s}}e^{y_{cm}}\simeq
\frac{m_D}{\sqrt{s}}e^{y_{cm}}\, ,
\end{equation}
in terms of the hadronic center of mass rapidity,
which leads to
\begin{equation}
y_{cm}> \frac{1}{2}\ln \Biggl(\frac{x_c\, 2 m_p E}{m_D^2}\Biggr)\equiv y_{cm}^c
\end{equation}
for 90\% of the $Z$-moment evaluation. For $E=10^6$ GeV, this
indicates that the $Z$-moment is dominated by $y_{cm}>4.9$ with
$\sqrt{s}=1.4-7.3$ TeV. For $E=10^7$ GeV, $y_{cm}>5.7$ and
$\sqrt{s}=4.4-35$ TeV. These approximate results show that the LHCb
results directly constrain only a small portion of the contribution of charm
production to the prompt neutrino flux.

Finally, let us note that there could potentially be another source of charm in the proton. It has been suggested  \cite{Brodsky:1980pb,Hobbs:2013bia} that there could exist heavy quark pairs in the Fock state decomposition of bound hadrons. This would be an additional non-perturbative contribution and  is usually referred to as intrinsic charm
to distinguish it from the perturbative and radiatively generated component considered in this work. 
Intrinsic charm parameterized in
PDFs has been explored in e.g., refs.\ \cite{Dulat:2013hea,Ball:2016neh}.
There are  studies that explore how to probe intrinsic charm in direct and indirect ways \cite{Boettcher:2015sqn,Halzen:2013bqa}.
Intrinsic charm, if it exists, would be mostly concentrated at high values of $x$ of the proton and may therefore be another contribution to the very forward production
relevant to the prompt lepton flux. Its unique features
were recently discussed in \cite{Halzen:2016pwl,Halzen:2016thi,Laha:2016dri}.
We note however, that the current IceCube limit on the prompt flux
is already quite constraining and leaves a rather narrow window for a sizeable intrinsic charm component.
Eventual IceCube observations (rather than limits) of the prompt atmospheric lepton flux,
may be unique in its ability to measure or constrain the   physics at high rapidities.

\subsection{Summary}

In this work we have presented  results for prompt neutrino flux using several
QCD approaches: an NLO perturbative QCD calculation including
nuclear effects, three different dipole models and the
$k_T$ factorization approach with the unintegrated gluon density from
the unified BFKL-DGLAP framework with and without saturation
effects.  Numerical
results are listed in tables
\ref{table:flux-ncteq15}, \ref{table:flux-ct14eps}, \ref{table:flux-dmkt}
and \ref{table:flux-ncteq15-nutau}.
The energy dependence of the total charm cross section,
measured from low energies ($100$ GeV)
to LHC energies ($13$ TeV),
is best described with NLO pQCD approach.  On the other hand
 the dipole and $k_T$ factorization
approaches are theoretically suited to describe
heavy quark production at high energies, however,  they need
additional corrections at lower energies.

We have included theoretical uncertainties due to the
choice of PDFs, choice of scales and nuclear effects,
constrained by the total charm cross section measurements
for energies between $100$ GeV up to $13$ TeV.  We have shown that
differential cross sections for charmed mesons obtained with
these QCD parameters are in good agreement
with the LHCb data.   We have found that
the prompt neutrino flux is higher in case of the dipole
model and the $k_T$ factorization model (without saturation) than the NLO pQCD
case.  The former seem to be numerically consistent
with the previous ERS \cite{ERS} results, while NLO pQCD is
smaller than BERSS \cite{Bhattacharya:2015jpa}. For the nCTEQ15-14
evaluation,
this is mostly due to the nuclear effects.
In particular, we have found that the nuclear effect on the prompt neutrino
flux is large in case
of the pQCD approach with nCTEQ15-14 PDFs, as large as $30\%$ at
high energies, while this effect is smaller ($\sim 20\%$)
for the dipole model approach. The EPS09 nuclear corrections suppress
the pQCD flux calculations with free nucleons by only $\sim 10\%$.
The nuclear corrections are also significant in the $k_T$ 
factorization approach, as large as $50\%$ at high energies, 
thus lowering the flux to the level comparable with that 
obtained using the NLO pQCD with nuclear PDFs. 

Contributions from $b\bar{b}$ are on the order of 5-10\% to the prompt
flux
of $\nu_\mu+\bar{\nu}_\mu$ in the energy range of interest to
IceCube. For completeness, we have also evaluated the flux of
$\nu_\tau + \bar{\nu}_\tau$ from $D_s$ and $B$ decays.

We have also shown results for different cosmic ray
primary fluxes and show how the shape of the
particular choice affects the neutrino flux.
As before, the updated fluxes for the primary CR give much lower results than the simple broken power law used in many previous estimates.

Finally we have compared our predictions with the IceCube
limit \cite{IceCube}.  We have found that the current
IceCube limit seems to exclude some dipole models and
the upper limit of the
$k_T$ factorization model (without any nuclear shadowing), while our results obtained
with the NLO pQCD approach with nCTEQ15-14 and the calculations based on the $k_T$ factorization with nuclear corrections included are substantially lower and thus
evade this limit.

Since it is very important to determine the energy at
which prompt neutrinos become dominant over the
conventional neutrino flux, we expect that the
calculation of the conventional flux might be improved
 by using the two experiments at the LHC that have detectors
in the forward region, the Large Hadron Collider forward (LHCf)
experiment \cite{LHCf} and
the Total, Elastic and Diffractive Cross-section Measurement
experiment (TOTEM) \cite{TOTEM}.
The LHCf experiment measures neutral particles emitted in
the very forward region ($ 8.8 < y < 10.7$), where particles carry a large
fraction of the collision energy, of relevance to the
better understanding of the development of showers of
particles produced in the atmosphere by high-energy cosmic rays.
The TOTEM experiment takes precise measurements of
protons as they emerge from collisions in the LHC
at small angles to the beampipe, thus in the forward region.
In addition to measuring the total and elastic cross section,
TOTEM has measured the pseudorapidity distribution of charged
particles at $\sqrt s = 8$ TeV in the forward region
($ 5.3 < |\eta| < 6.4$).  These measurements could be
used to constrain models of particle production in cosmic
ray interactions with the atmosphere and potentially affect
the conventional neutrino flux, which is coming from
pion decays.

Future IceCube measurements have a good chance
of providing valuable information about the elusive physics at very small $x$,
in the kinematic range which is beyond the reach of  the present colliders.  Keeping in mind
 the caveats involved in the current IceCube
 treatment of the atmospheric cascade and the incoming
cosmic ray fluxes, the observation or non-observation
of the prompt flux may give important insight into the
QCD mechanism for heavy quark production.

First, the nuclear gluon distribution in the region $x \le 0.01$ is currently not  constrained
with collider or fixed target experiments. On the other hand, we expect
that the upcoming 6 year IceCube data will
be sensitive to our pQCD flux results, especially those obtained with the EPS framework that includes nuclear effects.

Second, from our study we find that the IceCube limit shown in fig.~\ref{fig:flux-icecube}
already severely constrains the dipole model approach, even with the lowest cosmic
ray flux (H3p). While it is possible that a modified dipole approach, such as a next-to-leading order calculation, would yield a lower charm cross section that is not in tension with IceCube, the
dipole model calculation used here is not flexible enough to modify so that it evades the limit, i.e.,
this tension cannot be solved by adjusting dipole parameters, because they are constrained by the LHCb data.

\subsection*{Note added}
After submitting our paper, the IceCube Collaboration released their 
upper limit on the prompt atmospheric muon neutrino flux as 
1.06 times the ERS flux based on the 6 year data \cite{Aartsen:2016xlq}. 
We find that our results presented in this work are below this 
new IceCube limit.

\acknowledgments

We thank Krzysztof Kutak for making available the code for the calculation of the unintegrated gluon distribution, and Fred Olness and Aleksander Kusina for providing the low-$x$ grids for the nCTEQ15-14 PDFs. We thank Olga Botner, Jean-Ren\'{e} Cudell, Sebastian Euler, Allan Hallgren, Gunnar Ingelman, Leif R\"adel, Sebastian Schoenen, and Torbj\"orn Sj\"ostrand for helpful comments and discussion.

This research was supported in part by
the US Department of Energy contracts DE-SC-0010113, DE-SC-0010114, DE-SC-0002145, DE-SC0009913,
the National Research Foundation of Korea (RF) grant funded by the Korea government of the Ministry of Education, Science and Technology (MEST) No.\ 2011-0017430 and No.\ 2011-0020333,
the National Science Center, Poland, grant No.\ 2015/17/B/ST2/01838,
the Swedish Research Council contract No.\ 621-2011-5107 and the Fonds de la Recherche Scientifique-FNRS, Belgium,
under grant No.\ 4.4501.15.

\appendix

\section{\label{sec:appendix}Appendix}

\subsection{\label{ssec:appfrag}Fragmentation}

The parameters for the charm fragmentation functions are shown in
Table \ref{table:fragkk} for the fragmentation function of energy
fraction $z$, of the form discussed by Kniehl and Kramer in ref.~\cite{Kniehl:2006mw},
\begin{equation}
\label{eq:frag}
D_c^h(z) = \frac{Nz(1-z)^2}{[(1-z)^2+\epsilon z]^2}\ .
\end{equation}
We use the LO parameters of ref.~\cite{Kniehl:2006mw}, with the overall normalization $N$ rescaled
 to account for updated fragmentation fractions
determined from a recent review of charm production data in
ref.~\cite{Lisovyi:2015uqa}. The corresponding fragmentation fractions
for
$c=D^0,\ D^+,\ D_s^+$ and $\Lambda_c^+$ are listed in the column labeled $B_c$ in
table \ref{table:fragkk}. This is our default set of fragmentation
functions (KK).

\begin{table}[h]
\begin{center}
\begin{tabular}{|l|c|c|c|}
\hline
Particle & $N $& $B_c$&   $\epsilon$  \\
\hline
$D^0$ & 0.577 &0.606 &   0.101 \\
\hline
$D^+$ & 0.238 &0.244 &   0.104 \\
\hline
$D_s^+$ & 0.0327  & 0.081 &  0.0322 \\
\hline
$\Lambda_c^+$ & 0.0067  &0.061  & 0.00418  \\
\hline
\end{tabular}
\caption{Parameters for the charm quark fragmentation from \cite{Kniehl:2006mw}, with the normalization $N$
rescaled to match the
fragmentation fractions in ref.~\cite{Lisovyi:2015uqa}.}
\label{table:fragkk}
\end{center}
\vspace{-0.6cm}
\end{table}

Alternate fragmentation functions for charmed hadrons is provided by
Braaten et al.~\cite{Braaten:1994bz} (BCFY) fragmentation
functions for
quark fragmentation to pseudoscalar and vector states, with input from
 Cacciari and Nason in ref.~\cite{Cacciari:2003zu}  for $D^0$ and $D^\pm$.
The parameter $r$ in their fragmentation functions
\begin{equation}
D_c^h(z,r)  = N_P^h D^{(P)}(z,r) + N_V^h \tilde{D}^{(V)}(z,r)
\label{eq:bcfy}
\end{equation}
depends on the charm mass and hadron mass. Cacciari et al.\ in ref.~\cite{Cacciari:2003zu} suggest a central
value of
$r=0.1$ for $D^0$ and $D^+$, however they note that $r=0.06$ is a
better choice when $m_c=1.27$ GeV. Detailed formulas for the direct $(P)$ and vector meson $(V)$
contributions are in ref.~\cite{Braaten:1994bz}.
In a manner similar to
ref.~\cite{Adam:2015jda},
we have adapted the BCFY fragmentation functions to $c\to D_s$, with
two reasonable choices for $r$ to account for the larger $D_s$ mass:
$r=0.15$ and $r=0.09$. For the $\Lambda_c$ with the BCFY fragmentation functions,
we use a delta function. While a delta function used with the BCFY
meson fragmentation functions overestimates the
$\Lambda_c$ contribution to the prompt flux, it is nevertheless only
$\sim 2\%$ to the total prompt muon neutrino plus antineutrino flux at
$E=10^6$ GeV. The fragmentation fractions are normalized to
the  same fragmentation fractions in  table
\ref{table:fragkk}.
The coefficients used for the direct and vector contributions in eq.~(\ref{eq:bcfy})
for charmed meson fragmentation are listed in table \ref{table:bcfy}.
Our evaluation of the fluxes with the BCFY fragmentation functions
shows an enhancement by a factor of $\sim 1.3-1.4$ to $1.4-1.5$,
depending on the value of $r$ chosen.

\begin{table}[h]
\begin{center}
\begin{tabular}{|l|c|c|c|}
\hline
Particle $h$ & $N _P^h$& $N_V^h $& $r$ \\
\hline
$D^0$ & 0.214 &0.391 &0.1\ (0.06)\\
\hline
$D^+$ & 0.164 &0.0808 &0.1\ (0.06)\\
\hline
$D_s^+$ & 0.0252  & 0.0555 &0.15\ (0.09) \\
\hline
\end{tabular}
\caption{Following ref.~\cite{Cacciari:2003zu} for  the BCFY
  fragmentation functions of the form eq.~(\ref{eq:bcfy}), normalized to the
fragmentation fractions of ref.~\cite{Lisovyi:2015uqa}.}
\label{table:bcfy}
\end{center}
\vspace{-0.6cm}
\end{table}

For the $B$ meson fragmentation, we use the power law form of Kniehl
et al.\ from ref.~\cite{Kniehl:2008zza},
rescaled to match the fragmentation fractions of ref.~\cite{Aaij:2011jp}. The functional form of the
fragmentation functions for $B$ mesons is
\begin{equation}
D_b(z) = N z^\alpha (1-z)^\beta\ ,
\label{eq:fragfunb}
\end{equation}
with the parameters in table  \ref{table:fragkkb}.
Because the fragmentation fraction for $b\to \Lambda_b$ is so large,
we do not use a delta function for its fragmentation fraction. We use
the same form of eq.~(\ref{eq:fragfunb}), normalized to $B_{\Lambda_b}=0.236$.

\begin{table}[h]
\begin{center}
\begin{tabular}{|l|c|c|c|c|}
\hline
Particle & $N $& $B_b$& $\alpha$   & $\beta$   \\
\hline
$B^0$ & 3991 &0.337 &  16.87 & 2.63 \\
\hline
$B^+$ & 3991  &0.337 &  16.87 & 2.63 \\
\hline
$B_s^+$ & 1066 & 0.090 & 16.87 & 2.63 \\
\hline
$\Lambda_b$ & 2795 &0.236 & 16.87 & 2.63 \\
\hline
\end{tabular}
\caption{Parameters for the $b$ quark fragmentation from Kniehl et
  al.
\cite{Kniehl:2008zza}, with the normalization
rescaled to match the
fragmentation fractions in \cite{Aaij:2011jp,Aaij:2013noa}. The $\Lambda_b$ fragmentation function is
approximated by the meson fragmentation function, normalized to a fragmentation fraction $B_b=0.236$.}
\label{table:fragkkb}
\end{center}
\vspace{-0.6cm}
\end{table}

\subsection{\label{ssec:appdecay}Decay distributions}

Meson decay moments are evaluated using the decay distributions
\begin{equation}
\frac{dn}{dE_\nu} = \frac{1}{E_h}B_{h\to \nu} F_{h\to \nu}(z)
\end{equation}
where $z=E_\nu/E_h$, the fraction of the hadron energy carried by the neutrino,
and $B_{h\to \nu}$ is the branching fraction. Following ref.~\cite{Bugaev:1998bi}, we approximate
charmed meson semileptonic decay distributions as a function of
neutrino energies  by the three-body decay distribution. (See also
ref.~\cite{Lipari:1993hd}.)  This fractional energy distribution comes from evaluating the
pseudoscalar three-body semileptonic decay to a lighter pseudoscalar
meson (e.g., $D\to K\ell \nu_\ell$)
with form factor
$f_+(q^2)\simeq f_+$ approximately constant
and $f_-(q^2)\simeq 0$, keeping the mass of the meson in the final
state but neglecting the lepton mass. The distribution is the same for
$\ell=\mu,e$ and $\nu_\ell$ and is
\begin{eqnarray}
\nonumber
F_{h\to \nu_\ell} &=& \frac{1}{D(r)}\Bigl[ 6(1-2r)(1-r)^2 - 4(1-r)^3
-12 r^2(1-r)\\
&+& 12r^2y-6(1-2r)y^2 + 4y^3+12r^2\ln\bigl(\frac{1-y}{r}\bigr) ]\Bigr]\\
D(r) &=& 1-8r-12 r^2\ln r + 8 r^3 - r^4 
\ ,
\end{eqnarray}
where $r=m^2/m_h^2$ for $m$ the mass of the hadron in the final state
and $m_h$ the mass of the decaying particle.
Details for the decay in rest frame are
described in ref.~\cite{Pietschmann:1984en}, and the procedure to
convert the distribution to the frame where the decaying particle has
energy $E_h$ can be found in ref.~\cite{Gaisser:1990vg}.
We use the same form of the decay distribution for $B$ meson semileptonic
decays, and to simplify the evaluation, for $\Lambda_c$ and $\Lambda_b$ decays to $\nu_\mu$ and
$\nu_e$.

In ref.~\cite{Bugaev:1998bi}, an effective hadron mass
$\sqrt{s^{eff}_X}$ is used for charm decays to account for the
contributions of both pseudoscalar mesons, vector mesons and two
mesons in the final state. The effective hadron masses are used for
$r ={s^{eff}_X}/{m_h^2}$.
The branching fractions and effective hadron masses used in our evaluations of
the prompt fluxes are listed in table \ref{table:hbranching}.

\begin{figure}[tb]
\centering
\includegraphics[scale=0.5]{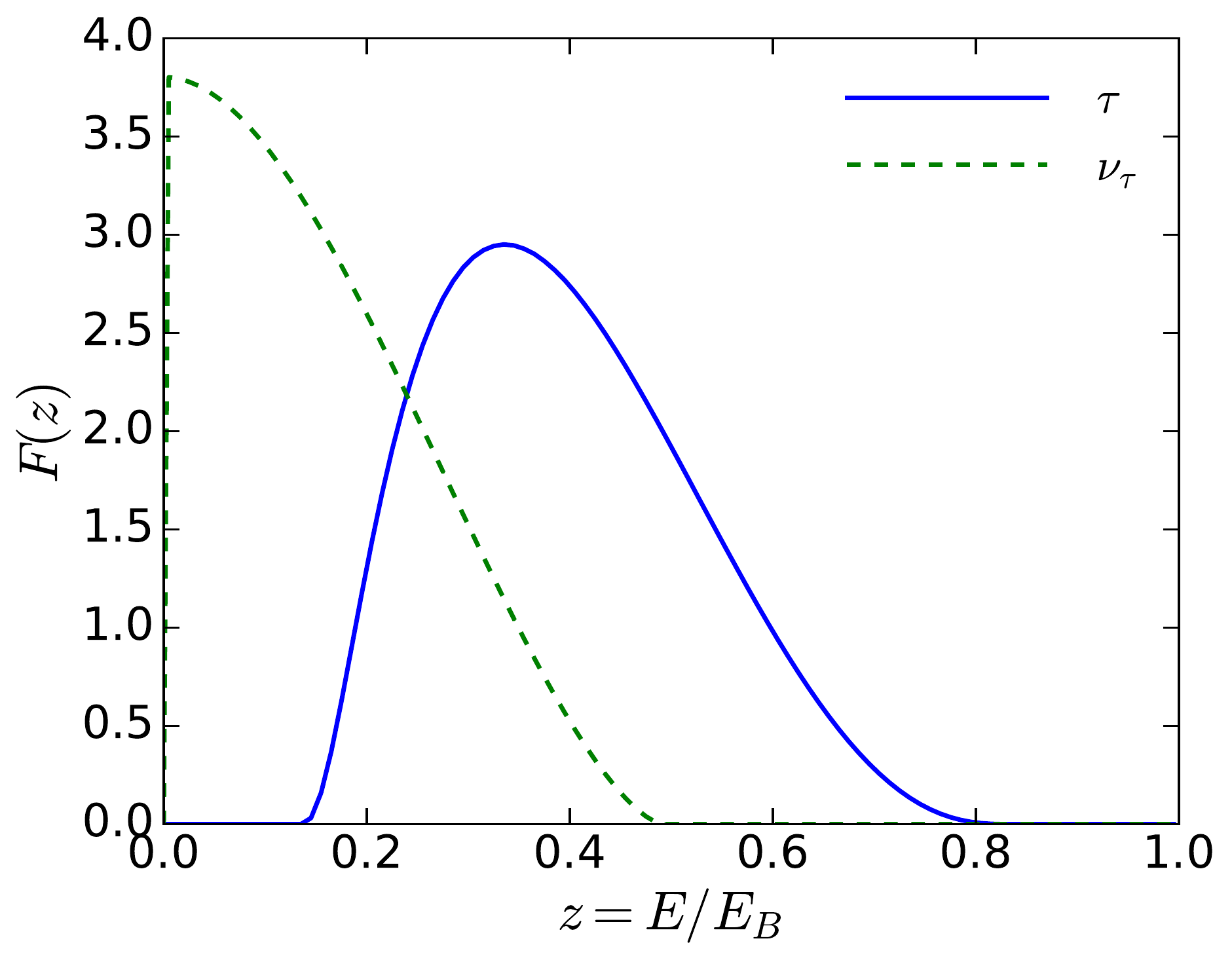}
\caption{The distribution of $\tau$ and $\nu_\tau$ in $B$ decays.}
\label{fig:dndz-b}
\end{figure}

The three body decays of $B$ mesons to $\tau\nu_\tau$ require
additional mass terms proportional to $m_\tau^2/m_B^2$. Using the same
approximations as above except for keeping the tau mass, we find
\begin{eqnarray}
\nonumber
F_{B\to \nu_\tau} &=& N_\nu\int_z^{1-r_{sum}}\, dx\,
  w(1-x,r_\pi,r_\tau )
\frac{x}{(1-x)^2}\\
&\times & [4(1-x)(1-x-r_\pi)+r_\tau(r_\tau-r_\pi-3(1-x))]
\end{eqnarray}
for the neutrino distribution, and
\begin{eqnarray}
\nonumber
F_{B\to \tau} & = & N_\tau \int_{z+r_\tau/z}
^{1+r_\tau-r_{eff}}\, dx
\Biggl(  \frac{1+r_\tau-r_\pi-x}{1+r_\tau-x}   \Biggr) ^2  \\
&\times & [4x(1-x)+5 r_\tau x - 2r_\tau^2]
\end{eqnarray}
for the tau distribution,
where $r_{sum}= (m_\tau+\sqrt{s^{eff}_X})^2/m_b^2$ and
$w(a,b,c)=[a^2+b^2+c^2-2(ab+bc+ac)]^{1/2}$. Numerically, for the
effective hadron masses in table \ref{table:hbranching}, the
normalization constants are
$N_\nu=N_\tau= 2.72\times 10^{-2}$. The kinematic limits for $y$ are
\begin{eqnarray}
z_\nu^{min} = 0  & &  \quad z_\nu^{max} = 1-r_{sum}\\
z_\tau^{min} = \sqrt{r_\tau}  & &\quad  z_\tau^{max} = 0.5
(x_{max}+(x_{max}^2-4 r_\tau)^{1/2})
\end{eqnarray}
for $x_{max} = 1+r_\tau - r_{eff} $. For reference, the $z$
distributions for $B$ decays to $\tau\nu_\tau$ are shown in
fig.~\ref{fig:dndz-b}.

\begin{table}[tb]
\begin{center}
\begin{tabular}{|l|c|c|c|}
\hline
Process & $B $ & $\sqrt{s^{eff}_X}$ [GeV] & $c\tau$ [$\mu$m]\\
\hline
$D^0\to \nu_\ell$ & 0.067 & 0.67 & 122.9\\
\hline
$D^+\to \nu_\ell$ & 0.176 & 0.63 & 311.8\\
\hline
$D_s^+\to \nu_\ell$ & 0.065 & 0.84 & 149.9\\
\hline
$\Lambda_c^+\to \nu_\ell$ & 0.045& 1.3& 59.9\\
\hline
$B^0\to \nu_\ell$ & 0.10& 2.0 & 455.7\\
\hline
$B^0\to \nu_\tau$ & 0.029& 2.0 & 455.7\\
\hline
$B^+ \to \nu_\ell$ & 0.11 & 2.0 & 491.1\\
\hline
$B^+ \to \nu_\tau$ & 0.019 & 2.0 & 491.1\\
\hline
$B_s \to \nu_\ell$ & 0.10 & 2.1 & 452.7\\
\hline
$\Lambda_b\to \nu_\ell$ & 0.11 & 2.4 & 439.5\\
\hline
\end{tabular}
\caption{ Parameters for the branching fractions and effective hadronic mass for charmed and $b$
hadron decays, following ref.~\cite{Bugaev:1998bi} for charmed meson decay distributions. }
\label{table:hbranching}
\end{center}
\vspace{-0.6cm}
\end{table}
%

The tau neutrino flux is dominated by the  $D_s\to \tau \nu_\tau$ process. The two-body decay distribution,
for $z=E_{\nu_\tau}/E_{D_s}$,
is
\begin{equation}
\label{eq:dstonu}
\frac{dn_{D_s\to \nu_\tau}}{dE_\nu}= \frac{1}{E_D}\frac{B_{D_s\to \nu_\tau \tau}}{1-r_\tau}\theta ((1-r_\tau)- z)\ ,
\end{equation}
for $r_\tau=m_\tau^2/m_{D_s}^2=0.815$ and
branching ratio $B_{D_s\to \tau \nutau}=(5.54\pm 0.24)\% $
\cite{Agashe:2014kda}.

The energy distribution of the taus in $D_s$ decays, in terms of $z_\tau\equiv E_\tau/E_{D_s}$, is
of the same form as eq.~\ref{eq:dstonu}, however, with a different range for $z_\tau$ that
gives the $\tau$ a larger fraction of $E_{D_s}$ than the $\nu_\tau$ energy fraction:
\begin{equation}
\label{eq:dstotau}
\frac{dn_{D_s\to \tau}}{dE_\tau}= \frac{1}{E_{D_s}}\frac{B(D_s\to \nu_\tau \tau)}{1-r_\tau}\theta (1 - z_\tau)\theta (z_\tau - r_\tau)\ .
\end{equation}

The energy distribution of tau neutrinos in tau decays in terms of $y\equiv E_{\nu}/E_\tau$
depends on the tau polarization. It
is
\begin{equation}
\frac{dn_{\tau \to \nu_\tau}}{dE_\nu}= \frac{1}{E_\tau}\sum_i\ B_i\, [g_0^i(y)-P_\tau g_1^i(y)]\
\end{equation}
for $\tau \to i$ decays.
The branching fractions $B_i$ and functions $g_0^i, \ g_1^i$ are listed in
Table \ref{table:taudecays}. Here, the function  $P_\tau=P_\tau(E_{D_s},E_\tau)$
is the polarization of the tau from the $D_s$ decay:
\begin{equation}
P_\tau = \frac{2 E_{D_s} r_\tau}{E_\tau (1-r_\tau)} - \frac{1+r_\tau}{1-r_\tau}\ ,
\end{equation}
in the relativistic limit. The energy distributions of $\nu_\tau$ and $\bar{\nu}_\tau$ are the same, a result of the opposite polarizations
of the $\tau$ and $\bar{\tau}$ and the opposite neutrino and antineutrino helicities.
We have approximated $P_\tau$ for taus coming from the
semi-leptonic $B$ decays with the polarization of taus coming from $D_s$ decays.

\begin{table}[h]
\begin{tabular}{|l|c|c|c|c|}
\hline
Process & $B_i$ & $g_0^i$ & $g_1^i$  & $y_{max}$ \\ \hline
$\tau\rightarrow \nu_\tau \mu \nu_\mu$ & 0.18  &  ${5/ 3} -
3y^2+{4}y^3/3$
& ${1/ 3} - 3y^2+{8}y^3/3 $  & 1\\ \hline
$\tau\rightarrow \nu_\tau e\nu_e$ & 0.18  &  ${5/ 3} -
3y^2+{4}y^3/3$
& ${1/ 3} - 3y^2+{8}y^3/3 $ & 1 \\ \hline
$\tau\rightarrow \nu_\tau \pi$ & 0.12 & ${ (1-r_\pi)^{-1}}$
& $-{(2y -1+r_\pi)/ (1-r_\pi)^2}$ &$ (1-r_\pi)$\\ \hline
$\tau\rightarrow \nu_\tau \rho$ & 0.26 & ${( 1-r_\rho)^{-1}}$
& $-(2y-1+r_\rho)(1-2r_\rho)$ & $ (1-r_\rho)$ \\
& & & $\times[ (1-r_\rho)^{2}
(1+2r_\rho)]^{-1}$ & \\ \hline
$\tau \rightarrow  \nu_\tau a_1$ & 0.13 & ${(1-r_{a_1})^{-1}}$
& $-(2y-1+r_{a_1})(1-2r_{a_1}) $ & $ (1-r_{a_1})$\\
 & & & $\times[ (1-r_{a_1})^2
( 1+2r_{a_1})]^{-1}$ & \\ \hline
\end{tabular}
\caption{Functions $g_0^i$ and $g_1^i$ and the branching fractions in the tau neutrino energy distribution
from relativistic $\tau$ decays, in terms of $y=E_\nu/E_\tau$ and $r_i=m_i^2/m_\tau^2$
for purely leptonic decays and for $i=\pi,\rho$ and $a_1$. }
\label{table:taudecays}
\end{table}

\subsection{Flux Tables}

We provide here in tables \ref{table:flux-ncteq15}, \ref{table:flux-ct14eps} and \ref{table:flux-dmkt} numerical values of the distributions in the vertical
direction for
$E^3\phi$ for $\nu_\mu+\bar{\nu}_\mu$ from $c\bar{c}$ and $b\bar{b}$ from atmospheric production by cosmic rays. Up to $E_\nu\sim 10^7$ GeV, these fluxes are isotropic. The prompt fluxes for $\ell = \nu_\mu,\ \nu_e$ and $\mu$ are equal. Table \ref{table:flux-ncteq15-nutau} shows the vertical $\nu_\tau
+\bar{\nu}_\tau$ flux evaluated using the nCTEQ15-14 PDFs.

\begin{table}[tb]
\begin{center}
\vskip 0.5in
\begin{tabular}{|c|c|c|c|}
\hline
log$_{10}$($E$/GeV) & $E^3\phi$ &  $E^3\phi$(lower) & $E^3\phi$(upper)\\
\hline

   3.00  &  2.22E-05  &  1.75E-05  &  2.62E-05  \\  \hline
   3.20  &  2.83E-05  &  2.18E-05  &  3.42E-05  \\  \hline
   3.40  &  3.58E-05  &  2.70E-05  &  4.41E-05  \\  \hline
   3.60  &  4.49E-05  &  3.33E-05  &  5.63E-05  \\  \hline
   3.80  &  5.57E-05  &  4.06E-05  &  7.07E-05  \\  \hline
   4.00  &  6.79E-05  &  4.88E-05  &  8.72E-05  \\  \hline
   4.20  &  8.09E-05  &  5.72E-05  &  1.05E-04  \\  \hline
   4.40  &  9.37E-05  &  6.53E-05  &  1.23E-04  \\  \hline
   4.60  &  1.05E-04  &  7.24E-05  &  1.40E-04  \\  \hline
   4.80  &  1.14E-04  &  7.73E-05  &  1.53E-04  \\  \hline
   5.00  &  1.18E-04  &  7.92E-05  &  1.61E-04  \\  \hline
   5.20  &  1.17E-04  &  7.76E-05  &  1.61E-04  \\  \hline
   5.40  &  1.11E-04  &  7.29E-05  &  1.53E-04  \\  \hline
   5.60  &  1.02E-04  &  6.61E-05  &  1.40E-04  \\  \hline
   5.80  &  9.15E-05  &  5.90E-05  &  1.26E-04  \\  \hline
   6.00  &  8.27E-05  &  5.30E-05  &  1.13E-04  \\  \hline
   6.20  &  7.64E-05  &  4.85E-05  &  1.04E-04  \\  \hline
   6.40  &  7.29E-05  &  4.61E-05  &  9.98E-05  \\  \hline
   6.60  &  7.30E-05  &  4.57E-05  &  1.01E-04  \\  \hline
   6.80  &  7.65E-05  &  4.75E-05  &  1.08E-04  \\  \hline
   7.00  &  8.18E-05  &  5.03E-05  &  1.18E-04  \\  \hline
   7.20  &  8.48E-05  &  5.16E-05  &  1.25E-04  \\  \hline
   7.40  &  8.43E-05  &  5.10E-05  &  1.27E-04  \\  \hline
   7.60  &  7.93E-05  &  4.78E-05  &  1.22E-04  \\  \hline
   7.80  &  6.90E-05  &  4.15E-05  &  1.07E-04  \\  \hline
   8.00  &  5.75E-05  &  3.45E-05  &  8.94E-05  \\  \hline

\end{tabular}
\end{center}
\caption{
The NLO pQCD vertical flux of $\nu_\mu+\bar{\nu}_\mu$ scaled by $E^3$
in units of Gev$^2$/cm$^2$/s/sr using the  nCTEQ15-14
PDFs with low-$x$ grids, evaluated with $m_T$ dependent renormalization and factorization
scales, with the H3p cosmic ray flux. }
\label{table:flux-ncteq15}
\end{table}%

\begin{table}[tb]
\begin{center}
\vskip 0.5in
\begin{tabular}{|c|c|c|c|}
\hline
log$_{10}$($E$/GeV) &$E^3\phi$ &  $E^3\phi$(lower) & $E^3\phi$(upper) \\
\hline
   
   3.00  &  2.34E-05  &  1.81E-05  &  2.72E-05  \\  \hline
   3.20  &  3.00E-05  &  2.26E-05  &  3.56E-05  \\  \hline
   3.40  &  3.82E-05  &  2.79E-05  &  4.63E-05  \\  \hline
   3.60  &  4.83E-05  &  3.43E-05  &  5.95E-05  \\  \hline
   3.80  &  6.02E-05  &  4.14E-05  &  7.57E-05  \\  \hline
   4.00  &  7.38E-05  &  4.90E-05  &  9.51E-05  \\  \hline
   4.20  &  8.84E-05  &  5.76E-05  &  1.15E-04  \\  \hline
   4.40  &  1.03E-04  &  6.57E-05  &  1.36E-04  \\  \hline
   4.60  &  1.16E-04  &  7.22E-05  &  1.57E-04  \\  \hline
   4.80  &  1.27E-04  &  7.59E-05  &  1.74E-04  \\  \hline
   5.00  &  1.32E-04  &  7.54E-05  &  1.87E-04  \\  \hline
   5.20  &  1.31E-04  &  7.35E-05  &  1.87E-04  \\  \hline
   5.40  &  1.25E-04  &  6.82E-05  &  1.79E-04  \\  \hline
   5.60  &  1.14E-04  &  6.08E-05  &  1.65E-04  \\  \hline
   5.80  &  1.03E-04  &  5.27E-05  &  1.51E-04  \\  \hline
   6.00  &  9.29E-05  &  4.52E-05  &  1.40E-04  \\  \hline
   6.20  &  8.57E-05  &  4.10E-05  &  1.29E-04  \\  \hline
   6.40  &  8.16E-05  &  3.81E-05  &  1.24E-04  \\  \hline
   6.60  &  8.09E-05  &  3.65E-05  &  1.26E-04  \\  \hline
   6.80  &  8.43E-05  &  3.61E-05  &  1.35E-04  \\  \hline
   7.00  &  9.06E-05  &  3.62E-05  &  1.50E-04  \\  \hline
   7.20  &  9.63E-05  &  3.74E-05  &  1.62E-04  \\  \hline
   7.40  &  9.75E-05  &  3.68E-05  &  1.67E-04  \\  \hline
   7.60  &  9.25E-05  &  3.38E-05  &  1.63E-04  \\  \hline
   7.80  &  8.09E-05  &  2.82E-05  &  1.47E-04  \\  \hline
   8.00  &  6.74E-05  &  2.19E-05  &  1.26E-04  \\  \hline
   
\end{tabular}
\end{center}

\caption{
The NLO pQCD vertical flux of $\nu_\mu+\bar{\nu}_\mu$ scaled by $E^3$
in units of Gev$^2$/cm$^2$/s/sr using the  CT-14
PDFs with EPS09, evaluated with $m_T$ dependent renormalization and factorization
scales, with the H3p cosmic ray flux. }
\label{table:flux-ct14eps}
\end{table}%

\begin{table}[tb]
\begin{center}
\vskip 0.5in
\begin{tabular}{|c|c|c|c|c|}
\hline
\multirow{2}{*}{log$_{10}$($E$/GeV)} &
\multicolumn{2} {c|}{Dipole Model} & \multicolumn{2} {c|} {$k_T$ factorization} \\
 \cline{2-5}
 &  $E^3\phi$(lower)  &  $E^3\phi$(upper)  &  $E^3\phi$(lower)  & $E^3\phi$(upper) \\
\hline
   3.00  &  1.61E-05  &  6.04E-05  &  1.17E-05  &  1.35E-05  \\  \hline
   3.20  &  2.14E-05  &  8.30E-05  &  1.55E-05  &  1.84E-05  \\  \hline
   3.40  &  2.80E-05  &  1.12E-04  &  2.01E-05  &  2.46E-05  \\  \hline
   3.60  &  3.60E-05  &  1.48E-04  &  2.55E-05  &  3.26E-05  \\  \hline
   3.80  &  4.54E-05  &  1.93E-04  &  3.18E-05  &  4.23E-05  \\  \hline
   4.00  &  5.63E-05  &  2.47E-04  &  3.86E-05  &  5.39E-05  \\  \hline
   4.20  &  6.82E-05  &  3.09E-04  &  4.55E-05  &  6.69E-05  \\  \hline
   4.40  &  8.04E-05  &  3.77E-04  &  5.19E-05  &  8.08E-05  \\  \hline
   4.60  &  9.19E-05  &  4.47E-04  &  5.72E-05  &  9.49E-05  \\  \hline
   4.80  &  1.01E-04  &  5.12E-04  &  6.06E-05  &  1.08E-04  \\  \hline
   5.00  &  1.06E-04  &  5.61E-04  &  6.13E-05  &  1.17E-04  \\  \hline
   5.20  &  1.07E-04  &  5.83E-04  &  5.91E-05  &  1.21E-04  \\  \hline
   5.40  &  1.02E-04  &  5.73E-04  &  5.46E-05  &  1.21E-04  \\  \hline
   5.60  &  9.25E-05  &  5.29E-04  &  4.85E-05  &  1.16E-04  \\  \hline
   5.80  &  8.17E-05  &  4.65E-04  &  4.22E-05  &  1.10E-04  \\  \hline
   6.00  &  7.17E-05  &  3.96E-04  &  3.66E-05  &  1.05E-04  \\  \hline
   6.20  &  6.38E-05  &  3.36E-04  &  3.21E-05  &  1.01E-04  \\  \hline
   6.40  &  5.82E-05  &  2.89E-04  &  2.89E-05  &  1.00E-04  \\  \hline
   6.60  &  5.47E-05  &  2.52E-04  &  2.68E-05  &  1.03E-04  \\  \hline
   6.80  &  5.36E-05  &  2.22E-04  &  2.62E-05  &  1.12E-04  \\  \hline
   7.00  &  5.48E-05  &  2.02E-04  &  2.66E-05  &  1.27E-04  \\  \hline
   7.20  &  5.68E-05  &  1.91E-04  &  2.69E-05  &  1.42E-04  \\  \hline
   7.40  &  5.71E-05  &  1.80E-04  &  2.63E-05  &  1.54E-04  \\  \hline
   7.60  &  5.52E-05  &  1.68E-04  &  2.48E-05  &  1.60E-04  \\  \hline
   7.80  &  5.19E-05  &  1.55E-04  &  2.32E-05  &  1.62E-04  \\  \hline
   8.00  &  4.72E-05  &  1.39E-04  &  2.17E-05  &  1.58E-04  \\  \hline
\end{tabular}
\end{center}

\caption{
The vertical flux of $\nu_\mu+\bar{\nu}_\mu$ scaled by $E^3$ from the dipole model using the  CT-14 PDFs and $k_T$ factorization
in units of Gev$^2$/cm$^2$/s/sr with the H3p cosmic ray flux. }
\label{table:flux-dmkt}
\end{table}%

\begin{table}[tb]
\begin{center}
\vskip 0.5in
\begin{tabular}{|c|c|c|c|}
\hline
log$_{10}$($E$/GeV) & $E^3\phi$ &  $E^3\phi$(lower) & $E^3\phi$(upper)\\
\hline
   3.00  &  2.04E-06  &  1.65E-06  &  2.35E-06  \\  \hline
   3.20  &  2.61E-06  &  2.06E-06  &  3.09E-06  \\  \hline
   3.40  &  3.31E-06  &  2.56E-06  &  4.00E-06  \\  \hline
   3.60  &  4.17E-06  &  3.17E-06  &  5.13E-06  \\  \hline
   3.80  &  5.19E-06  &  3.88E-06  &  6.47E-06  \\  \hline
   4.00  &  6.34E-06  &  4.68E-06  &  8.01E-06  \\  \hline
   4.20  &  7.61E-06  &  5.54E-06  &  9.71E-06  \\  \hline
   4.40  &  8.87E-06  &  6.38E-06  &  1.14E-05  \\  \hline
   4.60  &  1.01E-05  &  7.18E-06  &  1.32E-05  \\  \hline
   4.80  &  1.11E-05  &  7.78E-06  &  1.46E-05  \\  \hline
   5.00  &  1.17E-05  &  8.14E-06  &  1.56E-05  \\  \hline
   5.20  &  1.18E-05  &  8.14E-06  &  1.59E-05  \\  \hline
   5.40  &  1.15E-05  &  7.83E-06  &  1.55E-05  \\  \hline
   5.60  &  1.07E-05  &  7.23E-06  &  1.45E-05  \\  \hline
   5.80  &  9.71E-06  &  6.52E-06  &  1.31E-05  \\  \hline
   6.00  &  8.79E-06  &  5.84E-06  &  1.17E-05  \\  \hline
   6.20  &  8.06E-06  &  5.33E-06  &  1.06E-05  \\  \hline
   6.40  &  7.65E-06  &  5.03E-06  &  1.00E-05  \\  \hline
   6.60  &  7.55E-06  &  4.94E-06  &  9.93E-06  \\  \hline
   6.80  &  7.84E-06  &  5.08E-06  &  1.05E-05  \\  \hline
   7.00  &  8.42E-06  &  5.41E-06  &  1.14E-05  \\  \hline
   7.20  &  8.88E-06  &  5.64E-06  &  1.23E-05  \\  \hline
   7.40  &  8.96E-06  &  5.63E-06  &  1.26E-05  \\  \hline
   7.60  &  8.73E-06  &  5.43E-06  &  1.24E-05  \\  \hline
   7.80  &  8.03E-06  &  4.96E-06  &  1.15E-05  \\  \hline
   8.00  &  6.77E-06  &  4.18E-06  &  9.75E-06  \\  \hline
\end{tabular}
\end{center}

\caption{
The NLO pQCD vertical flux of $\nu_\tau+\bar{\nu}_\tau$ scaled by $E^3$
in units of Gev$^2$/cm$^2$/s/sr using the  nCTEQ15-14
PDFs with low-$x$ grids, evaluated with $m_T$ dependent renormalization and factorization
scales, with the H3p cosmic ray flux. }
\label{table:flux-ncteq15-nutau}
\end{table}%

\clearpage
\bibliographystyle{JHEP}
\bibliography{refs}

\end{document}